\documentclass[11pt, a4paper]{article}

\usepackage{amsthm}
\usepackage{amsmath}
\usepackage{amssymb}
\usepackage{amsfonts}
\usepackage{latexsym}
\usepackage{booktabs}
\usepackage{multirow}
\usepackage{graphicx}
\usepackage{natbib}
\usepackage[left = 2.5cm, right = 2.5cm, top = 3cm, bottom = 3cm]{geometry}
\usepackage[colorlinks,citecolor=black,urlcolor=black]{hyperref}

\newcommand{\reali}{{\rm I}\negthinspace {\rm R}}
\newcommand{\T}{\mathrm{T}}

\newcommand{\mpls}{l_{M}(\psi)}
\newcommand{\mplmc}{l_{M^*}(\psi)}

\newcommand{\ilst}{I^*_{\lambda_i\lambda_i}(\hat{\theta};\hat{\theta}_{\psi})}

\title{Monte Carlo modified profile likelihood\\in models for clustered data}

\author{Claudia Di Caterina \\ \texttt{dicaterina@stat.unipd.it} \smallskip \\
  Department of Statistical Sciences, University of Padova \\  Via
  Cesare Battisti 241, 35121 Padova, Italy \\ \bigskip \\
  Giuliana Cortese \\ \texttt{gcortese@stat.unipd.it} \smallskip \\
  Department of Statistical Sciences, University of Padova \\  Via
  Cesare Battisti 241, 35121 Padova, Italy \\ \bigskip \\
  Nicola Sartori \\ \texttt{sartori@stat.unipd.it} \smallskip \\
  Department of Statistical Sciences, University of Padova \\  Via
  Cesare Battisti 241, 35121 Padova, Italy \\ \bigskip \\
}

\begin{document}

\maketitle

\begin{abstract}\noindent
The main focus of the analysts who deal with clustered data is usually not on the clustering variables, and hence the group-specific parameters are treated as nuisance. If a fixed effects formulation is preferred and the total number of clusters is large relative to the single-group sizes, classical frequentist techniques relying on
the profile likelihood are often misleading. The use of alternative tools, such as modifications to the profile likelihood or integrated likelihoods, for making accurate inference on a parameter of interest can be complicated by the presence of nonstandard modelling and/or sampling assumptions.
We show here how to employ Monte Carlo simulation in order to approximate the modified profile likelihood in some of these unconventional frameworks. The proposed solution is widely applicable and is shown to retain the usual properties of the modified profile likelihood.
The approach is examined in two instances particularly relevant in applications, i.e. missing-data models and survival models with unspecified censoring distribution. The effectiveness of the proposed solution is validated  via simulation studies and two clinical trial applications.
\end{abstract}

\section{Introduction}
\label{sec:intro}
Clustered data, either cross-sectional or longitudinal observations which may be arranged in groups, are nowadays encountered in all applied areas. 
Their main characteristic is the unobserved heterogeneity across clusters, with the consequence that units within a cluster might be correlated. How to deal with such correlation depends strongly on the purpose of the study. When the interest is on group-specific effects and also on estimation of the intra-cluster correlation, a frequent practice is to assume a random effects model. In this setting, cluster-specific covariate effects depend on unobservable latent variables, named random effects. Such a modelling strategy leads to the so-called conditional approach. 
Alternatively, if the interest centers on comparing the response variable of units across groups, it is preferable to adopt so-called marginal models, where the clustering structure is ignored for estimation of the regression coefficients, and it is only employed to ensure correct inference on the standard errors. These specifications are usually estimated using generalized estimating equations \citep{liang86}. 
Generally, interpretation of the regression coefficients in conditional and marginal models is different. Therefore the corresponding estimates are not directly comparable (see, for instance, \citealp{leenelder2004}, and \citealp[Chapter~9]{agr15}). In this paper, we will focus mainly on  the conditional approach.

Random effects models require to assume some suitable underlying distribution for the random effects, and even their incorrelation with the covariates in the model \citep{lancaster00}. The latter unrealistic hypothesis frequently drives the decision to opt for a more flexible fixed effects approach, a choice particularly popular in the econometric literature. Fixed effects models capture the heterogeneity among clusters via the inclusion of nuisance parameters, one for every group. These are also referred to as incidental parameters  \citep{lancaster00}, since each of them appears only in the distribution of the observations in one given cluster. 
Under fixed effects models, as well as under marginal models, inferential results are free from any assumption on the probabilistic distribution regarding the dependence structure within clusters. In addition, if also relevant for the analysis, the fixed effects strategy allows to quantify the heterogeneity across groups by comparing the estimates of the cluster-specific parameters.	

The increased robustness of the fixed effects solution over the random effects one in the conditional approach is balanced by the drawback that when clusters have small to moderate size, likelihood inference is prone to suffer from the incidental parameters problem \citep{neymanscott48}. Such problem depends on the fact that the bias of the profile score function for the parameter of interest increases along with the dimension of the nuisance component in the model (see, e.g., \citealp{mccullagh90}), invalidating usual asymptotic results if the number of groups, $N$, is much larger than the single group size, $T$. Reliable inference on the parameter of interest needs thus to be carried out via alternative pseudo-likelihoods which are unaffected by this issue. 

Exact pseudo-likelihoods leading to extremely accurate conclusions are only available in specific  model classes \citep[Chapter~8]{severini00}. Correcting the profile likelihood for the presence of incidental parameters represents instead a more general strategy. Among the several adjustments found in the literature, a prominent position is held by the modified profile likelihood (MPL) of Barndorff-Nielsen (\citeyear{barndorffnielsen80}, \citeyear{barndorffnielsen83}). Specifically, \cite{sartori03} proved its inferential superiority with respect to the ordinary profile likelihood within the $(T\times N)$-asymptotic setting that characterizes clustered data with independent units.

Under the same frequentist paradigm, another possible approach to avoid the incidental parameters problem is the integrated likelihood for the component of interest \citep{sev07}, where elimination of the fixed effects is achieved by integration in an appropriate parametrization. \cite{debin15} have shown that this function is asymptotically equivalent to the MPL and enjoys analogue properties in the two-index asymptotics for clustered data.

The first formulation of the MPL involves statistical quantities which are easily obtainable only for exponential or group family models \citep[Chapters~5,~7]{pacesalvan97}. Such computational difficulties can be overcome by the approximate MPL owed to \cite{severini98} just within a limited range of statistical problems. \cite{bartetal16} use Monte Carlo simulation in order to compute Severini's version of the MPL when its exact calculation, although possible in principle, is especially tedious, given the assumed dependence structure of the data. The same complication arises under the nonstationary autoregressive model for normally distributed observations, discussed in \cite[Example~4.3]{debin15} as an example of application for the integrated likelihood. In the Supplementary material, it is possible to see how the MPL can be conveniently computed via Monte Carlo simulation even in that setting.

The aim of this paper is to extend the Monte Carlo approach 
to situations where the analytical calculation of the MPL is not only tedious, but can also be infeasible 
due to peculiar modelling and/or sampling hypotheses. We illustrate the potential usefulness of the method in two frameworks highly relevant for applications. In particular, the first considers binary regression with nonignorable missing response, while the second deals with survival data with unspecified censoring distribution. In both cases, although not directly covered by the theory in \cite{sartori03}, the usual good inferential properties of the MPL are empirically confirmed, suggesting the proposed Monte Carlo solution as the default choice in applications.

The structure of the paper is as follows. Section~\ref{sec:PL} introduces the basic notation and defines the profile and modified profile log-likelihoods in models for clustered data. The procedure for computing the MPL through Monte Carlo approximation is detailed in Section~\ref{sec:MPL}, and then its use is illustrated by means of simulations in different nonstandard contexts for grouped observations. In particular, datasets with possibly missing binary response are considered by Section~\ref{sec:MD}, whereas Section~\ref{sec:surv} is dedicated to survival data with unspecified censoring mechanism. The issues related to the calculation of the MPL are ascribable to the model complexity implied by the incompleteness of the data in the first case, and to the lack of parametric assumptions on the distribution of the censoring times in the second. In both frameworks we also consider an application to clinical trial data that illustrates the practical effectiveness of the method. Main results are summarised and commented in Section~\ref{sec:disc}, which also mentions potential developments of the present work.

\section{Profile and modified profile likelihood}
\label{sec:PL}
For clustered observations $y_{it}$ subdivided in $N$ groups of sizes $T_i$, suppose the parametric statistical model
\begin{equation}\label{stratmod2}
Y_{it}|X_{it}=x_{it}\sim p(y_{it}|x_{it}; \psi, \lambda_i)\,, \qquad i=1,\dots,N, \ \ \; t=1,\dots,T_i,
\end{equation}
which accommodates also dynamic specifications where the index $t$ runs over consecutive time periods and the temporal evolution of the dependent variable is explained by including in the $p$-dimensional vector of covariates $x_{it}$ responses previously recorded in the same cluster (see Section S2 of the Supplementary material). The global parameter is $\theta=(\psi, \lambda)$, where $\psi \in \Psi \subseteq \reali^k$ denotes the component of interest and $\lambda=(\lambda_1, \ldots, \lambda_N) \in \Lambda$ indicates the vector containing the incidental parameters. Note that, here and henceforth, in order to avoid clutter we omit the transpose symbol acting on vectors unless such an omission could result in ambiguity. 
In the following, the assumptions of balanced groups and scalar nuisance components, i.e. $T_i=T$ and dim$(\lambda_i)=1$ for each $i=1,\ldots,n$, respectively, shall be used  without loss of generality, for the sake of notational simplicity only. 

Under the hypothesis of independent groups, the log-likelihood function about $\theta$ can be expressed by
\[ l(\theta)=\sum_{i=1}^{N} l^i(\theta)=\sum_{i=1}^{N} l^i(\psi,\lambda_i)\,, \]
with $l^i(\psi,\lambda_i)=\sum_{t=1}^{T} \log p(y_{it}|x_{it}; \psi, \lambda_i)$
being the log-likelihood contribution for the $i$th cluster. Let us define the full maximum likelihood (ML) estimate for model (\ref{stratmod2}) as $\hat{\theta}=(\hat \psi, \hat \lambda)=\arg \max_{\theta} l(\theta)$. Standard likelihood inference on the parameter of interest is typically based on the profile log-likelihood
\begin{equation}\label{profclusters}
l_P(\psi)=\sum_{i=1}^{N} l^i\big(\psi,\hat{\lambda}_{i\psi}\big)=\sum_{i=1}^{N}l^i_P(\psi)\,,
\end{equation}
where $\hat{\lambda}_{i\psi}$ is the constrained ML estimate of $\lambda_i$ for fixed $\psi$ obtained, under standard regularity conditions, by equating to zero the score
\begin{equation}\label{score}
l_{\lambda_i}(\theta)=\frac{\partial l^i(\psi,\lambda_i)}{\partial \lambda_i} 
\end{equation}
and solving for $\lambda_i$ $(i=1,\dots,N)$. Given $\hat{\lambda}_\psi=(\hat{\lambda}_{1\psi},\dots,\hat{\lambda}_{N\psi})$, the full constrained ML estimate for fixed $\psi$ is denoted by $\hat{\theta}_\psi=(\psi, \hat{\lambda}_\psi)$.

The general expression taken by the logarithmic version of the MPL is
\begin{equation}\label{mpl}
l_{M}(\psi)=l_P(\psi)+M(\psi)\,,
\end{equation}
where the modification term $M(\psi)$ serves to remedy the effect 
of replacing the unknown nuisance parameter $\lambda$ with the estimate $\hat\lambda_\psi$ in the profile likelihood. Such plug-in effect typically translates in bias of the profile score function. The expression of $M(\psi)$ largely corrects this bias, making the MPL much closer to a proper likelihood \citep{mccullagh90,diciccioetal1996}.  

The independence hypothesis among clusters implies the additive form $M(\psi)=\sum_{i=1}^{N}M_i(\psi)$. 
By using  Severini's formulation of the MPL \citep{severini98},  the $i$th summand in the modification term equals
\begin{equation}\label{misev}
M_i(\psi)=\frac{1}{2}\log j_{\lambda_i\lambda_i}(\hat{\theta}_{\psi}) - \log I_{\lambda_i\lambda_i}(\hat{\theta};\hat{\theta}_{\psi})\,, \qquad i=1,\dots,N.
\end{equation}
In (\ref{misev}), $j_{\lambda_i\lambda_i}(\theta)=-\partial^2 l^i(\psi,\lambda_i)/(\partial \lambda_i\partial \lambda_i)$ is evaluated at the constrained ML estimate $\hat{\theta}_{\psi}$ and $I_{\lambda_i\lambda_i}(\hat{\theta};\hat{\theta}_{\psi})$ is an approximation of a term involving sample space derivatives in the original Barndorff-Nielsen's MPL. In particular, $I_{\lambda_i\lambda_i}(\hat{\theta};\hat{\theta}_{\psi})=
E_{\theta_0} \big\{l_{\lambda_i}(\theta_0)l_{\lambda_i}(\theta_1)\big\}\big|_{\theta_0=\hat{\theta}, \theta_1=\hat{\theta}_\psi}$ 
indicates the scalar expected value calculated with regard to the full ML estimate $\hat{\theta}$ of the product of partial score functions defined in (\ref{score}) evaluated at two different points in the parameter space, i.e. $\hat{\theta}$ and $\hat{\theta}_{\psi}$.
In contrast to the original formulation introduced by  \cite{barndorffnielsen80, barndorffnielsen83}, Severini's variant of $M(\psi)$  is computable even when the conditional probability density or mass function of $y_{it}$ given $x_{it}$ does not belong to full exponential or composite group families. 

\cite{sartori03} gives sufficient conditions under which inferences on $\psi$ conducted via the profile or the modified profile likelihood are adequate when dealing with independent  clustered data. In more detail, usual results apply in the $(T\times N)$-asymptotics for quantities based on $l_P(\psi)$ if $N=o(T)$, while it suffices that $N=o(T^3)$ to achieve reliable conclusions using the MPL in (\ref{mpl}). This explains why the employment of $l_M(\psi)$ should be preferred in the presence of highly stratified datasets where the quantity of groups is much larger than the amount of observations per group.

\section{Monte Carlo modified profile likelihood}
\label{sec:MPL}
Analytical computation of (\ref{misev}) is fairly simple in a number of widespread statistical models (see, e.g., \citealp{bellio03}, and \citealp{bellio06}). However, the expected value $I_{\lambda_i\lambda_i}(\hat{\theta};\hat{\theta}_{\psi})$ cannot be readily obtained in circumstances that demand special assumptions to correctly model the relevant aspects of the phenomenon under study. Sometimes its exact calculation is too cumbersome, sometimes infeasible.

One convenient expedient to compute Severini's MPL even when the necessary expectation is not available in closed form 
foresees to approximate $I_{\lambda_i\lambda_i}(\hat{\theta};\hat{\theta}_{\psi})$ by the following empirical quantity based on $R$ Monte Carlo replicates:
\begin{equation}\label{istar}
I^{*}_{\lambda_i\lambda_i}(\hat{\theta};\hat{\theta}_{\psi})=\dfrac{1}{R}\sum_{r=1}^Rl^{r}_{\lambda_i}(\hat{\theta})l^{r}_{\lambda_i}(\hat{\theta}_{\psi})\,,\qquad i=1,\dots,N, 
\end{equation}
where $l^{r}_{\lambda_i}(\cdot)$ is the score (\ref{score}) computed on observations $y_{it}^r$ of the $r$th sample $(r=1,\dots,R)$ randomly generated under the ML fit of model (\ref{stratmod2}), thus setting $(\psi,\lambda)=(\hat{\psi},\hat{\lambda})$. It is worth mentioning that such a strategy only requires to derive the score function $l_{\lambda_i}(\theta)$ and to simulate from the assumed distribution of the data, with no additional fitting. 
Indeed, $\hat{\theta}$ and $\hat{\theta}_\psi$ in (\ref{istar}) are the estimates derived from the observed dataset.
This makes the approximation far less expensive than a standard bootstrap from a computational standpoint. Moreover, the execution time is not particularly 
influenced by the value of $T$ and the number of replications $R$ usually does not need to exceed 500 for a reasonably accurate estimation of $\psi$, as attested by preliminary sensitivity analyses.

The principal quality of this Monte Carlo solution is its broad applicability. \cite{bartetal16} already experimented it, proving its competitiveness with econometric inferential methods in the estimation of dynamic fixed effects models for binary panel data. Here, we propose to adopt and extend the same technique in order to investigate the superiority of $\mpls$ with respect to ML procedures in alternative scenarios. To this end, we will consider regression models for missing binary data (Section~\ref{sec:MD}) and survival models for right-censored data (Section~\ref{sec:surv}), two practically relevant settings in which explicit formulation of (\ref{misev}) is either computationally involving or impossible due to the particular modelling framework. Specifically, in the survival analysis case we will
use a fitted semiparametric model for generating the Monte Carlo samples to calculate (\ref{istar}), making inference robust with respect to a possibly  misspecified censoring distribution.


For ease of reference, from now on Severini's version of the MPL obtained by making use of Monte Carlo simulation will be called Monte Carlo MPL (MCMPL). The corresponding log-likelihood function is
$\mplmc=l_P(\psi)+M^*(\psi)$, where the modification term takes the form
\begin{align*}\label{m*}
M^*(\psi)=\sum_{i=1}^{N}M^*_i(\psi)
=\sum_{i=1}^{N}\bigg\{\frac{1}{2}\log j_{\lambda_i\lambda_i}(\hat{\theta}_{\psi}) - \log I^*_{\lambda_i\lambda_i}(\hat{\theta};\hat{\theta}_{\psi})\bigg\}\,,
\end{align*}
with $I^*_{\lambda_i\lambda_i}(\hat{\theta};\hat{\theta}_{\psi})$ defined in (\ref{istar}).

Generally, both $l_P(\psi)$ and $\mplmc$ are maximized numerically.
Of course, the higher the dimension of $\psi$, the larger the number of iterations for the numerical optimization could be. Nevertheless, for fixed $R$, the overall computational effort required by $\mplmc$ increases linearly with the number of iterations and, in our experience, hardly becomes too costly with respect to $l_P(\psi)$.

\section{Regression models for missing binary data}
\label{sec:MD}
\subsection{Introduction}\label{sec:md}\noindent
The lacking registration of some data is the rule rather than the exception in quantitative research analysis. 
\cite{rubin76} developed the first basic classification of data still in use today: missing completely at random (MCAR), missing at random (MAR) and missing not at random (MNAR). 
While the first two categories are associated with an ignorable mechanism of missingness, 
when data are MNAR the probability of missing observations also depends on values that are unobserved, and thus the supposed model must 
take into account the missingness process for providing valid results \citep[Section~15.1]{littlerub02}.

Among the various approaches proposed to deal with the nonignorable incompleteness of the data, 
selection models and pattern-mixture models play a major role \citep[Chapter~18]{fitzm08}.
Let us consider independent possibly missing clustered observations $y_{it}$
and define the corresponding missingness indicators $M_{it}$ such that $M_{it}=1$ if $y_{it}$ is unobserved and $M_{it}=0$ otherwise $(i=1,\ldots,N, t=1,\dots,T)$. From a likelihood viewpoint, the joint distribution of $Y_{it}$ and $M_{it}$ in some global parametrization $\varphi$ has to be specified.
Following the classical formulation of selection models, we shall assume a marginal distribution for $Y_{it}$ depending on the parameter $\theta$ and a conditional distribution of $M_{it}$ given $Y_{it}=y_{it}$ depending on $\gamma$, so that
\begin{equation}\label{sel-mod}
p_{Y,M}(y_{it}, m_{it}
|x_{it}
;\varphi)=p_Y(y_{it}
|x_{it}
;\theta) p_{M|Y}(m_{it}|y_{it}
, x_{it}
; \gamma)\,, 
\end{equation}
with $\varphi=(\theta,\gamma)$.

Computationally speaking, in moderately complex models for incomplete datasets, maximization of the log-likelihood function incorporating all the available information is often quite an arduous task. Indeed this function, named observed log-likelihood, involves integrals or summations over the distribution of the missing data which can be hardly tractable.
It is well-known that the EM algorithm \citep{demp77} is a possibly advantageous strategy for ML estimation whenever data either are partially not observed or may be viewed as such. This method is pervasive in the literature of missing data, and many extensions to the original version have been posited 
in the years \citep[Section~8.5]{littlerub02}. 
Optimization problems in likelihood inference may also be solved
by numerical iterative algorithms different from the EM.
For example, we recall the Nelder-Mead simplex method \citep{nelder65} applied by \cite{troxel1-98} and \cite{troxel2-98} in presence of arbitrarily MNAR clustered observations, and the popular Newton-Raphson algorithm employed by both \cite{parzen06} and \cite{sinha11}.

A universally optimal solution to maximize the log-likelihood in studies with incomplete observations is impossible to prescribe. It is yet important to point out that, regardless of the selected technique, nonignorable missing-data models need to be fitted with special care because the available information may be insufficient to estimate all parameters \citep{ibra01}.

\subsection{Setup and Monte Carlo modified profile likelihood}\label{bin:md}\noindent
We focus here on possibly missing clustered binary observations.
Adopting the typical factorization of selection models defined in (\ref{sel-mod}), for independent data $y_{it}$ one can write the marginal probability mass function
\begin{equation} \label{mod-bern}
Y_{it}
\sim Bern(\pi_{it}), \; \pi_{it}=\pi_{it}(\theta)=F(\lambda_i+\beta^\T x_{it})\,, \; i=1,\dots,N,\ t=1,\dots,T,
\end{equation} 
with $F(\cdot)$ a suitable cumulative distribution function (CDF) and $\beta=(\beta_1,\ldots,\beta_p)$ vector of regression parameters, whereas
the conditional model for the missingness indicator introduced in Section~\ref{sec:md} may be expressed by
\begin{equation}\label{mod-m}
M_{it}|Y_{it}=y_{it}
\sim Bern(\zeta_{it})\,,  \qquad i=1,\dots,N,\ \ \; t=1,\dots,T,
\end{equation}
where $\zeta_{it}\in (0,1)$. Notice that, since covariates are considered given and entirely observed, in writing the two distributions we neglect the conditioning on the $p$-vector $x_{it}$ for succinctness.

A general formulation for $\zeta_{it}$ is
\begin{equation}\label{pit}
\zeta_{it}=\zeta_{it}(\gamma)=P(M_{it}=1|Y_{it}=y_{it}
)=G(
\gamma_1^\T x_{it}+\gamma_2y_{it})\,,
\end{equation}
where $G(\cdot)$ is a CDF and $\gamma_1=(\gamma_{11}, \ldots, \gamma_{1p})$.
The parameter of primary interest in the joint model described by (\ref{mod-bern})--(\ref{pit}) is the regression coefficient $\beta \in\reali^p$, and the incidental parameters are grouped in $\lambda=(\lambda_1, \dots,\lambda_N)\in \reali^N$, so that $\theta=(\beta,\lambda)\in \Theta \subseteq \reali^{p+N}$. In the binary regression with the indicator of missingness as response, the coefficients are $\gamma=(\gamma_1,\gamma_2)\in \Gamma \subseteq \reali^{p+1}$, thus the overall parameter is given by $\varphi=(\theta,\gamma) \in \varPhi \subseteq \reali^{2p+N+1}$. To simplify reference, let us gather the parameters common to all clusters in one vector and denote it by $\psi=(\beta, \gamma) \in \reali^{2p+1}$. We finally stress that expression (\ref{pit}) does not contemplate the presence of an intercept, either common or cluster-specific, in the model for $M_{it}$ in order to avoid identifiability issues during the fitting phase (see, e.g., the discussion in \citealp[Section~6]{parzen06}).

According to the assumption about the missing-data mechanism, it is possible to identify different relations between the probability of missingness and the variables in the study. Such relations, in their turn, translate into constraints on the model parameters \citep{parzen06}. 
Here, since covariates are nonrandom, from specification (\ref{pit}) follows that data can be either MCAR, when $\gamma_2=0$, or MNAR otherwise \citep{bak95}. 


Models like (\ref{mod-bern}) for complete datasets were already investigated in \cite{bellio03}, who showed how to analytically derive Severini's MPL in order to consistently estimate $\beta$ when $N$ is much larger than $T$. The presence of missing values, however, creates trouble in the explicit calculation of the adjustment term. The expectation therein should be evaluated with regard to the joint distribution $p_{Y,M}(y_{it}, m_{it};\hat{\varphi})$, taking also the missing-data mechanism into account, but the correct way of doing so is not without ambiguity. More specifically, in the light of the arguments made by \cite{ken98}, one expects to be allowed to neglect the missingness process only when data are MCAR. 

Consider now the most general MNAR framework and, for the sake of clarity, denote by $y^{obs}$ the vector of observed entries in the dataset $y=(y_{it})$ and by $y^{mis}$ the vector of remaining missing elements. As highlighted in \cite[Section~6.2]{littlerub02}, the actual data consist of $y^{obs}$ and of the vector containing the indicators of missingness, $m=(m_{it})$. The observed log-likelihood about $\varphi$ is obtained by summing over all possible values of $y^{mis}$ the joint probability mass function of $Y=(Y^{obs}, Y^{mis})$ and $M$, so that
\[
l(\varphi)=\sum_{i=1}^N l^i(\varphi)=
\log \bigg\{\sum_{y^{mis}} p_Y\big(y^{obs}, y^{mis};\theta\big)p_{M|Y}\big(m|y^{obs}, y^{mis};\gamma\big)\bigg\}
\]
has the global ML estimate $\hat{\varphi}$ as maximizer and can be decomposed in the $N$
cluster-specific contributions taking the form
\begin{align}\label{obs-ll}
l^i(\varphi)=\sum_{t=1}^{T} \bigg[ &m_{it} \log \big\{ (1-\pi_{it}) \zeta^0_{it}+\pi_{it}\zeta^1_{it}\big\} \\ 
&+(1-m_{it}) \big\{ y_{it}\log\pi_{it}
+ (1-y_{it})\log(1-\pi_{it})+\log(1-\zeta_{it}) \big\} \bigg]\,,\nonumber
\end{align}
where $\zeta^0_{it}=G(\gamma_1^\T x_{it})$ and $\zeta^1_{it}=G(\gamma_1^\T x_{it}+\gamma_2)$ $(i=1,\dots,N)$.
The score function (\ref{score}) in the global parametrization $\varphi$ equals here
\begin{align}\label{obs-score}
l_{\lambda_i}(\varphi)=\sum_{t=1}^{T} \bigg\{ m_{it} 
\log \dfrac{f_{it}(\zeta^1_{it}-\zeta^0_{it})}{\pi_{it}\zeta^1_{it}+(1-\pi_{it})\zeta^0_{it}}+ (1-m_{it})\dfrac{(y_{it}-\pi_{it})f_{it}}{\pi_{it}(1-\pi_{it})} \bigg\}\,,
\end{align}
where $f_{it}=f_{it}(\theta)=\partial F(\lambda_i+\beta^\T x_{it})/\partial \lambda_i$.
Then, differentiating one more time with respect to $\lambda_i$ and changing the sign of the obtained derivative lead to
\begin{align}\label{bern:j}
j_{\lambda_i\lambda_i}(\varphi)=\sum_{t=1}^{T} \bigg[ &m_{it} \bigg\{ \dfrac{f'_{it}}{f_{it}}- \dfrac{(\zeta^1_{it}-\zeta^0_{it})f_{it} }{ \nonumber \pi_{it}\zeta^1_{it}+(1-\pi_{it})\zeta^0_{it}}\bigg\} \\
&+ (1-m_{it}) 
(y_{it}-\pi_{it})\bigg\{ \dfrac{f'_{it}-f_{it}^2}{\pi_{it}(1-\pi_{it})} - \dfrac{f_{it}(1-2\pi_{it})}{\pi_{it}^2(1-\pi_{it})^2}
\bigg\}\bigg]\,,
\end{align}
where $f'_{it}=f'_{it}(\theta)=\partial^2 F(\lambda_i+\beta^\T x_{it})/\partial \lambda^2_i$. The constrained estimate $\hat{\lambda}_{i\psi}$ which solves the equation $l_{\lambda_i}(\varphi)=0$ can be found numerically and
its substitution for $\lambda_i$ $(i=1,\dots,N)$ in (\ref{obs-ll}) permits to obtain the MNAR profile log-likelihood,  $l_P(\psi)=\sum_{i=1}^Nl^i_P(\psi)$. Defined $\hat{\varphi}_{\psi}=(\psi, \hat{\lambda}_\psi)$, the same replacement in formula (\ref{bern:j}) gives instead $j_{\lambda_i\lambda_i}(\hat{\varphi}_{\psi})$.

Now computing $I_{\lambda_i\lambda_i}(\hat{\varphi};\hat{\varphi}_{\psi})=E_{\varphi_0}\big\{l_{\lambda_i}(\varphi_0) l_{\lambda_i}(\varphi_1)\big\}\big|_{\varphi_0=\hat{\varphi}, \varphi_1=\hat{\varphi}_\psi}$ over the unconditional sampling distribution, using the terminology of \cite{ken98}, is not obvious. Indeed, the joint distribution of $(Y_{it}, M_{it})$ was not specified directly, but divided in the two factors (\ref{mod-bern}) and (\ref{mod-m}). The mean of the product of scores should then be calculated with respect firstly to $p_{M|Y}(m_{it}|y_{it};\hat\gamma)$ and secondly to $p_Y(y_{it};\hat\theta)$, with sufficiently intricate computational steps.
Viceversa, the Monte Carlo solution presented in Section~\ref{sec:MPL} may be applied quite plainly. Particularly, the approximation (\ref{istar}) in the MNAR case takes the form
\begin{equation} \label{i-MNAR}
I^{*}_{\lambda_i\lambda_i}(\hat{\varphi};\hat{\varphi}_{\psi})=\dfrac{1}{R}\sum_{r=1}^Rl^{r}_{\lambda_i}(\hat{\varphi})l^{r}_{\lambda_i}(\hat{\varphi}_{\psi})\,,\qquad i=1,\dots,N, 
\end{equation}
where $l^{r}_{\lambda_i}(\cdot)$ is the score (\ref{obs-score}) of the $r$th partially observed sample $y_{it}^{r}$ $(r=1,\dots,R)$ obtained in two stages: first, a complete dataset $y_{it}^{r, C}$ is simulated under model (\ref{mod-bern}) with $\theta=\hat{\theta}$ and second, some entries in this dataset are deleted and considered missing according to the specification (\ref{mod-m}) with MNAR probability $\zeta_{it}=\zeta_{it}(\hat{\gamma})=G(\hat{\gamma}_1^\T x_{it}+\hat{\gamma}_2 y^{r, C}_{it})$. Note that $\hat{\psi}=(\hat{\theta},\hat{\gamma})$ is the global maximizer of the MNAR profile log-likelihood $l_P(\psi)$ which also takes the missingness process into consideration. Therefore, the average of score products over the $R$ incomplete samples properly estimates the unconditional expectation required. 

Before proceeding, it seems worthwhile making a few more comments about the general formula (\ref{obs-ll}). Supposing an ignorable MCAR missing-data mechanism by imposing $\gamma_2=0$ in (\ref{pit}) yields clearly to
$\zeta^0_{it}=\zeta^1_{it}=\zeta_{it}=G(\gamma_1^\T x_{it})$,
and hence (\ref{obs-ll}) simplifies to
\begin{align*}
l^i(\varphi)\!=\!\sum_{t=1}^{T}\! \big[ m_{it}\! \log \zeta_{it} + (1\!-\!m_{it}) \big\{ y_{it}\!\log\pi_{it}
+ (1\!-\!y_{it})\!\log(1\!-\!\pi_{it})+\log(1\!-\!\zeta_{it}) \big\} \big]\,.
\end{align*}
Since our interest is only on the parameter $\beta$, and $\zeta_{it}$ does not carry any useful information about it, we can rely on the equivalent function
\begin{align}\label{loglikMCAR}
l^i(\theta)=
\sum_{t :\, y_{it}\in y^{obs}} \big\{y_{it}\log\pi_{it}
+ (1-y_{it})\log(1-\pi_{it})\big\}\,,
\end{align}
which is the ordinary group-related log-likelihood in binary regression computed only on the recorded data. Indeed, when the missingness mechanism is MCAR, a complete-case analysis discarding units with missing values is unbiased, as the wholly observed cases are basically a random sample from the reference population \citep[Section~3.2]{littlerub02}. For this specific model, it is also fully efficient because $\theta$ and $\gamma$ are distinct, provided that the full parameter space is $\varPhi= \Theta \times \Gamma$ \citep[p.~120]{littlerub02}. This means that likelihood inference can be conducted disregarding the process which generates the missing observations. As a major implication for our study, the expected value involved in Severini's MPL may be derived from the conditional distribution of $Y_{it}$ given $M_{it}=0$. Specifically, it can be shown \citep{bellio03} that such expectation has the closed-form expression
\begin{equation}\label{i-an}
I_{\lambda_i\lambda_i}(\hat{\theta};\hat{\theta}_{\beta})= \sum_{t :\, y_{it}\in y^{obs}}
\dfrac{f_{it}\big(\hat{\theta}_{\beta}\big)f_{it}\big(\hat{\theta}\big)}{\big\{1-\pi_{it}\big(\hat{\theta}_{\beta}\big)\big\}\pi_{it}\big(\hat{\theta}_{\beta}\big)}\,,\qquad i=1,\dots,N, 
\end{equation}
where estimates $\hat{\theta}=(\hat{\beta}, \hat{\lambda})$ and $\hat{\theta}_\beta=(\beta, \hat{\lambda}_\beta)$ descend from ordinary ML inference on the parameter of interest $\beta$ via the MCAR profile log-likelihood $l_P(\beta)$ based on (\ref{loglikMCAR}).
Furthermore, inasmuch as under the hypothesis of ignorable missingness it is possible to use the function $l(\theta)$ with components like that in (\ref{loglikMCAR}), the general Monte Carlo approximation reported in (\ref{i-MNAR}) admits to be reformulated in the MCAR case as
\begin{align}\label{ist-MCAR}
I^{*}_{\lambda_i\lambda_i}(\hat{\theta};\hat{\theta}_{\beta})=\dfrac{1}{R}\sum_{r=1}^Rl^{r}_{\lambda_i}(\hat{\theta})l^{r}_{\lambda_i}(\hat{\theta}_{\beta})\,, \qquad i=1,\dots,N, 
\end{align}
where $l^r_{\lambda_i}(\theta)=\sum_{t :\, y_{it}\in y^{obs}}(y^r_{it}-\pi_{it})f_{it}/\{\pi_{it}(1-\pi_{it})\}$ is the score of the incomplete sample $y^r_{it}$ simulated by the two-step procedure above, but with an important difference: now $\hat{\theta}$ results from the maximization of $l(\theta)$, while $\hat{\gamma}=\hat{\gamma}_1$ is obtained by a separate ML fit of the binary regression based on (\ref{pit}) subject to the constraint $\gamma_2=0$, with the missingness indicator as dependent variable and the vector of covariates $x_{it}$ as unique predictor.

Below, the utility of Monte Carlo approximation in the presence of incomplete data will be evaluated through simulation experiments referring to binary regression with different missingness processes. Specifically, objects of comparison shall be the unadjusted profile log-likelihood, either the MCAR $l_P(\beta)$ or the MNAR $l_P(\psi)$, the modification proposed by Severini $l_{M}(\beta)$ that ignores the missing values and is analytically computed by formula (\ref{i-an}), and the MCMPL
that accounts for some presumed missingness mechanism.
In order to avoid confusion, its logarithmic MCAR variant employing the estimate (\ref{ist-MCAR}) will be denoted by $l_{M^*}(\beta)$, whereas $l_{M^*}(\psi)$ shall indicate the MNAR MCMPL with habitual expectation approximated by (\ref{i-MNAR}).

\subsection{Logistic regression: simulation studies}\label{sec:logsimu}
\noindent
The following analyses are performed supposing a logistic link between the mean of the response and the predictors, meaning $F(\cdot)=\mathrm{logit}^{-1}(\cdot)$ in model (\ref{mod-bern}), along with $G(\cdot)=\mathrm{logit}^{-1}(\cdot)$ in the expression for the probability of missingness (\ref{pit}), where logit$^{-1}(\cdot)$ denotes the CDF of the logistic random variable. Pairing these assumptions with that of an MCAR mechanism brings about the equality
\begin{align*}
I_{\lambda_i\lambda_i}(\hat{\theta};\hat{\theta}_{\beta})
= \sum_{t :\, y_{it}\in y^{obs}} \big[1- \mathrm{logit}^{-1}\big(\hat{\lambda}_i+\hat{\beta}^\T x_{it}\big)\big]\,, \qquad i=1,\dots,N, \nonumber
\end{align*}
whose right-hand side does not depend on the parameter of interest. 
Hence the only part of Severini's modification term relevant to estimating $\beta$ is $\log\! |j_{\lambda\lambda}\big(\!\hat{\theta}_\beta\!\big)\!|/2$, and one can write
\begin{align}\label{msev-MCARlog}
M(\beta)
=\frac{1}{2}\sum_{i=1}^{N} \log \bigg[\sum_{t :\, y_{it}\in y^{obs}} \mathrm{logit}^{-1}(\hat{\lambda}_{i\beta}+\beta^\T x_{it})\big\{1-\mathrm{logit}^{-1}\big(\hat{\lambda}_{i\beta}+\beta^\T x_{it}\big)\big\} \bigg]\,.
\end{align}
As the single cluster contribution to the profile log-likelihood $l_P(\beta)$ equals (\ref{loglikMCAR}) with $\pi_{it}$ replaced by $\mathrm{logit}^{-1}(\lambda_i + \beta^\T x _{it})$, it is simple to show that in such a setting the score related to the $i$th incidental parameter equals
\begin{align*}
l_{\lambda_i}(\theta)&
=\sum_{t :\, y_{it}\in y^{obs}}
\big\{y_{it}-\mathrm{logit}^{-1}(\lambda_{i}+\beta^\T x_{it})\big\}\,
, \qquad i=1,\dots,N, 
\end{align*}
thus the expression of the MCAR Monte Carlo estimate $I^{*}_{\lambda_i\lambda_i}(\hat{\theta};\hat{\theta}_{\beta})$ follows immediately from the previous formula and (\ref{ist-MCAR}). Loosely speaking, if observations are MCAR, $l_{M}(\beta)$ takes the same form as in general logistic regression for clustered data with no missing values, yet is computed only on the complete units. Its numerical maximization can then be automatically implemented in the R software \citep{R} exploiting the code of the current version of the package \texttt{panelMPL} \citep{panelmpl}, which can handle binary regression with logit or probit links.


For the reasons discussed above, one analytical formulation of Severini's MPL is not immediately obtainable when missingness of the data is hypothesized to be nonignorable. On the contrary, $M^*(\psi)$ can be calculated via Monte Carlo simulation through (\ref{i-MNAR}) simply by recalling that in expressions (\ref{obs-score}) and (\ref{bern:j}) one has
\begin{eqnarray*}
	f_{it}&=&\big\{1-\mathrm{logit}^{-1}(\lambda_{i}+\beta^\T x_{it})\big\}^2\,, \\
	f'_{it}&=&-2\,\mathrm{logit}^{-1}(\lambda_{i}+\beta^\T x_{it})\big\{1-\mathrm{logit}^{-1}(\lambda_{i}+\beta^\T x_{it})\big\}\,.
\end{eqnarray*}

In the MNAR scenario,
the functions $l_P(\psi)$ and $l_{M^*}(\psi)$ are optimized numerically using a quasi-Newton method (in the R  function \texttt{nlminb}). 
Standard errors of the parameters' estimates are calculated using the second numerical derivative of the functions at their maxima.
Notice that in the MNAR case the argument $\psi=(\beta, \gamma)$ of the objective functions to be optimized has dimension equal to $2p+1$, whereas in the MCAR case the argument $\beta$ is only $p$-dimensional. The higher complexity in the maximization problem is reflected by longer execution times and by some numerical instabilities, especially due to the estimation of $\gamma$ and its variance, which however are not of direct interest. 

It is worth recalling that, as is common practice for binary longitudinal regression, the optimization stage needs to be anticipated by the omission of non-informative groups \citep{bellio03} from the sample under analysis. In missing-data situations, whatever the supposed mechanism, the clusters which cannot contribute to estimate $\beta$ are those with $y^{obs}_{it}=0$ or $y^{obs}_{it}=1$ for every $t=1,\dots,T$ and those which are totally unobserved, i.e. where $y_{it}=y^{mis}_{it}$ for each $t=1,\dots,T$ $(i=1,\dots,N)$.

\begin{table}[t]\centering
	\caption{Inference on $\beta=1$ in the logistic regression for MCAR longitudinal data. The compared functions are the MCAR profile log-likelihood $l_P(\beta)$, Severini's exact MCAR MPL $l_{M}(\beta)$, and the MCAR MCMPL $l_{M^*}(\beta)$ computed with $R=500$. Results based on a simulation study with 2000 trials.}
	\label{tab:MCARlog}
	\medskip	
	\begin{tabular}{cccccccccc}
		\hline
		$N$ & $T$ & Method & B & MB & SD & RMSE & MAE & SE/SD & 0.95 CI\\ 
		\hline
		50 & 4 & $l_P(\beta)$ & 0.827 & 0.677 & 0.929 & 1.244 & 0.693 & 0.693 & 0.789 \\
		& & $l_{M}(\beta)$ & 0.193 & 0.160 & 0.482 & 0.519 & 0.323 & 0.979 & 0.965 \\
		& & $l_{M^*}(\beta)$ & 0.194 & 0.161 & 0.481 & 0.519 & 0.323 & 0.980 & 0.965 \\ 
		
		& 6 & $l_P(\beta)$ & 0.450 & 0.409 & 0.537 & 0.701 & 0.437 & 0.767 & 0.825 \\ 
		& & $l_{M}(\beta)$ & 0.099 & 0.084 & 0.364 & 0.377 & 0.237 & 0.942 & 0.953 \\
		& & $l_{M^*}(\beta)$ & 0.101 & 0.085 & 0.365 & 0.378 & 0.237 & 0.941 & 0.953 \\
		
		& 10 & $l_P(\beta)$ & 0.242 & 0.215 & 0.309 & 0.393 & 0.244 & 0.851 & 0.848 \\
		& & $l_{M}(\beta)$ & 0.049 & 0.031 & 0.250 & 0.255 & 0.165 & 0.947 & 0.946 \\
		& & $l_{M^*}(\beta)$ & 0.050 & 0.031 & 0.250 & 0.255 & 0.164 & 0.947 & 0.945 \\
		\hline
		100 & 4 & $l_P(\beta)$ & 0.682 & 0.615 & 0.584 & 0.898 & 0.618 & 0.730 & 0.663 \\
		& & $l_{M}(\beta)$ & 0.136 & 0.118 & 0.331 & 0.358 & 0.223 & 0.979 & 0.946 \\
		& & $l_{M^*}(\beta)$ & 0.137 & 0.119 & 0.331 & 0.358 & 0.223 & 0.980 & 0.948 \\
		
		& 6 & $l_P(\beta)$ & 0.428 & 0.400 & 0.355 & 0.556 & 0.404 & 0.814 & 0.707 \\
		& & $l_{M}(\beta)$ & 0.091 & 0.078 & 0.249 & 0.265 & 0.164 & 0.976 & 0.948 \\
		& & $l_{M^*}(\beta)$ & 0.092 & 0.080 & 0.249 & 0.266 & 0.164 & 0.975 & 0.948 \\
		
		& 10 & $l_P(\beta)$ & 0.231 & 0.221 & 0.214 & 0.314 & 0.226 & 0.879 & 0.765 \\
		& & $l_{M}(\beta)$ & 0.042 & 0.034 & 0.174 & 0.179 & 0.119 & 0.975 & 0.949 \\
		& & $l_{M^*}(\beta)$ & 0.042 & 0.034 & 0.174 & 0.179 & 0.118 & 0.975 & 0.949 \\
		\hline
		250 & 4 & $l_P(\beta)$ & 0.619 & 0.597 & 0.351 & 0.712 & 0.597 & 0.757 & 0.390 \\
		& & $l_{M}(\beta)$ & 0.117 & 0.111 & 0.210 & 0.241 & 0.159 & 0.980 & 0.924 \\ 
		& & $l_{M^*}(\beta)$ & 0.118 & 0.111 & 0.211 & 0.241 & 0.160 & 0.979 & 0.923 \\
		
		& 6 & $l_P(\beta)$ & 0.388 & 0.384 & 0.220 & 0.446 & 0.384 & 0.821 & 0.441 \\
		& & $l_{M}(\beta)$ & 0.068 & 0.067 & 0.157 & 0.171 & 0.113 & 0.973 & 0.937 \\ 
		& & $l_{M^*}(\beta)$ & 0.069 & 0.067 & 0.157 & 0.171 & 0.113 & 0.972 & 0.935 \\
		
		& 10 & $l_P(\beta)$ & 0.215 & 0.213 & 0.130 & 0.251 & 0.213 & 0.905 & 0.564 \\ 
		& & $l_{M}(\beta)$ & 0.029 & 0.028 & 0.106 & 0.110 & 0.072 & 1.004 & 0.948 \\ 
		& & $l_{M^*}(\beta)$ & 0.029 & 0.028 & 0.106 & 0.110 & 0.072 & 1.003 & 0.947 \\
		\hline
	\end{tabular}
	
\end{table}

\begin{table}[h!]\centering
	\caption{Inference on $\beta=1$ in the logistic regression for MCAR longitudinal data. The compared methods are the MNAR profile log-likelihood $l_P(\psi)$, the MNAR MCMPL $l_{M^*}(\psi)$ computed with $R=500$ and GEE. Results based on a simulation study with 2000 trials.}
	\label{tab:MCARlog2}
	\medskip	
	\begin{tabular}{cccccccccc}
		\hline
		$N$ & $T$ &Method & B & MB & SD & RMSE & MAE & SE/SD & 0.95 CI\\ 
		\hline
		50 & 4 & $l_P(\psi)$ & 0.582 & 0.446 & 1.035 & 1.187 & 0.655 & 0.661 & 0.823 \\ 
		&  & $l_{M^*}(\psi)$ & 0.008 & 0.002 & 0.619 & 0.619 & 0.417 & 0.895 & 0.914 \\ 
		&  & GEE & 0.032 & 0.023 & 0.306 & 0.308 & 0.205 & 1.010 & 0.961 \\
		
		& 6 & $l_P(\psi)$ & 0.346 & 0.310 & 0.583 & 0.677 & 0.415 & 0.752 & 0.862 \\ 
		&  & $l_{M^*}(\psi)$ & 0.008 & 0.014 & 0.430 & 0.430 & 0.267 & 0.912 & 0.939 \\ 
		&  & GEE & 0.015 & 0.009 & 0.259 & 0.259 & 0.171 & 0.980 & 0.949 \\
		
		& 10 & $l_P(\psi)$ & 0.213 & 0.193 & 0.333 & 0.395 & 0.249 & 0.840 & 0.862 \\ 
		&  & $l_{M^*}(\psi)$ & 0.030 & 0.019 & 0.272 & 0.273 & 0.179 & 0.958 & 0.948 \\ 
		&  & GEE & 0.007 & 0.006 & 0.193 & 0.193 & 0.128 & 0.994 & 0.950 \\
		
		\hline
		100 & 4 & $l_P(\psi)$ & 0.458 & 0.411 & 0.663 & 0.805 & 0.501 & 0.703 & 0.783 \\ 
		&  & $l_{M^*}(\psi)$ & -0.059 & -0.042 & 0.443 & 0.447 & 0.285 & 0.902 & 0.911 \\ 
		&  & GEE & 0.008 & -0.000 & 0.218 & 0.218 & 0.146 & 0.995 & 0.951 \\
		
		& 6 & $l_P(\psi)$ & 0.333 & 0.315 & 0.386 & 0.510 & 0.340 & 0.798 & 0.796 \\ 
		&  & $l_{M^*}(\psi)$ & 0.016 & 0.015 & 0.289 & 0.289 & 0.178 & 0.965 & 0.953 \\ 
		&  & GEE & 0.005 & 0.000 & 0.176 & 0.176 & 0.119 & 1.018 & 0.956 \\ 
		
		& 10& $l_P(\psi)$ & 0.204 & 0.197 & 0.230 & 0.307 & 0.214 & 0.868 & 0.809 \\ 
		&  & $l_{M^*}(\psi)$ & 0.024 & 0.019 & 0.188 & 0.190 & 0.126 & 0.985 & 0.949 \\ 
		&  & GEE & 0.001 & -0.008 & 0.137 & 0.137 & 0.092 & 0.995 & 0.954 \\
		
		\hline
		250 & 4 & $l_P(\psi)$ & 0.389 & 0.375 & 0.405 & 0.562 & 0.397 & 0.724 & 0.688 \\ 
		&  & $l_{M^*}(\psi)$ & -0.091 & -0.064 & 0.311 & 0.325 & 0.189 & 0.878 & 0.909 \\ 
		&  & GEE & -0.001 & -0.005 & 0.136 & 0.136 & 0.092 & 1.002 & 0.950 \\
		
		& 6 & $l_P(\psi)$ & 0.287 & 0.286 & 0.238 & 0.373 & 0.289 & 0.809 & 0.661 \\ 
		&  & $l_{M^*}(\psi)$ & -0.009 & -0.004 & 0.179 & 0.179 & 0.123 & 0.982 & 0.956 \\ 
		&  & GEE & -0.006 & -0.010 & 0.110 & 0.111 & 0.075 & 1.005 & 0.951 \\
		
		& 10 & $l_P(\psi)$ & 0.188 & 0.187 & 0.142 & 0.236 & 0.188 & 0.882 & 0.673 \\ 
		&  & $l_{M^*}(\psi)$ & 0.012 & 0.012 & 0.117 & 0.117 & 0.078 & 0.998 & 0.947 \\ 
		&  & GEE & -0.009 & -0.010 & 0.084 & 0.085 & 0.058 & 1.011 & 0.951 \\
		\hline
	\end{tabular}
	
\end{table}

\begin{table}[h!]\centering
	\caption{Inference on $\beta=1$ in the logistic regression for MNAR longitudinal data. The compared methods are the MNAR profile log-likelihood $l_P(\psi)$, Severini's exact MCAR MPL $l_{M}(\beta)$, the MNAR MCMPL $l_{M^*}(\psi)$ computed with $R=500$ and GEE. Results based on a simulation study with 2000 trials.}
	\label{tab:MNARlog}
	\medskip	
	\begin{tabular}{cccccccccc}
		\hline
		$N$ & $T$ &Method & B & MB & SD & RMSE & MAE & SE/SD & 0.95 CI\\ 
		\hline
		50 & 4 & $l_P(\psi)$ & 0.327 & 0.252 & 0.949 & 1.004 & 0.566 & 0.727 & 0.915 \\ 
		& & $l_{M}(\beta)$ & -0.061 & -0.096 & 0.571 & 0.574 & 0.386 & 0.954 & 0.944 \\ 
		& & $\mplmc$ & -0.097 & -0.104 & 0.595 & 0.603 & 0.386 & 0.947 & 0.941 \\ 
		& & GEE & -0.241 & -0.250 & 0.342 & 0.418 & 0.303 & 1.036 & 0.883 \\ 
		
		& 6 & $l_P(\psi)$ & 0.227 & 0.183 & 0.547 & 0.592 & 0.364 & 0.829 & 0.922 \\ 
		& & $l_{M}(\beta)$ & -0.160 & -0.193 & 0.406 & 0.436 & 0.308 & 0.959 & 0.904 \\ 
		& & $\mplmc$ & -0.047 & -0.067 & 0.406 & 0.408 & 0.265 & 0.977 & 0.953 \\ 
		& & GEE & -0.245 & -0.256 & 0.276 & 0.369 & 0.278 & 1.043 & 0.847 \\ 
		
		& 10 & $l_P(\psi)$ & 0.150 & 0.136 & 0.340 & 0.371 & 0.233 & 0.884 & 0.905 \\ 
		& & $l_{M}(\beta)$ & -0.217 & -0.230 & 0.275 & 0.350 & 0.267 & 0.974 & 0.840 \\ 
		& & $\mplmc$ & -0.016 & -0.024 & 0.281 & 0.281 & 0.188 & 0.981 & 0.946 \\ 
		& & GEE & -0.255 & -0.258 & 0.212 & 0.332 & 0.266 & 1.040 & 0.773 \\
		
		\hline
		100 & 4 & $l_P(\psi)$ & 0.295 & 0.270 & 0.605 & 0.673 & 0.419 & 0.767 & 0.886 \\ 
		& & $l_{M}(\beta)$ & -0.095 & -0.107 & 0.390 & 0.402 & 0.265 & 0.942 & 0.915 \\ 
		& & $\mplmc$ & -0.094 & -0.092 & 0.395 & 0.406 & 0.259 & 0.976 & 0.938 \\ 
		& & GEE & -0.250 & -0.261 & 0.229 & 0.339 & 0.268 & 1.075 & 0.823 \\
		
		& 6 & $l_P(\psi)$ & 0.217 & 0.192 & 0.378 & 0.436 & 0.271 & 0.849 & 0.887 \\ 
		& & $l_{M}(\beta)$ & -0.167 & -0.181 & 0.279 & 0.325 & 0.234 & 0.987 & 0.883 \\ 
		& & $\mplmc$ & -0.046 & -0.060 & 0.280 & 0.284 & 0.193 & 1.000 & 0.950 \\ 
		& & GEE & -0.256 & -0.262 & 0.193 & 0.321 & 0.266 & 1.064 & 0.759 \\
		
		& 10 & $l_P(\psi)$ & 0.149 & 0.142 & 0.240 & 0.282 & 0.186 & 0.893 & 0.882 \\ 
		& & $l_{M}(\beta)$ & -0.223 & -0.226 & 0.192 & 0.295 & 0.232 & 0.996 & 0.759 \\ 
		& & $\mplmc$ & -0.015 & -0.018 & 0.199 & 0.199 & 0.133 & 0.988 & 0.943 \\ 
		& & GEE & -0.264 & -0.267 & 0.150 & 0.303 & 0.267 & 1.036 & 0.585 \\
		
		\hline
		250 & 4 & $l_P(\psi)$ & 0.239 & 0.223 & 0.368 & 0.438 & 0.279 & 0.800 & 0.841 \\ 
		& & $l_{M}(\beta)$ & -0.135 & -0.144 & 0.242 & 0.277 & 0.195 & 0.973 & 0.887 \\ 
		& & $\mplmc$ & -0.112 & -0.115 & 0.246 & 0.270 & 0.183 & 0.999 & 0.920 \\ 
		& & GEE & -0.263 & -0.264 & 0.145 & 0.300 & 0.264 & 1.070 & 0.596 \\
		
		& 6 & $l_P(\psi)$ & 0.174 & 0.169 & 0.230 & 0.289 & 0.199 & 0.873 & 0.858 \\ 
		& & $l_{M}(\beta)$ & -0.198 & -0.197 & 0.171 & 0.262 & 0.205 & 1.006 & 0.775 \\ 
		& & $\mplmc$ & -0.074 & -0.077 & 0.171 & 0.186 & 0.126 & 1.021 & 0.920 \\ 
		& & GEE & -0.274 & -0.276 & 0.120 & 0.299 & 0.276 & 1.051 & 0.411 \\
		
		& 10 & $l_P(\psi)$ & 0.128 & 0.128 & 0.144 & 0.193 & 0.138 & 0.937 & 0.838 \\ 
		& & $l_{M}(\beta)$ & -0.246 & -0.248 & 0.115 & 0.272 & 0.248 & 1.043 & 0.459 \\ 
		& & $\mplmc$ & -0.032 & -0.031 & 0.119 & 0.123 & 0.082 & 1.038 & 0.948 \\ 
		& & GEE & -0.282 & -0.285 & 0.091 & 0.297 & 0.285 & 1.071 & 0.159 \\
		\hline
	\end{tabular}
\end{table}

The two principal simulation experiments are recognisable according to the model used to select the missing values in the experimental datasets of dimensions $T=4,6,10$ and $N=50,100,250$. 
In both scenarios, we consider $p=1$ and the covariate $x_{it}$ is simulated by means of independent draws from the $N(-0.35,1)$ distribution, with intercepts $\lambda_i$ $(i=1,\dots,N)$ also independently generated as $N(-0.35,1)$.
The global parameters in (\ref{mod-bern}) and (\ref{pit}) for generating the $S=2000$ samples with MCAR observations are set equal to $\beta=1$, $\gamma_1=2.5$ and $\gamma_2=0$. 
Instead, simulation of MNAR data is carried out with $\beta=1$, $\gamma_1=5$, $\gamma_2=1$.
Such settings were chosen in order to observe
a percentage of missing observations in the resulting datasets varying between 35\% and 40\%.

This simulation setup is taken from a conditional model, with a random effects specification. Such a choice also allows the comparison with inference from a marginal model with generalized estimating equations (GEE). Indeed, although the two approaches are not directly comparable, if the random effects model is correctly specified the corresponding coefficient of $x_{it}$ in the marginal model would approximately equal $\beta_m=\beta/\sqrt{1+\sigma^2_\lambda/c^2}$, where $c=1.7$ and $\sigma^2_\lambda$ is the variance of the random effects' distribution \citep[Section~9.4.1]{agr15}. Here $\sigma^2_\lambda=1$, and therefore $\beta_m=0.862$. In addition, results for the case where $\lambda_i=\sum_{t=1}^T x_{it}/T+u_i$, with $u_i \sim N(0,1)$, are made available in the Supplementary material. In that setting, incidental parameters are correlated with the covariate, thus a random effects model would be wrongly specified and the comparison with GEE unfeasible. Instead, the MCMPL approach guarantees the same qualitative results under both frameworks.

Tables \ref{tab:MCARlog}, \ref{tab:MCARlog2} and \ref{tab:MNARlog} report the performance of the compared inferential functions 
in respect of bias (B), median bias (MB), root mean squared error (RMSE) and median absolute error (MAE) of the corresponding estimators. Precisely, with specific reference to $\hat{\beta}$ we compute
\begin{align*}\label{simuq}
\mathrm{B}_{\hat{\beta}}&=\sum_{s=1}^S \big(\hat{\beta}^s-\beta\big)/S\,,\\
\mathrm{MB}_{\hat{\beta}}&=\left(\hat{\beta}^{\left(S/2\right)}+\hat{\beta}^{\left(S/2+1\right)}\right)/2 - \beta\,,\\
\mathrm{RMSE}_{\hat{\beta}}&=\sqrt{\sum_{s=1}^S \big(\hat{\beta}^s-\beta\big)^2/S}\,,\\
\mathrm{MAE}_{\hat{\beta}}&=\left(|\hat{\beta}-\beta|^{\left(S/2\right)}+|\hat{\beta}-\beta|^{\left(S/2+1\right)}\right)/2\,,
\end{align*}
where $\beta$ is the value of the regression coefficient used to simulate the $S$ datasets, $\hat{\beta}^s$ is its ML estimate on the $s$th sample $(s=1,\dots, S)$ and $x^{(s)}$ denotes the $s$th element in the vector of order statistics $(x^{(1)},\dots,x^{(S)})$.
The empirical standard deviation (SD) of the estimates is also reported. 
Considering again $\hat{\beta}$ for illustration, one may write
\begin{equation*}\label{sd}
\mathrm{SD}_{\hat{\beta}}=\sum_{s=1}^S \big(\hat{\beta}^s-\bar{\hat{\beta}}\big)^2/(S-1)\,, \qquad \bar{\hat{\beta}}=\sum_{s=1}^S \hat{\beta}^s/S\,.
\end{equation*}
In addition, the ratio SE/SD, where SE stands for the average over simulations of likelihood-based estimated standard errors, and empirical coverages of 0.95 Wald confidence intervals (CI) for $\beta$ are shown. Note that
the large values of $N$ examined here ensure adequacy of the quadratic approximation around the maximum of the various log-likelihoods, hence the generally more accurate coverages derived by inversion of the likelihood ratio statistic would be practically identical.

Behaviours of the likelihoods built under the correct MCAR hypothesis are shown in Table \ref{tab:MCARlog}. The latter attests the inadequacy of inference on $\beta$ deriving from the profile likelihood in this incidental parameters setting. The introduction of the modification term, either explicitly calculated or approximated by Monte Carlo simulation with $R=500$, conspicuously refines the point estimation and actual coverage of confidence intervals. 
In particular, as happens for complete stratified data, the bias of the ML estimator is of order $O(1/T)$ regardless of the value of $N$, as opposed to that of its modified version which is of order  $O(1/T^2)$. On the contrary, for fixed $T$, confidence intervals become less precise as $N$ increases, since standard deviations get smaller.  
The most important evidence supplied here by Table \ref{tab:MCARlog} is the absence of the need to take the MCAR mechanism into consideration when computing the MPL. Indeed, the performance of $l_{M}(\beta)$ is essentially identical to that of $l_{M^*}(\beta)$ for all the sample sizes considered. This finding confirms what argued by \cite{ken98}. 

Inference on the same MCAR datasets can also be made via $l_P(\psi)$ and $\mplmc$, which assume a general nonignorable model of missingness. Moreover, GEE provides a further alternative for inference, given that its consistency is guaranteed under MAR, and therefore MCAR, mechanisms \citep[Section 9.6.4]{agr15}.
Experimental outcomes of such analysis are presented in Table \ref{tab:MCARlog2}. 
Despite some undercoverage of Wald intervals when $T=4$, the global accuracy of the MNAR MCMPL is considerable and definitely higher than that of the corresponding unmodified profile likelihood. The latter proves to be more reliable than its MCAR counterpart in Table \ref{tab:MCARlog}, while $\mplmc$ is generally superior in terms of bias but inferior in terms of coverage to $l_{M}(\beta)$ and $l_{M^*}(\beta)$, which efficiently avoid unnecessary estimation of the missingness parameters. However, as will be seen in Table~\ref{tab:MNARlog}, $l_{M^*}(\psi)$ balances this loss of efficiency with its robustness to the underlying missingness mechanism.

Fit of the model via GEE was implemented through the \texttt{gee} library in R \cite{gee}, specifying either an independence, exchangeable or unstructured within-cluster correlation. 
The quasi-likelihood approach seems to work very well under the MCAR assumption. It is yet important to bear in mind that MPL and GEE are estimating two distinct models here (conditional and marginal, respectively, with different true parameter value). Note that Table \ref{tab:MCARlog2} reports the most favorable results for GEE, obtained assuming independence of observations and using non-robust standard errors of the estimates.

Table \ref{tab:MNARlog} refers to the second experiment based on datasets generated with MNAR observations. 
Classical inference through the MNAR profile log-likelihood is found imprecise, as expected. 
The most interesting simulation outcome concerns the pattern of inferential results reached by the two versions of the MPL considered. Indeed, 
for any given number of clusters, as $T$ increases the accuracy of $l_{M}(\beta)$ deteriorates both in terms of bias and of confidence intervals' coverage, whereas that of $\mplmc$ improves. 
The fact that the MPL by Severini leads to worse results for large $T$ may seem counterintuitive at first. In fact, this makes sense since incompleteness of the data is more perceived in larger groups and thus the harmful impact of the wrong MCAR assumption reveals itself as $T$ grows. Nevertheless, apart from some numerical instabilities that may occur occasionally when $T$ is small (mainly with $T=4$), the MCMPL ensures better inference on $\beta$ than its analytical version based on the wrong model. 
In this setting, $l_{M^*}(\beta)$ is still found equivalent to $l_{M}(\beta)$ and therefore is not shown in the table. Finally, the GEE method based on independent observations and non-robust standard errors proves, as expected, to be inconsistent when data are MNAR \citep[Section 9.6.4]{agr15}, suffering from severe bias in all simulation setups, with accuracy getting worse as the number of clusters $N$ increases.

In outline, the Monte Carlo strategy is particularly convenient in this missing-data scenario. It allows indeed to easily calculate the MNAR MCMPL which appears robust to the missingness mechanism, where the price to pay for this robustness is only a minor loss in efficiency. As proved by Section S3 of the Supplementary material, the same consideration can be made when a probit link function is used in model (\ref{mod-bern}).

\begin{table}[t]\centering
	\setlength{\tabcolsep}{1.5pt}
	\caption{Estimates and related standard errors (in parenthesis) in the logistic regression for the toenail data with missing response. The methods considered are MCAR profile log-likelihood $l_P(\beta)$, Severini's exact MCAR MPL $l_{M}(\beta)$, MCAR MCMPL $l_{M^*}(\beta)$, MNAR profile log-likelihood $l_P(\psi)$ and MNAR MCMPL $l_{M^*}(\psi)$, computed with $R=500$. }
	\label{tab:toefit1}
	\medskip	
	\begin{tabular}{lcccccccc}
		\hline
		& \multicolumn{2}{c}{$l_P(\beta)$} & \multicolumn{2}{c}{$l_M(\beta)$ and $l_{M^*}(\beta)$} & \multicolumn{2}{c}{$l_P(\psi)$} & \multicolumn{2}{c}{$l_{M^*}(\psi)$} \\ 
		
		& Estimate & $p$-value & Estimate & $p$-value & Estimate & $p$-value & Estimate & $p$-value \\
		\hline
		$\beta_1$ & -0.482\,(0.057) & 0.000 & -0.396\,(0.048) & 0.000 & -0.495\,(0.056) &0.000 & -0.409\,(0.048) & 0.000 \\
		$\beta_2$ & -0.184\,(0.094) & 0.050 & -0.122\,(0.077) & 0.110 & -0.197\,(0.095) & 0.038 & -0.140\,(0.079) & 0.077  \\
		\hline
	\end{tabular}
\end{table}

%

\subsection{Application to a toenail infection study}
The solution detailed in Section~\ref{bin:md} can be applied to the toenail data, carefully described in \citet[Section 2.3]{molver05}. This dataset was collected upon a two-armed clinical trial in $N=294$ patients treated for toenail infection and followed-up at $T=7$ time occasions. The outcome variable codes whether the infection was severe ($y_{it}=1$) or not ($y_{it}=0$), for $t=1, \ldots, T$. 
Each patient is here identified as a cluster, containing 7 observations at the different follow-up visits. This is a clear example of situations where the number of clusters ($N=294$) is much higher than the cluster size ($T=7$). 
Two covariates were also recorded: number of months from the first visit and the oral treatment, A or B, used. Note that only the former is a time-varying covariate, as every subject was randomly assigned to only one treatment for the whole duration of the study. The main interest was to understand how the percentage of severe infections evolved over time and if such evolution was affected by the treatment. Due to a variety of reasons, the response is missing at some time points for several subjects. The percentage of missing response in the sample is 7.29\%. Although this value is lower than those considered in the simulation studies, the results below indicate a perceivable difference between inference based on the various approaches.

Model (\ref{mod-bern}) is fitted to the data based on the assumption of the logistic link 
\[ 
\pi_{it}=\mathrm{logit}^{-1}(\lambda_i + \beta_1 x_{1it} + \beta_2 x_{2it}),
\]
where $x_{1it} \in \{0, 1, 2, 3, 6, 9, 12\}$ measures the time in months from the first visit of the $i$th patient and $x_{2it}$ is the interaction term between time and treatment. We remark that the treatment is not included in the regression because the presence of the fixed effects prevents those coefficients referred to covariates with no within-cluster variation from being identified. According to the mechanism of missingness supposed in (\ref{mod-m}), inference about $\beta=(\beta_1, \beta_2)$ can be conducted by either MCAR or MNAR methods. Results are partially shown in Table \ref{tab:toefit1}. Notice that the output obtained by using the MCAR MCMPL $l_{M^*}(\beta)$ is not reported, being basically indistinguishable from that of $l_{M}(\beta)$.
Whatever the hypothesis about the missingness process, both standard and modified profile likelihood functions detect a strongly significant decrease over time in the percentage of severe infections among subjects who receive oral treatment A, with a smaller estimated effect given by the MPL. For what concerns $\beta_2$, which represents the difference in the evolution of infection between the two treatment arms, the conclusion is less clear. This is in line with previous analyses neglecting the missing data problem \cite[Section 10.3]{molver05}. Contrary to the ML fits, the use of the MPL suggests no effect of oral treatment B in improving the recovery process with respect to treatment A,  at a 5\% significance level. However, the $p$-value is well below 0.1 if the data are assumed MNAR. This last hypothesis seems indeed quite realistic, as the probability of missing a visit for one patient is likely to depend on his current toenail infection status. The estimate of the parameter associated with the response $y_{it}$ in the specification (\ref{mod-m})-(\ref{pit}), $\gamma_2$, 
equals $-\infty$ for both $l_P(\psi)$ and $l_{M^*}(\psi)$. This indicates the occurrence of separation in the data available for estimating the MNAR probability of missingness. Particularly, one obvious interpretation is that a patient with a severe toenail infection is much more motivated to undergo the scheduled visit than a patient who has healed. We note however that separation in the estimation of $\gamma_2$ does not affect the estimates of the remaining parameters, as usually happens in binary regression without missing data. These considerations, along with the robustness confirmed by the simulations of Section \ref{sec:logsimu}, further make the MNAR MCMPL the most reliable inferential tool in this example.

\section{Survival models for right-censored data}
\label{sec:surv}

\subsection{Setup and background}\label{sec:stupsurv}\noindent
Let independent clustered failure times $\tilde{y}_{it}\geq 0$ be realizations of the random variables $\widetilde{Y}_{it}$ such that
\begin{equation}\label{modsurv}
\widetilde{Y}_{it} \sim p_{\widetilde{Y}_{it}}(\tilde{y}_{it}| x_{it}
; \psi;\lambda_i)\,, \qquad i=1,\dots,N, \ \ \; t=1,\dots,T,
\end{equation}
where $x_{it}$ is a $p$-dimensional vector of fixed covariates.
The survival function of $\widetilde{Y}_{it}$ is defined by $S_{\widetilde{Y}_{it}}(\tilde{y}_{it}
| x_{it}
; \psi;\lambda_i)=P_\theta(\widetilde Y_{it}>\tilde y _{it})$
and the hazard function equals
\begin{align*}\label{key}
h_{\widetilde{Y}_{it}}(\tilde{y}_{it}|x_{it}; \psi,\lambda_i)=\frac{p_{\widetilde{Y}_{it}}(\tilde{y}_{it}| x_{it}; \psi,\lambda_i)}{S_{\widetilde{Y}_{it}}(\tilde{y}_{it}| x_{it}; \psi,\lambda_i )}\,.
\end{align*}

Since observations may be right-censored, the sample actually consists of realizations of the pair $\big(Y_{it}, \Delta_{it}\big)$, where $Y_{it}=\min \big(\widetilde{Y}_{it}, C_{it}\big)$ with $C_{it}$ random censoring time
and $\Delta_{it}$ event indicator being $\Delta_{it}=1$ if $\widetilde{Y}_{it}\leq C_{it}$ and $\Delta_{it}=0$ otherwise. The censoring mechanism is hypothesized to be independent and non-informative, 
meaning that each $C_{it}$ is unrelated to the other survival or censoring times and its continuous distribution does not depend on $\theta$. We also suppose that the density of $C_{it}$ is the same in all $N$ groups.

Under this scenario, inferential solutions to the incidental parameters problem also need to cope with the presence of censored data. In the past years, the application of the MPL has been experimented only to a limited extent because its computation is not straightforward in regression frameworks like (\ref{modsurv}) with general censoring scheme. The technique proposed by \cite{pierce06} to overcome such complications relies on Monte Carlo simulations as well, but targets fully parametric settings where the distribution of censoring is completely defined. 
\cite{pierce15} considered also higher-order asymptotics for semiparametric Cox regression. In that case, an adjustment to the likelihood ratio statistic
was obtained either by implementation of a parametric bootstrap employing a reference censoring mechanism or by simulation. However, not only their proposal considers inference on scalar parameters of interest, but also it does not usually improve on the partial likelihood.

Model (\ref{modsurv}) can be viewed as an extension of the scenarios on which \cite{cor16} focused. Therein, the use of Severini's frequentist integrated likelihood for estimating $\psi$ was found to be superior to random effects models with seriously misspecified frailty distribution. However, the computational effort implied by their approach is remarkably sensitive to the number of predictors in the study, and indeed they only consider cases with $p=0$ or $p=1$. In addition, the authors specify some parametric distribution of $C_{it}$, whereas here we prefer to avoid such a restriction which might affect the inferential results. On the one hand, our choice relaxes the assumptions of the analysis, but on the other, it prevents the term (\ref{misev}) in Severini's MPL from being exactly calculated. In what follows, the Monte Carlo strategy presented in Section \ref{sec:MPL} will be shown general and flexible enough to tackle this difficulty.

\subsection{Monte Carlo modified profile likelihood}\label{mcsurv}\noindent
Consider the observed couple $\big(y_{it}, \delta_{it}\big)$ introduced in the previous section. If the censoring times $c_{it}$ are independent realizations of a continuous random variable with generic distribution $p_{C_{it}}(c_{it}; \varsigma)$ and survival function $S_{C_{it}}(c_{it}; \varsigma)=P_\varsigma(C_{it}>c_{it})$, then data are drawn from the joint density
\begin{equation}\label{joint}
p_{Y_{it}, \Delta_{it}}(y_{it}, \delta_{it}; \varphi) =  \left\{p_{\widetilde{Y}_{it}} (y_{it};\theta)S_{C_{it}}(y_{it}; \varsigma)\right\}^{\delta_{it}}
\left\{p_{C_{it}}(y_{it}; \varsigma) 
S_{\widetilde{Y}_{it}}(y_{it};\theta) \right\} ^{1-\delta_{it}}\,,
\end{equation}
where $\varphi=(\theta, \varsigma)$ and, in the interests of conciseness, dependence on covariates is omitted.
The distribution of $C_{it}$ is independent of the parameter $\theta$ and does not vary across clusters, thus the log-likelihood function about $\theta$ based on the whole dataset $\big(y_{it}, \delta_{it}\big)$ $(i=1,\dots,N, t=1,\dots,T)$ can be formulated by
\begin{equation}\label{llsurv}
l(\theta)=\sum_{i=1}^N\sum_{t=1}^T \left\{\delta_{it} \log p_{\widetilde{Y}_{it}} (y_{it};\theta) + (1-\delta_{it})
\log S_{\widetilde{Y}_{it}}(y_{it};\theta)\right\}\,.
\end{equation}
Starting from the previous expression, the profile log-likelihood for $\psi$ and the score in this setting can be derived following the general definitions in (\ref{profclusters}) and (\ref{score}).

The first quantity to be computed in (\ref{misev}), $j_{\lambda_i\lambda_i} (\hat \theta_\psi)$, is typically obtainable with ease even with right-censored data.
On the contrary, exact calculation of the expected value in (\ref{misev}) should be carried out with reference to a fully specified model, i.e. the joint probability density function (\ref{joint}) comprising also the distribution of the censoring times. However, in order to avoid unnecessary assumptions, we will not specify a parametric form for $p_{C_{it}}(c_{it}; \varsigma)$ and $S_{C_{it}}(c_{it}; \varsigma)$, as instead done by  \cite{cor16}. Indeed, such an assumption can be avoided for calculating the MPL via the Monte Carlo solution reported in Section \ref{sec:MPL}, because estimation of the censoring distribution can be implemented nonparametrically, making the resulting approximation more robust. 

With an unspecified density of $C_{it}$, it is still possible to simulate the Monte Carlo samples $(y^r_{it},\delta^r_{it})$ $(r=1,\dots,R)$ on which (\ref{istar}) is based.
Censoring times are not available for units with an observed failure, but they can be simulated by bootstrap techniques. The procedure is explained in the sequel.
First, failure times $\tilde{y}^r_{it}$ are generated from the ML fit of model (\ref{modsurv}). Second, new censoring times $c^r_{it}$ are determined by performing the conditional bootstrap described in Algorithm 3.1 of \citet[p. 85]{dav97}. In particular, 
if the original indicator $\delta_{it}$ equals zero we set $c^{r}_{it}=c_{it}$, otherwise we draw $c^{r}_{it}$ from the conditional distribution of $C_{it}|C_{it}>y_{it}$ computed as
\begin{equation*}
\widehat{S}_{C_{it}|C_{it}>y_{it}}(c_{it})=\frac{\widehat{S}_{C_{it}}(c_{it})}{\widehat{S}_{C_{it}}(y_{it})},
\end{equation*} 
where $\widehat{S}_{C_{it}}(\cdot)$ is the Kaplan-Meier nonparametric estimator of the survival function of $C_{it}$. Precisely, each $c^{r}_{it}$ corresponding to $\delta_{it}=1$ is found as the unique solution $c$ to the equation $\widehat{S}_{C_{it}}(c)=u^r_{it}\widehat{S}_{C_{it}}(y_{it})$, with $u^r_{it}\sim U(0,1)$. Finally, for $i=1,\dots,N$ and $t=1,\dots,T$, the observed survival times are $y^r_{it}=\min(\tilde{y}^r_{it}, c^r_{it})$ and hence the new event indicators are defined as $\delta^r_{it}=1$ if $\tilde{y}^r_{it}\leq c^r_{it}$ and $\delta^r_{it}=0$ if $\tilde{y}^r_{it} > c^r_{it}$.

\subsection{Weibull model}\label{sec:weib}\noindent
	As an illustration, assume now the Weibull distribution for the random variables $\widetilde{Y}_{it}$. Consequently, in model (\ref{modsurv}) the probability density function can be expressed as
\begin{equation}\label{mod:weib}
p_{\widetilde{Y}_{it}}(\tilde{y}_{it}| x_{it}
; \psi;\lambda_i)=\eta_{it}\xi\big(\eta_{it}\tilde{y}_{it}\big)^{\xi-1}
\!\exp\!\big\{\!\!-\!\big(\eta_{it}\tilde{y}_{it}\big)^{\xi}\big\}\,,
\,i=1,\dots,N,\,t=1,\dots,T,
\end{equation}
where $\eta_{it}\!=\!e^{-(\lambda_i+\beta ^\T x_{it})}$ controls the scale of the distribution.
The interest is on estimating the common shape parameter $\xi >0$ and the regression coefficients in $\beta=(\beta_1,\ldots,\beta_p) \in \reali^p$, while treating the vector of group-related intercepts $\lambda=(\lambda_1,\dots,\lambda_N) \in \reali^N$ as nuisance. We shall then write $\theta=(\psi,\lambda)$, with $\psi=(\xi,\beta) \in \reali^+ \times \reali^p$. 

The survival and hazard functions of $\widetilde{Y}_{it}$ are, respectively, $S_{\widetilde{Y}_{it}}(\tilde{y}_{it}
| x_{it}
; \psi;\lambda_i)=\exp\!\big\{\!\!-\!\big(\eta_{it}\tilde{y}_{it}\big)^{\xi}\big\}$ 
and
\begin{align} \label{hazi}
h_{\widetilde{Y}_{it}}(\tilde{y}_{it}|x_{it}; \psi,\lambda_i)=h_0(\tilde{y}_{it};\xi)\eta_{it}^\xi
=h_{0i}(\tilde{y}_{it};\xi,\lambda_i)e^{-\xi(\beta^\T x_{it})}\,,
\end{align}
where $h_0(\tilde{y}_{it};\xi)=\xi\tilde{y}_{it}^{\,\xi-1}$ is the baseline hazard parametrically modeled and shared by all groups, whereas $h_{0i}(\tilde{y}_{it};\xi,\lambda_i)=h_0(\tilde{y}_{it};\xi)e^{-\xi\lambda_i}$ can be seen as the equivalent for the $i$th cluster $(i=1,\dots,N)$. Thus (\ref{mod:weib}) is a stratified proportional hazards model, and its logarithmic transformation coincides with a so-called accelerated failure time model (see, for instance, \citealp[Section~6]{cor16}).

Denoting the number of failures recorded in the $i$th group by $\delta_{i \cdot}= \sum_{t=1}^T \delta_{it}$ $(i=1,\dots,N)$ allows to write (\ref{llsurv}) under the Weibull model as
\begin{equation}\label{survll}
l(\theta) = \sum_{i=1}^N \left\{\xi \sum_{t=1}^T \delta_{it} \log \eta_{it} + \delta_{i \cdot} \log \xi + (\xi -1) \sum_{t=1}^T \delta_{it} \log y_{it} 
- \sum_{t=1}^T (\eta_{it} y_{it} )^{\xi}\right\}\,.
\end{equation}
Furthermore, the score in formula (\ref{score}) equals
\begin{equation}
\label{l_l}
\l_{\lambda_i} (\theta) = -\xi \delta_{i \cdot} + \xi \sum_{t=1}^T (\eta_{it} y_{it} )^{\xi}\,, \qquad i=1,\dots,N,
\end{equation}
and the relating cluster-specific constrained ML estimate is explicitly found as
\begin{equation}\label{lhat}
\hat\lambda_{i\psi}=  \frac{1}{ \xi} \bigg\{\log \sum_{t=1}^T y_{it}^{\xi} e^{-\xi (\beta^\T x_{it})} -  \log \delta_{i \cdot}\bigg\}\,, \qquad i=1, \ldots,N.
\end{equation}
The profile log-likelihood function for $\psi$ presented in (\ref{profclusters}) has the expression
\begin{align}\label{pl}
\l_P(\psi) =  \sum_{i=1}^N \bigg[&\delta_{i \cdot} \bigg\{ \log\delta_{i \cdot} - \log\sum_{t=1}^T y_{it}^{\xi} e^{-\xi \nonumber (\beta^\T x_{it})} \bigg\} - \xi \sum_{t=1}^T \delta_{it} (\beta^\T x_{it}) \\ &+\delta_{i \cdot} (\log \xi -1)+ (\xi -1) \sum_{t=1}^T \delta_{it} \log y_{it}\bigg]\,,
\end{align}
and its maximizer $\hat\psi=(\hat \xi, \hat\beta)$ can be obtained numerically.

For what concerns the computation of the MPL, changing sign to the derivative of (\ref{l_l}) with regard to $\lambda_i$ gives
\begin{equation*}
j_{\lambda_i, \lambda_i} (\hat \theta_\psi) = \xi^2 \sum_{t=1}^T (\tilde{\eta}_{it}y_{it})^{\xi}\,, \qquad i=1, \ldots,N,
\end{equation*}
where $\tilde{\eta}_{it}=\exp\big\{\!\!-\!(\hat\lambda_{i\psi}+\beta^\T x_{it})\big\}$. In the second summand of (\ref{misev}), the usual expectation can be estimated via Monte Carlo by	\begin{equation}\label{istars}
\ilst=\dfrac{1}{R}\sum_{r=1}^R \bigg[\bigg\{\!-\xi \delta^r_{i \cdot} + \xi \sum_{t=1}^T (\tilde{\eta}_{it} y^r_{it} )^{\xi}\bigg\}
\bigg\{\!-\hat{\xi} \delta^r_{i \cdot} + \hat{\xi }\sum_{t=1}^T (\hat\eta_{it} y^r_{it} )^{\hat\xi}\bigg\}
\bigg],
\end{equation}
where $\hat\eta_{it}=\exp\big\{\!\!-\!(\hat\lambda_{i}+\hat\beta^\T x_{it})\big\}$, $\delta^r_{i \cdot}= \sum_{t=1}^T \delta^r_{it}$ and $(y^r_{it},\delta^r_{it})$ $(r=1,\dots,R)$ are the simulated datasets generated via the procedure described at the end of Section~\ref{mcsurv}.

The simulation results in the following section will shed light on the possibility to solve the incidental parameters problem using the MPL under the Weibull model for clustered time-to-event data with unspecified censoring distribution. Specifically, the studies will examine on a comparative basis the profile log-likelihood $l_P(\psi)$ in (\ref{pl}) and the MCMPL $\mplmc$ depending on the approximation (\ref{istars}). A comparison with a stratified Cox regression, which models nonparametrically $h_{0i}(\tilde{y}_{it};\xi,\lambda_i)$ in (\ref{hazi}), will also be considered.

\subsection{Simulation studies}\label{sec:survsim}\noindent
Two experiments of $S=2000$ simulations are conducted to study inference on $\psi$ in the Weibull model for right-censored observations presented in Section \ref{sec:weib}. The within-group size and the number of clusters in the artificial samples are set equal to $T=4,6,10$ and $N=50,100,250$, respectively. The regression model includes $p=2$ covariates. The first, $x_{1it}$, in each $i$th group $(i=1,\dots,N)$ is obtained by imposing $x_{1it}=0$ for $t=1,\dots,T/2$ and $x_{1it}=1$ for $t=T/2+1,\dots,T$. The second, $x_{2it}$, is drawn from the standard normal distribution. The common shape parameter is chosen as $\xi=1.5$ and the vector of regression coefficients as $\beta=(-1,1)$, while each cluster-related intercept is independently sampled as $\lambda_i \sim N(0.5,0.5^2)$.
Failures $\tilde{y}_{it}$ are simulated via the Weibull density function (\ref{mod:weib}). The censoring times $c_{it}$ can be obtained by random generation from the distribution $Exp(\varsigma)$, where the parameter is selected in such a way as to control the overall proportion $P_c$ of censored data. In detail, given the quantities above and for a certain $P_c$, $\varsigma$ is fixed to the value solving the equation
\begin{equation*}
\frac{1}{TN} \sum_{i=1}^N \sum_{t=1}^T P_\varrho( \widetilde Y_{it} > C_{it}) = 
\frac{1}{TN} \sum_{i=1}^N \sum_{t=1}^T \int_0^{+\infty} \!\!\!S_{\widetilde{Y}_{it}}(y|x_{it}; \psi,\lambda_i) p_{C_{it}}(y; \varsigma) dy
= P_c,
\end{equation*}
where $\varrho=(\theta,\varsigma)$ and $p_{C_{it}}(y; \varsigma)=\varsigma e^{-\varsigma y}$. Then, in each of the $S$ fictitious datasets, observations $\big(y_{it}, \delta_{it}\big)$ stem from the usual definitions of censored failures and event indicators, i.e. $y_{it}=\min(\tilde{y}_{it}, c_{it})$ and $\delta_{it}=1$ when $\tilde{y}_{it}\leq c_{it}$, otherwise $\delta_{it}=0$ $(i=1,\dots,N,\,t=1,\dots,T)$.

\begin{table}[t]\centering
	\caption{Inference on $\xi=1.5$ in the stratified Weibull regression model for right-censored survival data and probability of censoring $P_c=0.2$. The compared functions are the profile log-likelihood $l_P(\psi)$ and the MCMPL $l_{M^*}(\psi)$ computed with $R=500$. Results based on a simulation study with 2000 trials.}
	\label{tab:surv02xi}
	\medskip	
	\begin{tabular}{cccccccccc}
		\hline
		$N$ & $T$ &Method & B & MB & SD & RMSE & MAE & SE/SD & 0.95 CI\\ 
		\hline
		50 & 4 & $l_P(\psi)$ & 0.386 & 0.374 & 0.141 & 0.411 & 0.374 & 0.877 & 0.111 \\ 
		&  & $\mplmc$ & 0.006 & 0.001 & 0.109 & 0.109 & 0.073 & 1.005 & 0.958 \\ 
		
		& 6 & $l_P(\psi)$ & 0.229 & 0.227 & 0.100 & 0.250 & 0.227 & 0.907 & 0.287 \\ 
		&  & $\mplmc$ & 0.006 & 0.003 & 0.085 & 0.085 & 0.058 & 0.992 & 0.949 \\ 
		
		& 10 & $l_P(\psi)$ & 0.123 & 0.121 & 0.068 & 0.141 & 0.121 & 0.937 & 0.530 \\ 
		&  & $\mplmc$ & 0.004 & 0.002 & 0.063 & 0.063 & 0.041 & 0.988 & 0.950 \\ 
		\hline
		100& 4 & $l_P(\psi)$ & 0.378 & 0.375 & 0.101 & 0.391 & 0.375 & 0.859 & 0.006 \\ 
		&  & $\mplmc$ & -0.003 & -0.003 & 0.078 & 0.078 & 0.052 & 0.981 & 0.944 \\ 
		
		& 6 & $l_P(\psi)$ & 0.222 & 0.218 & 0.068 & 0.232 & 0.218 & 0.930 & 0.053 \\ 
		&  & $\mplmc$ & 0.001 & -0.001 & 0.058 & 0.058 & 0.039 & 1.017 & 0.955 \\ 
		
		& 10 & $l_P(\psi)$ & 0.118 & 0.118 & 0.047 & 0.127 & 0.118 & 0.958 & 0.257 \\ 
		&  & $\mplmc$ & -0.000 & 0.000 & 0.043 & 0.043 & 0.029 & 1.009 & 0.950 \\ 
		\hline
		250 & 4 & $l_P(\psi)$ & 0.365 & 0.361 & 0.064 & 0.370 & 0.361 & 0.853 & 0.000 \\ 
		&  & $\mplmc$ & -0.011 & -0.014 & 0.049 & 0.051 & 0.036 & 0.977 & 0.935 \\ 
		
		& 6 & $l_P(\psi)$ & 0.213 & 0.212 & 0.044 & 0.217 & 0.212 & 0.907 & 0.000 \\ 
		&  & $\mplmc$ & -0.006 & -0.007 & 0.038 & 0.038 & 0.026 & 0.987 & 0.944 \\ 
		
		& 10 & $l_P(\psi)$ & 0.116 & 0.115 & 0.030 & 0.120 & 0.115 & 0.953 & 0.018 \\ 
		& & $\mplmc$ & -0.002 & -0.003 & 0.028 & 0.028 & 0.019 & 1.004 & 0.953 \\
		\hline
	\end{tabular}
	
\end{table}

The first series of simulations considers data with censoring probability $P_c=0.2$, the second relates to situations with higher proportion of censored observations, namely $P_c=0.4$. Inferences from the profile likelihood and from the MCMPL on $\psi$ are investigated as done in Section~\ref{sec:logsimu}. Notice that, before proceeding to maximize the two functions for every simulated dataset, non-informative clusters with only censored failure times must be discarded from the study. Indeed, (\ref{lhat}) shows that $\hat{\lambda}_{i\psi}$ is not finite if $\delta_{i \cdot}=0$ and hence the $i$th group does not make any contribution to estimating $\psi$ $(i=1,\dots,N)$. Both estimates $\hat \psi$ and $\hat{\psi}_{M^*}=\big(\hat{\xi}_{M^*},\hat{\beta}_{M^*}\big)$ are found by joint numerical optimization of $l_P(\psi)$ and $\mplmc$, respectively.

It is well-known that the MPL can lead to both a location and a curvature adjustment of the profile likelihood. These imply, respectively, a correction of the bias and of the standard errors of the corresponding estimates. Typically, both effects are present. But there are instances in which only the curvature adjustment is needed for some components of $\psi$ \citep[][Section 3.3]{bellio06}. This is the case  in the current example. Indeed, the presence of many nuisance parameters does not imply a bias in the estimation of the regression coefficients. For this reason, although the MCMPL still remarkably refines the corresponding interval estimation, statistical indicators referred to $\beta$ are displayed in Section~S4 of the Supplementary material. On the other hand, both effects are present in the estimation of $\xi$. Results are summarized in Tables \ref{tab:surv02xi} and \ref{tab:surv04xi}.

\begin{table}[t]\centering
	\caption{Inference on $\xi=1.5$ in the stratified Weibull regression model for right-censored survival data and probability of censoring $P_c=0.4$. The compared functions are the profile log-likelihood $l_P(\psi)$ and the MCMPL $l_{M^*}(\psi)$ computed with $R=500$. Results based on a simulation study with 2000 trials.}
	\label{tab:surv04xi}
	\medskip	
	\begin{tabular}{cccccccccc}
		\hline
		$N$ & $T$ &Method & B & MB & SD & RMSE & MAE & SE/SD & 0.95 CI\\ 
		\hline
		50 & 4 & $l_P(\psi)$ & 0.453 & 0.440 & 0.177 & 0.486 & 0.440 & 0.836 & 0.131 \\ 
		&  & $\mplmc$ & -0.003 & -0.009 & 0.124 & 0.124 & 0.083 & 1.000 & 0.945 \\
		
		& 6 & $l_P(\psi)$ & 0.272 & 0.266 & 0.120 & 0.298 & 0.266 & 0.885 & 0.277 \\ 
		&  & $\mplmc$ & 0.002 & -0.003 & 0.096 & 0.096 & 0.062 & 0.996 & 0.947 \\ 
		
		& 10 & $l_P(\psi)$ & 0.144 & 0.140 & 0.081 & 0.165 & 0.140 & 0.916 & 0.525 \\ 
		&  & $\mplmc$ & 0.002 & -0.002 & 0.072 & 0.072 & 0.047 & 0.982 & 0.949 \\ 
		\hline
		100 & 4 & $l_P(\psi)$ & 0.450 & 0.444 & 0.130 & 0.468 & 0.444 & 0.805 & 0.011 \\ 
		&  & $\mplmc$ & -0.012 & -0.017 & 0.092 & 0.092 & 0.062 & 0.952 & 0.933 \\ 
		
		& 6 & $l_P(\psi)$ & 0.259 & 0.255 & 0.083 & 0.272 & 0.255 & 0.902 & 0.059 \\ 
		&  & $\mplmc$ & -0.006 & -0.009 & 0.067 & 0.067 & 0.046 & 1.010 & 0.951 \\ 
		
		& 10 & $l_P(\psi)$ & 0.137 & 0.135 & 0.056 & 0.148 & 0.135 & 0.928 & 0.259 \\ 
		&  & $\mplmc$ & -0.004 & -0.006 & 0.050 & 0.050 & 0.034 & 0.996 & 0.951 \\ 
		\hline
		250 & 4 & $l_P(\psi)$ & 0.431 & 0.428 & 0.081 & 0.439 & 0.428 & 0.812 & 0.000 \\ 
		&  & $\mplmc$ & -0.022 & -0.024 & 0.056 & 0.061 & 0.043 & 0.971 & 0.912 \\ 
		
		& 6 & $l_P(\psi)$ & 0.249 & 0.248 & 0.054 & 0.255 & 0.248 & 0.865 & 0.000 \\ 
		& & $\mplmc$ & -0.014 & -0.016 & 0.044 & 0.046 & 0.031 & 0.971 & 0.926 \\ 
		
		& 10 & $l_P(\psi)$ & 0.134 & 0.134 & 0.035 & 0.139 & 0.134 & 0.933 & 0.021 \\ 
		&  & $\mplmc$ & -0.006 & -0.006 & 0.032 & 0.032 & 0.022 & 0.996 & 0.944 \\
		\hline
	\end{tabular}
\end{table}

The accuracy of $\mplmc$ is extremely good for all unknown quantities and diverse dimensions of the data, yet inferential conclusions on $\xi$ drawn via $l_P(\psi)$ are found quite misguided. Table \ref{tab:surv02xi} testifies how the Monte Carlo modification is capable not only of greatly reducing the severe empirical bias of the ML estimator, but also of correcting the excessively low actual Wald coverages derived by the profile likelihood. Indeed, these can also be ascribed to the supplied standard errors of $\hat \xi$, prominently downward biased for smaller $T$, independently of $N$. Estimated variability of $\hat{\xi}_{M^*}$ is, conversely, much more trustworthy. 

Performances of the two inferential tools under examination in the second simulation study are summarized by Table \ref{tab:surv04xi}, for what concerns the shape parameter. The reported indexes prove the convenience of $\mplmc$ even when a greater amount of data is subject to censoring. When $P_c=0.4$ the empirical bias of $\hat{\xi}_{M^*}$ remains systematically lower than that of $\hat{\xi}$, reaching negligible values when $T$ and $N$ increase. In contrast, the imprecise point estimation provided by $l_P(\psi)$ is especially critical when the within-group size is smaller and stays basically constant as $N$ grows, coherently with the existing theoretical knowledge for models without censoring \citep{sartori03}. All the empirical coverage probabilities based on the MCMPL are very close to the nominal level, while those based on the profile likelihood are well below it, even for the aforementioned unreliable estimated standard errors of $\hat{\xi}$. 

\begin{table}[t]\centering
	\caption{Inference on the relative risk $\mathrm{RR}_1=e^{1.5}$ in the stratified regression for right-censored survival data and probability of censoring $P_c=0.2$. The compared functions are the profile log-likelihood $l_P(\psi)$ and the MCMPL $l_{M^*}(\psi)$ computed with $R=500$ under the Weibull model and the partial log-likelihood $l_{Cox}$ under the Cox proportional hazards model. Results based on a simulation study with 2000 trials.}
	\label{tab:surv02rr1}
	\medskip	
	\begin{tabular}{cccccccccc}
		\hline
		$N$ & $T$ &Method & B & MB & SD & RMSE & MAE & SE/SD & 0.95 CI\\ 
		\hline
		50 & 4 & $l_P(\psi)$ & 2.422 & 2.105 & 1.989 & 3.134 & 2.105 & 0.784 & 0.788 \\ 
		&  & $l_{M^*}(\psi)$ & 0.105 & -0.021 & 1.012 & 1.017 & 0.597 & 0.930 & 0.934 \\ 
		&  & $l_{Cox}$ & 0.333 & 0.022 & 1.568 & 1.602 & 0.869 & 0.928 & 0.948 \\ 
		& 6 & $l_P(\psi)$ & 1.287 & 1.161 & 1.141 & 1.720 & 1.165 & 0.855 & 0.827 \\ 
		&  & $l_{M^*}(\psi)$ & 0.074 & -0.013 & 0.771 & 0.775 & 0.510 & 0.954 & 0.943 \\ 
		&  & $l_{Cox}$ & 0.214 & 0.053 & 1.065 & 1.086 & 0.643 & 0.958 & 0.951 \\ 
		& 10 & $l_P(\psi)$ & 0.623 & 0.567 & 0.674 & 0.918 & 0.603 & 0.932 & 0.885 \\ 
		&  & $l_{M^*}(\psi)$ & 0.027 & -0.018 & 0.548 & 0.549 & 0.379 & 0.992 & 0.946 \\ 
		&  & $l_{Cox}$ & 0.067 & -0.013 & 0.685 & 0.688 & 0.458 & 0.985 & 0.953 \\ 
		
		\hline
		100 & 4 & $l_P(\psi)$ & 2.311 & 2.146 & 1.358 & 2.681 & 2.146 & 0.792 & 0.446 \\ 
		&  & $l_{M^*}(\psi)$ & 0.056 & -0.013 & 0.700 & 0.702 & 0.446 & 0.936 & 0.940 \\ 
		&  & $l_{Cox}$ & 0.201 & 0.040 & 1.019 & 1.038 & 0.610 & 0.969 & 0.950 \\
		& 6 & $l_P(\psi)$ & 1.203 & 1.147 & 0.756 & 1.421 & 1.147 & 0.894 & 0.608 \\ 
		&  & $l_{M^*}(\psi)$ & 0.029 & -0.009 & 0.516 & 0.517 & 0.351 & 0.994 & 0.950 \\ 
		&  & $l_{Cox}$ & 0.105 & 0.063 & 0.696 & 0.704 & 0.448 & 1.003 & 0.951 \\
		& 10 & $l_P(\psi)$ & 0.620 & 0.588 & 0.485 & 0.787 & 0.593 & 0.912 & 0.748 \\ 
		&  & $l_{M^*}(\psi)$ & 0.028 & 0.007 & 0.395 & 0.396 & 0.264 & 0.973 & 0.939 \\ 
		&  & $l_{Cox}$ & 0.059 & 0.015 & 0.482 & 0.486 & 0.311 & 0.984 & 0.943 \\
		
		\hline
		250 & 4 & $l_P(\psi)$ & 2.096 & 2.010 & 0.789 & 2.240 & 2.010 & 0.828 & 0.069 \\ 
		&  & $l_{M^*}(\psi)$ & -0.047 & -0.083 & 0.414 & 0.417 & 0.280 & 0.970 & 0.931 \\ 
		&  & $l_{Cox}$ & 0.054 & -0.006 & 0.594 & 0.597 & 0.397 & 0.994 & 0.954 \\
		& 6 & $l_P(\psi)$ & 1.139 & 1.124 & 0.467 & 1.231 & 1.124 & 0.902 & 0.215 \\ 
		&  & $l_{M^*}(\psi)$ & -0.016 & -0.028 & 0.320 & 0.320 & 0.221 & 1.002 & 0.946 \\ 
		&  & $l_{Cox}$ & 0.040 & 0.022 & 0.427 & 0.429 & 0.286 & 1.011 & 0.953 \\
		& 10 & $l_P(\psi)$ & 0.575 & 0.561 & 0.290 & 0.644 & 0.561 & 0.953 & 0.466 \\ 
		&  & $l_{M^*}(\psi)$ & -0.007 & -0.017 & 0.237 & 0.237 & 0.158 & 1.016 & 0.952 \\ 
		&  & $l_{Cox}$ & 0.014 & -0.005 & 0.288 & 0.289 & 0.197 & 1.025 & 0.957 \\ 
		\hline
	\end{tabular}
\end{table}

Both $l_P(\psi)$ and $l_{M^*}(\psi)$ are invariant under reparametrizations. Hence we can also consider inference on the relative risks referred to the two covariates, which are typically the measures of main interest in survival analysis. Under the Weibull model (\ref{mod:weib}), such quantities are expressed by $\mathrm{RR}_1=e^{-\xi\beta_1}=e^{1.5}$ and $\mathrm{RR}_2=e^{-\xi\beta_2}=e^{-1.5}$. Alternatively, these relative risks can be estimated by fitting a stratified Cox proportional hazards regression, where a separate baseline hazard function is supposed for each group. 
The function \texttt{coxph} in the R package \texttt{survival} \citep{survival} performs such analysis. In Tables~\ref{tab:surv02rr1} and \ref{tab:surv04rr1} for $\mathrm{RR}_1$ and Tables~\ref{tab:surv02rr2} and \ref{tab:surv04rr2} for $\mathrm{RR}_2$, we compare results from the fit of the Weibull regression via the profile likelihood and via MCMPL with those obtained assuming the semiparametric survival model. 
Reported 0.95 Wald coverages related with the Cox specification descend from the confidence intervals for the relative risks returned by \texttt{summary.coxph}. The profile likelihood under the Weibull model performs very poorly in estimating the relative risks, as a result of the imprecise ML inference provided on the shape parameter $\xi$. On the contrary, $l_{M^*}(\psi)$ proves to be extremely accurate in terms of both point and interval estimation. Empirical coverages derived through the partial likelihood of Cox are generally the closest to the nominal level, however this is due to the larger variability of the obtained estimates with respect to those descending from the MCMPL. Indeed, the latter exhibits the lowest RMSE and implies a gain in efficiency over the semiparametric approach, which in turn is more robust.

\begin{table}[t]\centering
	\caption{Inference on the relative risk $\mathrm{RR}_2=e^{-1.5}$ in the stratified regression for right-censored survival data and probability of censoring $P_c=0.2$. The compared functions are the profile log-likelihood $l_P(\psi)$ and the MCMPL $l_{M^*}(\psi)$ computed with $R=500$ under the Weibull model and the partial log-likelihood $l_{Cox}$ under the Cox proportional hazards model. Results based on a simulation study with 2000 trials.}
	\label{tab:surv02rr2}
	\medskip	
	\begin{tabular}{cccccccccc}
		\hline
		$N$ & $T$ &Method & B & MB & SD & RMSE & MAE & SE/SD & 0.95 CI\\ 
		\hline
		50 & 4 & $l_P(\psi)$ & -0.071 & -0.073 & 0.030 & 0.077 & 0.073 & 0.853 & 0.272 \\ 
		&  & $l_{M^*}(\psi)$& 0.000 & -0.001 & 0.034 & 0.034 & 0.023 & 0.987 & 0.944 \\ 
		&  & $l_{Cox}$ & -0.003 & -0.005 & 0.052 & 0.052 & 0.035 & 1.010 & 0.953 \\
		& 6 & $l_P(\psi)$ & -0.046 & -0.046 & 0.023 & 0.051 & 0.046 & 0.902 & 0.414 \\ 
		&  & $l_{M^*}(\psi)$ & -0.000 & -0.001 & 0.025 & 0.025 & 0.017 & 0.995 & 0.939 \\ 
		&  & $l_{Cox}$ & -0.003 & -0.003 & 0.034 & 0.034 & 0.023 & 1.014 & 0.950 \\
		& 10 & $l_P(\psi)$ & -0.026 & -0.026 & 0.018 & 0.031 & 0.026 & 0.927 & 0.624 \\ 
		&  & $l_{M^*}(\psi)$ & 0.000 & -0.000 & 0.019 & 0.019 & 0.013 & 0.984 & 0.941 \\ 
		&  & $l_{Cox}$ & -0.001 & -0.001 & 0.023 & 0.023 & 0.015 & 0.995 & 0.948 \\
		
		\hline
		100 & 4 & $l_P(\psi)$ & -0.071 & -0.071 & 0.021 & 0.074 & 0.071 & 0.828 & 0.071 \\ 
		&  & $l_{M^*}(\psi)$ & 0.002 & 0.002 & 0.024 & 0.024 & 0.017 & 0.960 & 0.942 \\ 
		&  & $l_{Cox}$ & -0.002 & -0.003 & 0.035 & 0.035 & 0.023 & 1.003 & 0.949 \\
		& 6 & $l_P(\psi)$ & -0.045 & -0.046 & 0.016 & 0.048 & 0.046 & 0.906 & 0.194 \\ 
		&  & $l_{M^*}(\psi)$ & 0.001 & -0.000 & 0.018 & 0.018 & 0.012 & 1.005 & 0.949 \\ 
		&  & $l_{Cox}$ & -0.002 & -0.003 & 0.024 & 0.024 & 0.016 & 1.022 & 0.954 \\
		& 10 & $l_P(\psi)$ & -0.026 & -0.026 & 0.012 & 0.028 & 0.026 & 0.939 & 0.409 \\ 
		&  & $l_{M^*}(\psi)$ & 0.000 & -0.000 & 0.013 & 0.013 & 0.009 & 0.999 & 0.948 \\ 
		&  & $l_{Cox}$ & -0.001 & -0.001 & 0.016 & 0.016 & 0.011 & 0.998 & 0.952 \\
		
		\hline
		250 & 4 & $l_P(\psi)$ & -0.070 & -0.071 & 0.013 & 0.071 & 0.071 & 0.847 & 0.001 \\ 
		&  & $l_{M^*}(\psi)$ & 0.003 & 0.003 & 0.015 & 0.015 & 0.010 & 0.985 & 0.942 \\ 
		&  & $l_{Cox}$ & -0.001 & -0.002 & 0.022 & 0.022 & 0.015 & 1.002 & 0.949 \\
		& 6 & $l_P(\psi)$ & -0.044 & -0.044 & 0.011 & 0.045 & 0.044 & 0.895 & 0.009 \\ 
		&  & $l_{M^*}(\psi)$ & 0.002 & 0.002 & 0.011 & 0.012 & 0.008 & 0.986 & 0.954 \\ 
		&  & $l_{Cox}$ & 0.000 & -0.000 & 0.015 & 0.015 & 0.010 & 1.003 & 0.954 \\
		& 10 & $l_P(\psi)$ & -0.025 & -0.025 & 0.008 & 0.026 & 0.025 & 0.945 & 0.099 \\ 
		&  & $l_{M^*}(\psi)$ & 0.001 & 0.001 & 0.008 & 0.008 & 0.006 & 1.003 & 0.947 \\ 
		& & $l_{Cox}$ & 0.000 & 0.000 & 0.010 & 0.010 & 0.007 & 1.026 & 0.959 \\
		
		\hline
	\end{tabular}
\end{table}

\begin{table}[t]\centering
	\caption{Inference on the relative risk $\mathrm{RR}_1=e^{1.5}$ in the stratified regression for right-censored survival data and probability of censoring $P_c=0.4$. The compared functions are the profile log-likelihood $l_P(\psi)$ and the MCMPL $l_{M^*}(\psi)$ computed with $R=500$ under the Weibull model and the partial log-likelihood $l_{Cox}$ under the Cox proportional hazards model. Results based on a simulation study with 2000 trials.}
	\label{tab:surv04rr1}
	\medskip	
	\begin{tabular}{cccccccccc}
		\hline
		$N$ & $T$ &Method & B & MB & SD & RMSE & MAE & SE/SD & 0.95 CI\\ 
		\hline
		50 & 4 & $l_P(\psi)$  & 3.175 & 2.620 & 2.784 & 4.222 & 2.620 & 0.746 & 0.829 \\ 
		& & $l_{M^*}(\psi)$ & 0.109 & -0.076 & 1.172 & 1.177 & 0.716 & 0.939 & 0.925 \\ 
		&  & $l_{Cox}$ & 0.477 & 0.050 & 1.978 & 2.035 & 0.989 & 0.893 & 0.950 \\ 
		& 6 & $l_P(\psi)$ & 1.651 & 1.419 & 1.501 & 2.231 & 1.433 & 0.819 & 0.836 \\ 
		&  & $l_{M^*}(\psi)$ & 0.080 & -0.030 & 0.912 & 0.915 & 0.584 & 0.944 & 0.939 \\ 
		&  & $l_{Cox}$ & 0.295 & 0.074 & 1.259 & 1.293 & 0.752 & 0.958 & 0.953 \\
		& 10 & $l_P(\psi)$& 0.764 & 0.675 & 0.830 & 1.128 & 0.720 & 0.901 & 0.884 \\ 
		&  & $l_{M^*}(\psi)$ & 0.023 & -0.041 & 0.640 & 0.640 & 0.433 & 0.979 & 0.942 \\ 
		&  & $l_{Cox}$ & 0.080 & -0.030 & 0.807 & 0.811 & 0.517 & 0.964 & 0.946 \\
		
		\hline
		100 & 4 & $l_P(\psi)$ & 2.987 & 2.698 & 1.922 & 3.551 & 2.698 & 0.739 & 0.473 \\ 
		&  & $l_{M^*}(\psi)$ & 0.019 & -0.067 & 0.822 & 0.822 & 0.529 & 0.923 & 0.925 \\ 
		&  & $l_{Cox}$ & 0.246 & 0.017 & 1.223 & 1.247 & 0.712 & 0.944 & 0.952 \\ 
		& 6 & $l_P(\psi)$& 1.203 & 1.147 & 0.756 & 1.421 & 1.147 & 0.894 & 0.608 \\ 
		&  & $l_{M^*}(\psi)$ & 0.029 & -0.009 & 0.516 & 0.517 & 0.351 & 0.994 & 0.950 \\ 
		&  & $l_{Cox}$ & 0.105 & 0.063 & 0.696 & 0.704 & 0.448 & 1.003 & 0.951 \\ 
		& 10 & $l_P(\psi)$ & 0.746 & 0.695 & 0.597 & 0.956 & 0.699 & 0.882 & 0.750 \\ 
		&  & $l_{M^*}(\psi)$& 0.016 & -0.016 & 0.461 & 0.461 & 0.298 & 0.960 & 0.936 \\ 
		&  & $l_{Cox}$ & 0.068 & 0.019 & 0.553 & 0.557 & 0.359 & 0.989 & 0.950 \\
		
		\hline
		250 & 4 & $l_P(\psi)$ & 2.683 & 2.547 & 1.086 & 2.894 & 2.547 & 0.784 & 0.079 \\ 
		&  & $l_{M^*}(\psi)$ & -0.089 & -0.126 & 0.481 & 0.489 & 0.347 & 0.968 & 0.921 \\ 
		&  & $l_{Cox}$ & 0.086 & 0.037 & 0.709 & 0.714 & 0.445 & 0.970 & 0.951 \\
		& 6 & $l_P(\psi)$ & 1.399 & 1.355 & 0.600 & 1.522 & 1.355 & 0.865 & 0.213 \\ 
		&  & $l_{M^*}(\psi)$ & -0.055 & -0.076 & 0.371 & 0.375 & 0.254 & 0.994 & 0.932 \\ 
		&  & $l_{Cox}$ & 0.040 & 0.006 & 0.497 & 0.499 & 0.308 & 1.004 & 0.946 \\
		& 10 & $l_P(\psi)$ & 0.697 & 0.678 & 0.354 & 0.782 & 0.678 & 0.931 & 0.457 \\ 
		&  & $l_{M^*}(\psi)$ & -0.021 & -0.033 & 0.274 & 0.275 & 0.182 & 1.009 & 0.945 \\ 
		&  & $l_{Cox}$ & 0.021 & 0.011 & 0.337 & 0.338 & 0.225 & 1.011 & 0.953 \\
		\hline
	\end{tabular}
\end{table}

\begin{table}[t]\centering
	\caption{Inference on the relative risk $\mathrm{RR}_2=e^{-1.5}$ in the stratified regression for right-censored survival data and probability of censoring $P_c=0.4$. The compared functions are the profile log-likelihood $l_P(\psi)$ and the MCMPL $l_{M^*}(\psi)$ computed with $R=500$ under the Weibull model and the partial log-likelihood $l_{Cox}$ under the Cox proportional hazards model. Results based on a simulation study with 2000 trials.}
	\label{tab:surv04rr2}
	\medskip	
	\begin{tabular}{cccccccccc}
		\hline
		$N$ & $T$ &Method & B & MB & SD & RMSE & MAE & SE/SD & 0.95 CI\\ 
		\hline
		50 & 4 & $l_P(\psi)$ & -0.082 & -0.085 & 0.035 & 0.089 & 0.085 & 0.808 & 0.264 \\ 
		&  & $l_{M^*}(\psi)$ & 0.003 & 0.001 & 0.040 & 0.041 & 0.028 & 0.981 & 0.943 \\ 
		& & $l_{Cox}$ & -0.004 & -0.007 & 0.060 & 0.060 & 0.040 & 0.993 & 0.949 \\ 
		& 6& $l_P(\psi)$ & -0.054 & -0.055 & 0.027 & 0.061 & 0.055 & 0.861 & 0.382 \\ 
		&  & $l_{M^*}(\psi)$ & 0.001 & 0.000 & 0.029 & 0.029 & 0.020 & 0.982 & 0.945 \\ 
		&  & $l_{Cox}$ & -0.004 & -0.005 & 0.039 & 0.040 & 0.027 & 1.000 & 0.954 \\
		& 10 & $l_P(\psi)$ & -0.030 & -0.030 & 0.021 & 0.037 & 0.030 & 0.896 & 0.603 \\ 
		&  & $l_{M^*}(\psi)$ & 0.001 & 0.001 & 0.021 & 0.022 & 0.014 & 0.968 & 0.940 \\ 
		&  & $l_{Cox}$ & -0.001 & -0.001 & 0.026 & 0.026 & 0.018 & 0.989 & 0.951 \\
		
		\hline
		100 & 4 & $l_P(\psi)$ & -0.083 & -0.083 & 0.024 & 0.086 & 0.083 & 0.790 & 0.066 \\ 
		&  & $l_{M^*}(\psi)$ & 0.005 & 0.004 & 0.028 & 0.029 & 0.019 & 0.958 & 0.942 \\ 
		&  & $l_{Cox}$ & -0.003 & -0.005 & 0.040 & 0.040 & 0.027 & 1.005 & 0.953 \\
		& 6 & $l_P(\psi)$ & -0.045 & -0.046 & 0.016 & 0.048 & 0.046 & 0.906 & 0.194 \\ 
		& & $l_{M^*}(\psi)$ & 0.001 & -0.000 & 0.018 & 0.018 & 0.012 & 1.005 & 0.949 \\ 
		&  & $l_{Cox}$ & -0.002 & -0.003 & 0.027 & 0.027 & 0.018 & 1.036 & 0.956 \\ 
		& 10 & $l_P(\psi)$ & -0.030 & -0.030 & 0.014 & 0.033 & 0.030 & 0.932 & 0.392 \\ 
		&  & $l_{M^*}(\psi)$ & 0.001 & 0.001 & 0.015 & 0.015 & 0.010 & 1.007 & 0.951 \\ 
		&  & $l_{Cox}$ & -0.001 & -0.001 & 0.019 & 0.019 & 0.013 & 1.005 & 0.955 \\
		
		\hline
		250 & 4 & $l_P(\psi)$ & -0.082 & -0.083 & 0.015 & 0.083 & 0.083 & 0.809 & 0.001 \\ 
		&  & $l_{M^*}(\psi)$ & 0.006 & 0.005 & 0.018 & 0.018 & 0.012 & 0.992 & 0.948 \\ 
		& & $l_{Cox}$ & -0.002 & -0.002 & 0.025 & 0.025 & 0.017 & 1.008 & 0.949 \\ 
		& 6 & $l_P(\psi)$ & -0.051 & -0.051 & 0.012 & 0.053 & 0.051 & 0.856 & 0.009 \\ 
		&  & $l_{M^*}(\psi)$ & 0.004 & 0.004 & 0.013 & 0.014 & 0.010 & 0.976 & 0.936 \\ 
		&  & $l_{Cox}$ & -0.000 & -0.000 & 0.018 & 0.018 & 0.012 & 0.996 & 0.947 \\ 
		& 10 & $l_P(\psi)$ & -0.029 & -0.029 & 0.009 & 0.031 & 0.029 & 0.936 & 0.080 \\ 
		&  & $l_{M^*}(\psi)$ & 0.002 & 0.002 & 0.009 & 0.009 & 0.006 & 1.010 & 0.950 \\ 
		& & $l_{Cox}$ & -0.000 & 0.000 & 0.012 & 0.012 & 0.008 & 1.016 & 0.955 \\
		
		\hline
	\end{tabular}
\end{table}

A thorough comparison between the outcomes of the two experiments above may be helpful to check whether and how the incidence of censored data in the sample affects the accuracy of the statistical techniques employed. 
Particularly, $l_P(\psi)$ appears to suffer more than $\mplmc$ from a high censoring probability. Indeed, in making inference on $\xi$ via the profile likelihood, only the coverages when $N=50$ are slightly more adequate with $P_c=0.4$. The same pattern is observed with regard to the estimated relative risks. On the contrary, conclusions descending from the MPL and the partial likelihood for the Cox model look less impacted by the percentage of observations subject to censoring. 

The empirical findings in this example are substantially in accordance with those relating to the contrast between the profile likelihood and the integrated likelihood in \cite{cor16}. Nonetheless, there exist three important motivations to prefer the MCMPL approach illustrated in Section~\ref{mcsurv}. Firstly, it is far less computationally expensive, as the effort entailed by the numerical integration to calculate Severini's integrated likelihood in the regression setting is considerable. Secondly, its basic procedure easily lends itself to encompass the bootstrap for nonparametric estimation of the censoring mechanism, protecting against misspecification risks. And thirdly, it can handle different distributions of the failure times $\widetilde{Y}_{it}$, such as logNormal or Gamma, whereas the method of \cite{cor16} demands to derive ad hoc formulae for finding a suitable reparametrization of the model \citep{sev07}.

\subsection{Application to an HIV clinical trial}
We now employ the proposed methodology in the analysis of a dataset from one clinical trial conducted to compare the time to death under  two different treatments for Mycobacterium avium complex, a frequent disease in late-stage HIV-infected people \citep{carhod99}. 
Data are the observed survival times and the corresponding event indicators, i.e. the realizations  $(y_{it}, \delta_{it} )$ from the pairs $(Y_{it}, \Delta_{it} )$, recorded along with the treatment used ($x_{it}=0,1$, respectively, in treatment groups ``Tx 1'' and ``Tx 2''), for a total of 69 patients enrolled by $N=11$ different medical centers. 
While such a number of clinics (i.e., groups) is small compared with the simulation settings of the previous section, the interesting aspect here is the relatively moderate amount of patients
(i.e., cluster sizes $T_i$, $i=1,\ldots, 11$) followed by most of the centers. Moreover, very few or no events of death are observed in each group (only 5 people died in group ``Tx 1'', and 13 in ``Tx 2''), with a global proportion of censored observations equal to 74\%. We remark that the simulation-based evidence attested that ordinary inferential techniques especially suffer from a high censoring probability.

Fitting the Weibull model \eqref{hazi} via the profile likelihood returns estimates $\hat{\xi}=1.150$ (s.e. $0.236$) and $\hat{\beta}=-1.012$ (s.e. $0.520$). Exploiting the invariance property, one can say that the estimated relative risk is $e^{-\hat{\xi}\hat{\beta}}=3.199$, indicating a clear higher mortality rate for patients under the second treatment ($x_{it}=1$). The likelihood ratio test for $H_0\!:\beta=0$ rejects at the 0.05 level the hypothesis of no treatment effect ($p$-value = 0.027). Confidence intervals for $\xi$ and $\beta$ based on the corresponding profile likelihood ratio statistics are (0.742, 1.669) and (-2.310, -0.111), respectively.

\begin{figure}[t]\centering
	\includegraphics[width=0.95\textwidth]{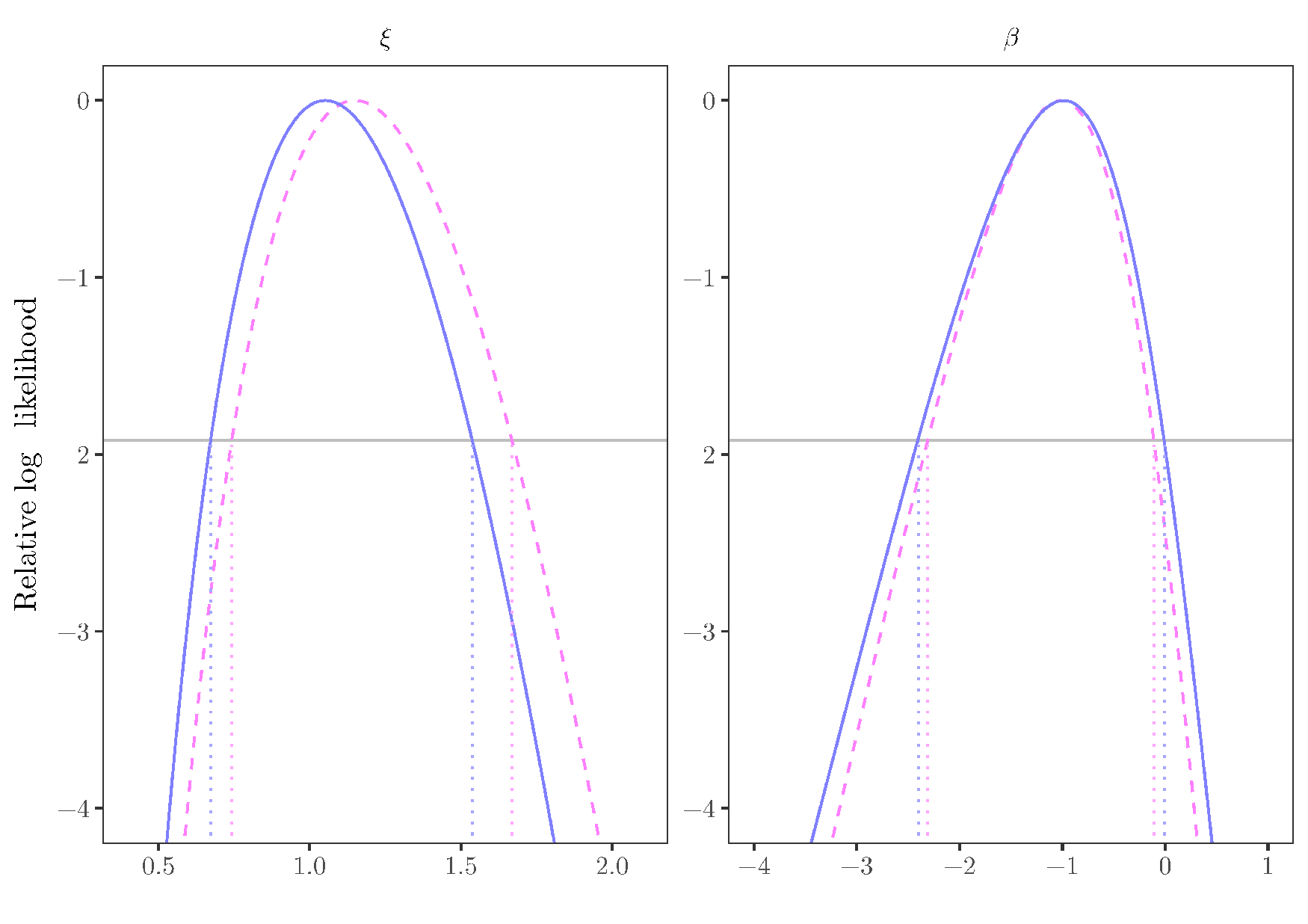}
	\caption{\label{fig:HIV} Relative profile (dashed) and Monte Carlo modified profile (solid) log-likelihoods of the HIV data for $\xi$ and $\beta$, with corresponding 0.95 confidence intervals.}
\end{figure}

The study performed through the MCMPL does not produce the same significant results in support of the first treatment. The estimates are $\hat{\xi}_{M^*}=1.051$ (s.e. $0.221$) and $\hat{\beta}_{M^*}=-0.981$ (s.e. $0.564$), implying an hazard ratio equal to $e^{-\hat{\xi}_{M^*}\hat{\beta}_{M^*}}=2.806$. Testing $H_0\!:\beta=0$ by means of the MPL ratio statistic leads now to a more dubious conclusion, since the $p$-value equals 0.049, and such uncertainty is reflected by the 0.95 confidence interval for $\beta$ obtained by inversion of the same quantity, namely (-2.400, -0.004). A similar interval for $\xi$ is instead (0.672, 1.538). Figure \ref{fig:HIV} provides a graphical representation of the inferential discrepancy existing between the two contrasted methods, even in this case with a moderate value of $N$.

Inference on the relative risk can also be made by fitting a stratified Cox proportional hazards regression by means of the {\tt R} function \texttt{coxph}, introduced in Section~\ref{sec:survsim}. Specifically, under this model the estimated risk ratio is equal to 1.953 (s.e. 1.054), with corresponding 0.95 confidence interval (0.641, 5.950) returned as output.  The analogue Wald confidence regions for the Weibull relative risk based on the profile likelihood and MCMPL, resulting upon application of the Delta method, are instead (2.697, 3.702) and (2.378, 3.234), respectively. Thus, in this specific example, the interval obtained under the semiparametric Cox model is significantly wider than both those obtained assuming a Weibull distribution for the survival times, since there is a substantial price to be paid for the weaker assumption with such a moderate sample size.

\section{Discussion}
\label{sec:disc}
This work shows how to exploit Monte Carlo simulation for widening the field of application of the MPL \citep{barndorffnielsen80, barndorffnielsen83}. 
\cite{severini98} made a first step in this direction, yet his approximation is still not approachable enough to deal with the today's degree of modelling sophistication for clustered units. Our solution, introduced in Section \ref{sec:MPL}, helps to fill such a gap in accessibility and to solve the incidental parameters problem even when the experimental design imposes quite complex assumptions on the analysis.
The suggested procedure is easy, implementable in broad generality and reasonably fast, and for these reasons could be thought of as the default choice in applications. Indeed, an extended version of the package \texttt{panelMPL} is currently under development and will also handle the automatic computation of the MCMPL for the model classes discussed here.

Section \ref{sec:MD} addresses issues in inferences on the parameter of interest related to the presence of missing values in binary grouped data. 
In this case analytical calculation of Severini's MPL is practicable but is not a simple task, while its approximation can be done by means of a simple two-step procedure to simulate the required Monte Carlo samples.
Results of simulation studies are presented here for the logistic regression scenario and for the probit regression in the Supplementary material. In the analysis of MCAR observations Monte Carlo simulation is found unnecessary to compute the MPL, as the inferential precision of the MCAR MCMPL appears equivalent to that of the analytical MPL which disregards the missing data.
Remarkably, the MNAR MCMPL sets an example of robustness to the ignorable incompleteness of the data.
When the true mechanism of missingness is nonignorable, the MNAR MCMPL proves to be generally more accurate than Severini's function, especially if $T$ is not too small, for any $N$. Justifications for this outcome are given in Section \ref{sec:logsimu}. 
To explore the usefulness of the Monte Carlo strategy when computing the MPL on partially MAR or partially missing always at random data for inference about $\psi$ \citep{litetal17} could be another possible direction of research.

Clustered survival times subject to right-censoring are discussed in Section \ref{sec:surv}. Under the Weibull regression model with group-related intercepts, our proposed approximation to the MPL is made necessary by the lack of distributional assumptions on the random censoring mechanism. An explicit calculation of Severini's modification requires full parametric specification of the density for the censoring times, whereas the Monte Carlo strategy enables to estimate it nonparametrically, using a conditional bootstrap \citep[Algorithm 3.1]{dav97}. Experimental outcomes examined in Section \ref{sec:survsim} corroborate the theory pertaining to inference in the standard two-index asymptotic setting. Estimation of the parameter of interest via the MCMPL is preferable to that via the profile likelihood in every relevant respect and is not affected by the proportion of censored data in the sample. Finally, the computational burden demanded by existing alternative statistical procedures \citep{cor16} is much heavier than that of the solution adopted here.

The potential room for future applications is vast, thanks to the generality of the methodology presented. One instance is given by semiparametric regression models where the incidental nuisance parameters are expressed as unknown real-valued functions \citep{he14}.

\section*{Acknowledgments}
The authors are grateful to Ruggero Bellio for helpful discussions and to Geert Dhaene and Annamaria Guolo for useful comments and observations that significantly helped to improve the value of this work.
They also thank Clovis Kenne Pagui for his contribution in writing the new code of the R package \texttt{panelMPL}.
The research of the first and third authors was supported by grant PRIN 2015 grant 2015EASZFS 003 from the  Ministry for Education, University and Research, while the second author was supported by `Progetto di Ateneo PRAT2015' (CPDA153257), University of Padova.

\bibliographystyle{chicago}
\bibliography{MPLref}

\end{document}


\maketitle

\section{Introduction}
Additional simulation results concerning possible applications of the methodology presented in Section~3 are available here.
Specifically, Section~\ref{sec:ar} deals with the autoregressive model for nonstationary normally-distributed panel data, where computation of the MPL is quite cumbersome. The probit regression model with missing response is instead considered in Section~\ref{sec:probit}, which complements the analysis for the logistic link shown in Section~4.3 of the paper. For both link functions, results obtained within the fixed effects simulation setup are also reported. Finally, Section~\ref{sec:surv} contains supplementary outcomes of simulations for the stratified Weibull regression under right censoring, described in Section~5.4.

\section{Nonstationary normal AR(1) model}
\label{sec:ar}
\subsection{Setup and background}
Let us consider the nonstationary version of the first-order autoregressive model for normal response
\begin{equation}\label{ar1}
Y_{it}|Y_{i,t-1}=y_{i,t-1}\sim N(\lambda_i+\rho y_{i,t-1}, \sigma^2), \qquad i=1,\dots,N,\ \ \; t=1,\dots,T,
\end{equation}
with $y_{0}=(y_{10},\dots,y_{N0})$ vector of unrestricted given initial conditions. The parameter of interest is $\psi=(\rho, \sigma^2) \in \reali \times \reali^+$ and $\lambda=(\lambda_1,\dots,\lambda_N) \in \reali ^N$ denotes the nuisance component of fixed effects. 
The lack of stationarity of the stochastic process $Y_{it}$ in each $i$th group $(i=1,\dots,N)$ implies the temporal variation of its mean or its autocovariance function, i.e. the covariance of the response with itself at pairs of time points.
As a consequence, the autoregressive parameter $\rho$ is left free to equal or exceed unity and the fixed vector $y_{0}$ does not need to meet any specific requirement, so that the log-likelihood function is expressed by conditioning on it. In order to facilitate the presentation, both exogenous covariates and further lagged responses $y_{i,t-l}\, (l>1)$ are excluded from the set of model regressors; however, no additional difficulties would be encountered in applying the proposed methodology otherwise.

The incidental parameters problem occurring in the analogue stationary AR(1) model has been addressed in the statistical literature several times. Particularly, \cite{cruddas89} proved that a marginal likelihood for $\psi$ exists and is equivalent to the MPL introduced by \cite{barndorffnielsen80, barndorffnielsen83}. Also econometricians showed interest in this issue and one latest proposition to improve ML inference in fixed effects dynamic models for stationary panel data is given by the bias-corrected estimator of \cite{dhaene16}, specially tailored for macroeconomic settings with $N=O(T)$.

Here, a great deal of attention is paid to the nonstationarity assumption of model (\ref{ar1}). Analytical derivation of $\il$ in this case would be possible but quite tedious. Instead Monte Carlo approximation dramatically reduces the amount of effort demanded to use Severini's modification. Moreover, we are specifically concerned with datasets where $T$ is much smaller than $N$, i.e. with situations where $l_P(\psi)$ exhibits its worst performance. Estimation of $\psi$ under these conditions was already investigated in the past. For example, inference in
autoregressions of order $l$ 
was thoroughly examined in \cite{dhaene13}, who obtained an adjusted profile log-likelihood through integration of a recentered score function. For $l=1$, their solution is essentially the same as
that found in \citet[Section~3]{lancaster02}.
From a purely Bayesian perspective, the latter proposed a strategy
to integrate out the incidental parameters from the likelihood in order to derive a marginal posterior density with consistent mode for $\psi$. 
Such approach is in substance equivalent to that adopted by \cite{debin15} to make inference on $\rho$ via the frequentist integrated likelihood of \cite{sev07}.

\subsection{Monte Carlo modified profile likelihood}\noindent
When groups are supposed independent, the log-likelihood of model (\ref{ar1}) conditioned on the initial vector $y_{0}$ is
\begin{equation}
\label{lar}
l(\theta) = - \sum_{i=1}^N \bigg\{\dfrac{T}{2}\mathrm{log} \sigma^2 + \dfrac{1}{2\sigma^2}\sum_{t=1}^T(y_{it}-\lambda_i-\rho y_{i,t-1})^2\bigg\}\,,
\end{equation}
and the score function 
takes the form
\[
l_{\lambda_i}(\theta)=\dfrac{1}{\sigma^2}\sum_{t=1}^T(y_{it}-\lambda_i-\rho y_{i,t-1})\,, \qquad i=1,\dots,N.
\]
Solution for $\lambda_i$ to the equation $l_{\lambda_i}(\theta)=0$ delivers the constrained ML estimate of $\lambda_i$ depending just on the autoregressive parameter
\begin{equation}\label{lambdarho}
\hat{\lambda}_{i\psi}=\bar{y}_i-\rho\bar{y}_{i,-1}=\hat{\lambda}_{i\rho}\,,
\end{equation}
where $\bar{y}_i=\sum_{t=1}^T y_{it}/T$ and $\bar{y}_{i,-1}=\sum_{t=0}^{T-1} y_{it}/T$. 
The profile log-likelihood is then obtained by replacement of $\lambda_i$ with $\hat{\lambda}_{i\rho}$ in expression (\ref{lar}) for each $i=1,\dots,N$. 

The first part in Severini's modification term is immediately available, since $j_{\lambda_i\lambda_i}(\hat{\theta}_\psi)= T/\sigma^2$. Yet, the derivation of
$\il$ requires more elaboration. The ML estimate of $\lambda_i$ equals $\hat{\lambda}_i=\hat{\lambda}_{i\hat{\rho}}=\bar{y}_i-\hat{\rho}\bar{y}_{i,-1}$, where
\begin{equation}
\label{rho:ols}
\hat{\rho}= \dfrac{\sum_{i=1}^N \sum_{t=1}^T y_{it}y_{i,t-1}-T\sum_{i=1}^N \bar{y}_i \bar{y}_{i,-1}}{\sum_{i=1}^N \sum_{t=1}^T y^2_{i,t-1}-T\sum_{i=1}^N \bar{y}^2_{i,-1}}\,.
\end{equation}
Therefore, by adding and subtracting the same quantity $\rho\bar{y}_{i,-1}$, one gets
\begin{align}
\hat{\lambda}_i&=\bar{y}_i-\rho\bar{y}_{i,-1}+\rho\bar{y}_{i,-1}-\hat{\rho}\bar{y}_{i,-1}\nonumber\\
&=\hat{\lambda}_{i\rho}-(\hat{\rho}-\rho)\bar{y}_{i,-1}\,. \label{eq:lhat}
\end{align}
The equation above lets us express the score evaluated at the constrained ML estimate in a convenient way. In particular, we start writing
\begin{align}\label{bah}
l_{\lambda_i}(\hat{\theta}_\psi)\nonumber
&=\dfrac{1}{\sigma^2} \sum_{t=1}^T\big(y_{it}-\hat{\lambda}_{i\rho}-\rho y_{i,t-1}\big)
\\
&=\dfrac{1}{\sigma^2} \sum_{t=1}^T\big(y_{it}-\hat{\lambda}_{i\rho}+\hat{\lambda}_{i}-\hat{\lambda}_{i}
-\rho y_{i,t-1}+\hat{\rho}y_{i,t-1}-\hat{\rho}y_{i,t-1}\big)\,,
\end{align}
where the second equality holds because we simultaneously sum to and subtract from the bracketed part both $\hat{\lambda}_i$ and $\hat{\rho} y_{i,t-1}$.
Now, since manipulating (\ref{eq:lhat}) leads to \begin{equation*}
\hat{\lambda}_{i\rho}= \hat{\lambda}_i + (\hat{\rho}-\rho)\bar{y}_{i,-1}\,, 
\end{equation*}
by substitution of the latter expression in (\ref{bah}) it is not hard to obtain
\begin{align*}
l_{\lambda_i}(\hat{\theta}_\psi)&=\dfrac{1}{\sigma^2} \bigg\{\sum_{t=1}^T\big(y_{it}-\hat{\lambda}_{i}-\hat{\rho}y_{i,t-1}\big)+T\big(\hat{\lambda}_{i}-\hat{\lambda}_{i\rho}\big)+\sum_{t=1}^T\big(\hat{\rho}-\rho\big)y_{i,t-1}\bigg\}
\\
&=\dfrac{1}{\sigma^2} \big\{\hat{\sigma}^2l_{\lambda_i}(\hat{\theta})+T\big(\hat{\lambda}_{i}-\hat{\lambda}_{i\rho}\big)+T\big(\hat{\rho}-\rho\big)\bar{y}_{i,-1}\big\}\,.
\end{align*}
Then, the necessary expected value is a linear function of $\rho$, and specifically
\begin{equation}
\begin{split}\label{iar1}
\il&=E_{\hat{\theta}} \big\{l_{\lambda_i}(\hat{\theta}_\psi)l_{\lambda_i}(\hat{\theta})\big\}\\
&=\dfrac{1}{\sigma^2}E_{\hat{\theta}}\big[\big\{\hat{\sigma}^2l_{\lambda_i}(\hat{\theta})+T\big(\hat{\lambda}_{i}-\hat{\lambda}_{i\rho}\big)+T\big(\hat{\rho}-\rho\big)\overline{Y}_{i,-1}\big\}l_{\lambda_i}(\hat{\theta})\big]\\
&=\dfrac{1}{\sigma^2} \big\{\hat{\sigma}^2\widehat{E}_1+T\big(\hat{\rho}-\rho\big) \widehat{E}_2\big\}\,,
\end{split}
\end{equation}
with $\widehat{E}_1=E_{\hat{\theta}}\big\{l^2_{\lambda_i}(\hat{\theta})\big\}$ and $\widehat{E}_2=E_{\hat{\theta}}\big\{\overline{Y}_{i,-1}l_{\lambda_i}(\hat{\theta})\big\}$. Note that expectation (\ref{iar1}) is computed with reference to the distribution $p(y_{it}|x_{it}; \hat{\psi}, \hat{\lambda}_i)$.

Although possible in principle, the analytical calculation of $\widehat{E}_1$ and $\widehat{E}_2$ is not straightforward. Conversely, estimating $\il$ via Monte Carlo simulation represents an 
easily implementable solution. Based on what illustrated in Section~3, the empirical mean for approximating (\ref{iar1}) becomes here
\begin{equation}\label{istar1}
\ilst=\dfrac{1}{R}\sum_{r=1}^R \bigg[\bigg\{\dfrac{1}{\sigma^2}\sum_{t=1}^T\big(y^{r}_{it}-\hat{\lambda}_{i\rho}-\rho y^{r}_{i,t-1}\big)\bigg\}
\bigg\{\dfrac{1}{\hat{\sigma}^2}\sum_{t=1}^T\big(y^{r}_{it}-\hat{\lambda}_i-\hat{\rho}y^{r}_{i,t-1}\big)\bigg\}
\bigg]\,,
\end{equation}
where $y^{r}_{it}\, (i=1,\dots,N, t=1,\dots,T)$ is generated by model (\ref{ar1}) with $(\psi, \lambda)=(\hat{\psi},\hat{\lambda})$, but the starting vector is kept unchanged, namely $y^r_{0}=y_{0}$ for each $r=1,\dots,R$. 
We acknowledge that, in this specific case, (\ref{iar1}) could be alternatively obtained through analogue Monte Carlo approximations to the expected values $\widehat{E}_1$ and $\widehat{E}_2$, which need to be derived just once because they involve $\hat{\theta}$ only. Nevertheless, for the reasons exposed in Section~3, the general approach detailed by (\ref{istar1}) is not much more costly in terms of computing effort.

\subsection{Computational aspects}\label{sec:compAR}\noindent
The estimate $\hat{\theta}$ can be written in closed form by applying the ordinary least squares method to the linear autoregression with normally distributed errors corresponding to (\ref{ar1}). As a consequence, the ML estimate of $\sigma^2$ is expressed by
\begin{equation}\label{sigmahat}
\hat{\sigma}^2 = \sum_{i=1}^N \sum_{t=1}^T \dfrac{\big(y_{it}-\hat{\lambda}_i-\hat{\rho}y_{i,t-1}\big)^2}{NT}\,,
\end{equation}
where formulations for $\hat{\rho}$ and $\hat{\lambda}_i$ follow directly from (\ref{rho:ols}).
On the contrary, maximization of $\mplmc$ for finding the estimate $\mplest=(\hat{\rho}_{M^*}, \hat{\sigma}^2_{M^*})$ usually has to be performed by means of numerical algorithms, and estimated standard errors are derived using the second derivative of the function at its maximum. Under this particular scenario, it is more convenient to derive $\hat{\sigma}^2_{M^*}$ by evaluation of the explicit constrained estimate
\[ \hat{\sigma}^2_{\rho,M^*}=\hat{\sigma}^2_{M^*}(\rho) =\sum_{i=1}^N \sum_{t=1}^T \dfrac{\big(y_{it}-\hat{\lambda}_{i\rho}-\rho y_{i,t-1}\big)^2}{N(T-1)}    \]
at $\hat{\rho}_{M^*}$, i.e. the scalar solution to the optimization problem with objective function $l^\rho_{M^*}(\rho)=l_{M^*}\big(\rho,\hat{\sigma}^2_{\rho,M^*}\big)$. Observe that also $l_P(\psi)$ can be further profiled in order to get $l^\rho_P(\rho)=l_P(\rho, \hat{\sigma}^2_\rho)$, where $\hat{\sigma}^2_\rho$ takes the form equivalent to (\ref{sigmahat}), but with estimates $\hat{\rho}$ and $\hat{\lambda}_i$ replaced by $\rho$ and $\hat{\lambda}_{i\rho}$ as in (\ref{lambdarho}), respectively. 

According to expression (\ref{iar1}), for values of the autoregressive parameter beyond a certain threshold depending on $\hat{\rho}$ the expectation $\il$ is negative and $\mpls$ is not computable, similarly to what happens with the integrated likelihood of \cite{debin15}.
In its turn, the approximation $\ilst$ can be smaller than or equal to zero for not too large values of $\rho$. 
A potentially undefined modification term poses a problem for the numerical optimization of $l^\rho_{M^*}(\rho)$. In addition, as will emerge from the figures available in Section~\ref{sec:arsimu}, the MCMPL is found to reach its global maximum as $\rho\rightarrow+\infty$ for any sample size, in accordance with the various functions for inference on $\psi$ studied in \cite{lancaster02}, \cite{dhaene13} and \cite{debin15}, respectively. On such grounds, we choose to maximize $l^\rho_{M^*}(\rho)$ by performing
a one-dimensional search in a real bounded interval $\varUpsilon$ through the algorithm implemented by the R function \texttt{optimize}. Specifically, adopting the same notation of \cite{lancaster02}, $\varUpsilon=(-\rho_l, \rho_u)$  with $\rho_l=\rho_u=1.5$, since in usual applications the autoregressive parameter is hardly observed to lie outside these extremes. The estimate resulting from the local maximization of $\mplmc$ is then uniquely defined as $\mplest\!=\!\big(\hat{\rho}_{M^*}, \hat{\sigma}^2_{M^*}\big)$, where $\hat{\rho}_{M^*}\!=\!\arg \max_{\rho\in \varUpsilon} l^\rho_{M^*}(\rho)$ and $\hat{\sigma}^2_{M^*}\!=\!\hat{\sigma}^2_{\hat{\rho}_{M^*},M^*}$. A careful discussion about the conditions under which consistency of this local maximizer is achieved is beyond the scope of the present work. We refer to \cite{dhaene13} for further details on the topic.

\begin{table}[!t]\centering\vspace{0.5cm}
	\caption{Inference on $\rho=0.5$ in the nonstationary AR(1) model for panel data. The compared functions are the profile log-likelihood $l_P(\psi)$ and the MCMPL $l_{M^*}(\psi)$ computed with $R=500$. Results based on a simulation study with 2000 trials.}
	\label{05rho}
	\medskip	
	\begin{tabular}{cccccccccc}
		\hline
		$N$ & $T$ &Method & B & MB & SD & RMSE & MAE & SE/SD & 0.95 CI\\ 
		\hline
		250 & 4 & $l_P(\psi)$& -0.186 & -0.186 & 0.025 & 0.187 & 0.186 & 0.879 & 0.000 \\ 
		& & $\mplmc$ & 0.020 & 0.018 & 0.037 & 0.042 & 0.028 & 0.921 & 0.915 \\ 
		
		& 8 & $l_P(\psi)$ & -0.114 & -0.115 & 0.018 & 0.116 & 0.115 & 0.921 & 0.000 \\ 
		& & $\mplmc$ & 0.002 & 0.002 & 0.020 & 0.020 & 0.013 & 0.989 & 0.942 \\ 
		
		& 16 & $l_P(\psi)$ & -0.070 & -0.070 & 0.013 & 0.071 & 0.070 & 0.960 & 0.000 \\ 
		& & $\mplmc$ & -0.000 & 0.000 & 0.014 & 0.014 & 0.009 & 1.002 & 0.944 \\ 
		\hline
		500 & 4 & $l_P(\psi)$ & -0.184 & -0.183 & 0.017 & 0.184 & 0.183 & 0.896 & 0.000 \\ 
		& & $\mplmc$ & 0.018 & 0.019 & 0.025 & 0.031 & 0.022 & 0.952 & 0.881 \\ 
		
		& 8 & $l_P(\psi)$ & -0.113 & -0.113 & 0.013 & 0.114 & 0.113 & 0.902 & 0.000 \\ 
		& & $\mplmc$& 0.002 & 0.002 & 0.014 & 0.014 & 0.010 & 0.972 & 0.943 \\ 
		
		& 16 & $l_P(\psi)$ & -0.069 & -0.069 & 0.009 & 0.069 & 0.069 & 0.983 & 0.000 \\ 
		& & $\mplmc$ & 0.000 & 0.000 & 0.009 & 0.009 & 0.007 & 1.029 & 0.959 \\ 
		\hline
		1000 & 4 & $l_P(\psi)$ & -0.187 & -0.187 & 0.013 & 0.187 & 0.187 & 0.879 & 0.000 \\ 
		& & $\mplmc$& 0.019 & 0.018 & 0.019 & 0.026 & 0.019 & 0.923 & 0.795 \\ 
		
		& 8 & $l_P(\psi)$ & -0.115 & -0.115 & 0.009 & 0.115 & 0.115 & 0.919 & 0.000 \\ 
		& & $\mplmc$ & 0.002 & 0.002 & 0.010 & 0.010 & 0.007 & 0.987 & 0.948 \\
		
		& 16 & $l_P(\psi)$ & -0.070 & -0.070 & 0.007 & 0.070 & 0.070 & 0.935 & 0.000 \\ 
		& & $\mplmc$ & 0.000 & 0.000 & 0.007 & 0.007 & 0.005 & 0.977 & 0.940 \\ 
		\hline
	\end{tabular}
	\vspace{0.5cm}
\end{table}

\subsection{Simulation studies and numerical examples} \label{sec:arsimu}\noindent
In this section, the accuracy of the MCMPL in drawing inferences on $\psi$ under the nonstationary normal AR(1) model is assessed with regard to that of the standard profile likelihood through a series of simulations. Two main experiments based on $S=2000$ iterations are performed, both considering datasets with $T=4,8,16$ and $N=250, 500, 1000$.
The two simulation setups differ only in the true value of the autoregressive parameter used to generate the samples from model (\ref{ar1}): in the first $\rho=0.5$, while in the second $\rho=0.9$. 
The conditional variance of the response variable is $\sigma^2=1$ and the fixed effects are independently drawn from a $N(1,1)$ distribution, following the example of \cite{lancaster02}. In every simulated dataset, all $N$ initial observations in the vector $y_{0}$ are fixed equal to zero with no loss of generality, as this is equivalent to interpret each $y_{it}$ as $y_{it}-y_{i0}$ and each $\lambda_i$ as $\lambda_i- y_{i0}(1-\rho)$ $(i=1,\ldots,N, t=1,\ldots,T)$ \citep{lancaster02}. Lastly, the number of Monte Carlo replicates employed to compute $\mplmc$ is $R=500$.

Inferential results of the first study for $\rho$ and $\sigma^2$ are displayed in Tables \ref{05rho} and \ref{05sigma}, respectively, using indexes defined in Section~4.3. Similar comments as in \cite{bartetal16} can be made. In all configurations, no significant differences between bias and median bias of the same estimator are observed, but the improvement determined by using the MCMPL in this sense is remarkable. Consistently with the theory for independent units \citep{sartori03}, the bias does not vary with $N$ but decreases as $T$ increases, whereas the root mean squared error depends on both indexes. Empirical coverage probabilities of 0.95 Wald confidence intervals based on $\mplmc$ are generally accurate for both components of interest, with larger departures from the nominal level occurring when $T=4$. Such conspicuous refinements to the poor interval estimation supplied by $l_P(\psi)$ mainly stem from bias reduction. Yet some correction in curvature also takes place, being SE/SD for the MCMPL typically closer to one than for the ordinary profile likelihood.

\begin{table}[!t]\centering\vspace{0.5cm}
	\caption{Inference on $\sigma^2=1$ in the nonstationary AR(1) model for panel data with $\rho=0.5$. The compared functions are the profile log-likelihood $l_P(\psi)$ and the MCMPL $l_{M^*}(\psi)$ computed with $R=500$. Results based on a simulation study with 2000 trials.}
	\label{05sigma}
	\medskip	
	\begin{tabular}{cccccccccc}
		\hline
		$N$ & $T$ &Method & B & MB & SD & RMSE & MAE & SE/SD & 0.95 CI\\
		\hline
		250 & 4 & $l_P(\psi)$ & -0.300 & -0.301 & 0.036 & 0.303 & 0.301  & 0.862 & 0.000\\ 
		& & $\mplmc$& 0.013 & 0.011 & 0.060 & 0.062 & 0.041 & 0.976 & 0.954\\
		
		& 8 & $l_P(\psi)$  & -0.147 & -0.148 & 0.029 & 0.150 & 0.148 & 0.938 & 0.001 \\
		& & $\mplmc$& -0.000 & -0.001 & 0.035 & 0.035 & 0.024 & 1.011 & 0.957 \\
		
		& 16 & $l_P(\psi)$ & -0.071 & -0.071 & 0.022 & 0.074 & 0.071 & 0.964 & 0.092 \\
		& & $\mplmc$ & -0.001 & -0.001 & 0.023 & 0.023 & 0.016 & 1.000 & 0.954 \\
		\hline
		500 & 4 & $l_P(\psi)$ & -0.299 & -0.299 & 0.026 & 0.300 & 0.299 & 0.853 & 0.000 \\
		& & $\mplmc$& 0.013 & 0.013 & 0.043 & 0.045 & 0.029 & 0.973 & 0.944 \\
		
		& 8 & $l_P(\psi)$ & -0.147 & -0.148 & 0.020 & 0.148 & 0.148 & 0.935 & 0.000 \\
		& & $\mplmc$& -0.000 & -0.001 & 0.024 & 0.024 & 0.017 & 1.008 & 0.950 \\
		
		& 16 & $l_P(\psi)$  & -0.070 & -0.070 & 0.015 & 0.072 & 0.070 & 0.959 & 0.006 \\
		& & $\mplmc$ & -0.000 & -0.000 & 0.017 & 0.017 & 0.011 & 0.995 & 0.949 \\
		\hline
		1000 & 4 & $l_P(\psi)$  & -0.300 & -0.299 & 0.018 & 0.301 & 0.299 & 0.867 & 0.000 \\
		& & $\mplmc$& 0.013 & 0.014 & 0.030 & 0.033 & 0.022 & 0.988 & 0.930 \\
		
		& 8 & $l_P(\psi)$  & -0.147 & -0.147 & 0.015 & 0.147 & 0.147 & 0.923 & 0.000 \\
		& & $\mplmc$ & 0.001 & 0.000 & 0.018 & 0.018 & 0.012 & 0.995 & 0.945 \\
		
		& 16 & $l_P(\psi)$ & -0.070 & -0.070 & 0.011 & 0.071 & 0.070 & 0.976 & 0.000 \\
		& & $\mplmc$ & -0.000 & -0.000 & 0.012 & 0.012 & 0.008 & 1.012 & 0.948 \\
		\hline
	\end{tabular}
	\vspace{0.5cm}
\end{table}

Tables \ref{09rho} and \ref{09sigma} illustrate results of the simulation experiment run with a true value of $\rho$ approaching the boundaries of the stationary region $(-1, 1)$, particularly $\rho=0.9$. Relative behaviours of the two methods for estimating the component of interest are basically in line with those viewed in the previous study. Perhaps here the general improvements originating from the employment of $\mplmc$ are somewhat milder than when the autoregressive parameter is farther away from nonstationariety. This observation can be referred both to bias and, mostly, to empirical coverages of confidence intervals for $\rho$. Nonetheless, the quality of MPL-based inference remains unquestionably higher than that reached through standard ML techniques.

\begin{table}[!t]\centering\vspace{0.5cm}
	\caption{Inference on $\rho=0.9$ in the nonstationary AR(1) model for panel data. The compared functions are the profile log-likelihood $l_P(\psi)$ and the MCMPL $l_{M^*}(\psi)$ computed with $R=500$. Results based on a simulation study with 2000 trials.}
	\label{09rho}
	\medskip	
	\begin{tabular}{cccccccccc}
		\hline
		$N$ & $T$ &Method & B & MB & SD & RMSE & MAE & SE/SD & 0.95 CI\\ 
		\hline
		250 & 4 & $l_P(\psi)$ & -0.130 & -0.130 & 0.018 & 0.131 & 0.130 & 0.894 & 0.000 \\ 
		& & $\mplmc$ & 0.022 & 0.021 & 0.028 & 0.036 & 0.024 & 0.899 & 0.871 \\ 
		
		& 8 & $l_P(\psi)$ & -0.051 & -0.051 & 0.008 & 0.052 & 0.051 & 0.922 & 0.000 \\ 
		& & $\mplmc$ & 0.003 & 0.003 & 0.010 & 0.010 & 0.007 & 0.976 & 0.933 \\ 
		
		& 16 & $l_P(\psi)$ & -0.022 & -0.023 & 0.004 & 0.023 & 0.023 & 0.957 & 0.001 \\ 
		& & $\mplmc$ & 0.000 & 0.000 & 0.005 & 0.005 & 0.003 & 1.003 & 0.950 \\
		\hline
		500 & 4 & $l_P(\psi)$ & -0.128 & -0.128 & 0.013 & 0.129 & 0.128 & 0.905 & 0.000 \\ 
		& & $\mplmc$ & 0.021 & 0.020 & 0.019 & 0.028 & 0.021 & 0.928 & 0.774 \\ 
		
		& 8 & $l_P(\psi)$ & -0.050 & -0.050 & 0.006 & 0.050 & 0.050 & 0.933 & 0.000 \\ 
		& & $\mplmc$ & 0.003 & 0.003 & 0.007 & 0.007 & 0.005 & 0.980 & 0.928 \\ 
		
		& 16 & $l_P(\psi)$ & -0.022 & -0.022 & 0.003 & 0.022 & 0.022 & 0.957 & 0.000 \\ 
		& & $\mplmc$ & 0.001 & 0.001 & 0.003 & 0.003 & 0.002 & 1.001 & 0.946 \\ 
		\hline
		1000 & 4 & $l_P(\psi)$ & -0.131 & -0.131 & 0.009 & 0.131 & 0.131 & 0.895 & 0.000 \\ 
		& & $\mplmc$ & 0.021 & 0.021 & 0.014 & 0.025 & 0.021 & 0.909 & 0.612 \\ 
		
		& 8 & $l_P(\psi)$ & -0.051 & -0.051 & 0.004 & 0.051 & 0.051 & 0.923 & 0.000 \\ 
		& & $\mplmc$ & 0.003 & 0.003 & 0.005 & 0.006 & 0.004 & 0.969 & 0.884 \\ 
		
		&16 & $l_P(\psi)$ & -0.022 & -0.022 & 0.002 & 0.022 & 0.022 & 0.923 & 0.000 \\ 
		& & $\mplmc$ & 0.001 & 0.001 & 0.002 & 0.002 & 0.002 & 0.968 & 0.930 \\ 
		\hline
	\end{tabular}
	\vspace{0.5cm}
\end{table}

\begin{figure}[!t]\centering
	\includegraphics[width=0.95\textwidth]{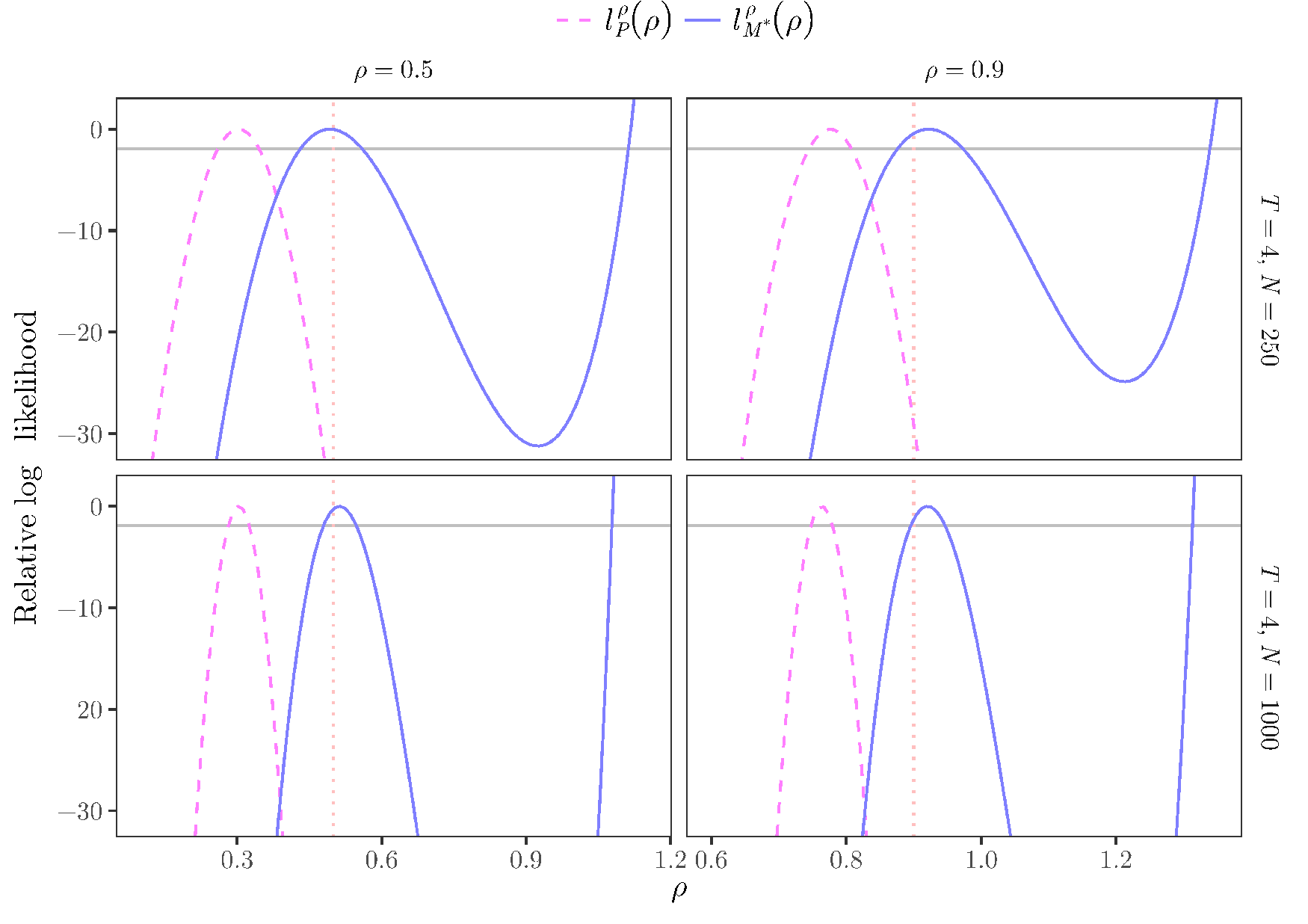}
	
	\caption{\label{fig:AR} Relative profile (dashed) and Monte Carlo modified profile (solid) log-likelihoods of four datasets generated under the nonstationary AR(1) model with $\rho = 0.5, 0.9$, $T=4$ and $N = 250, 1000$. The dotted vertical line indicates the true value of the autoregressive parameter, while the horizontal line gives the 0.95 confidence intervals for $\rho$ based on the likelihood ratio statistics.}
\end{figure}

Figure \ref{fig:AR} shows $l^\rho_P(\rho)$ and $l^\rho_{M^*}(\rho)$, as defined in Section~\ref{sec:compAR}, in their relative version. Specifically, the plots are referred to samples generated from model (\ref{ar1}) with $\rho = 0.5, 0.9$, $T=4$ and $N = 250, 1000$. All panels confirm the results of simulations discussed so far. The maximum of the profile log-likelihood is significantly 
smaller than the true value of the autoregressive parameter, corresponding to the vertical line. Because of this and the accentuated curvature of $l^\rho_P(\rho)$, such value never belongs to the 0.95 confidence region defined by inversion of the profile likelihood ratio statistic and marked by the horizontal line. Conversely, the local maximization of the MCMPL yields to adequate point and interval estimations of $\rho$. The unusual trend of $l^\rho_{M^*}(\rho)$, whose global maximizer lies at infinity, was already anticipated in Section~\ref{sec:compAR}. The absence of restrictions on the initial conditions in $y_{0}$ causes in fact the MCMPL to be re-increasing, sometimes already in the stationary parameter region \citep{dhaene13}.

\begin{table}[!t]\centering\vspace{0.5cm}
	\caption{Inference on $\sigma^2=1$ in the nonstationary AR(1) model for panel data with $\rho=0.9$. The compared functions are the profile log-likelihood $l_P(\psi)$ and the MCMPL $l_{M^*}(\psi)$ computed with $R=500$. Results based on a simulation study with 2000 trials.}
	\label{09sigma}
	\medskip	
	\begin{tabular}{cccccccccc}
		\hline
		N & T &Method & B & MB & SD & RMSE & MAE & SE/SD & 0.95 CI\\
		\hline
		250 & 4 & $l_P(\psi)$ & -0.297 & -0.298 & 0.036 & 0.299 & 0.298 & 0.870 & 0.000 \\ 
		& & $\mplmc$ & 0.021 & 0.017 & 0.062 & 0.066 & 0.043 & 0.956 & 0.944 \\  
		
		& 8 & $l_P(\psi)$ & -0.144 & -0.145 & 0.029 & 0.147 & 0.145 & 0.945 & 0.002\\ 
		& & $\mplmc$ & 0.002 & 0.000 & 0.035 & 0.035 & 0.024 & 1.007 & 0.956 \\ 
		
		& 16 & $l_P(\psi)$ & -0.070 & -0.070 & 0.021 & 0.074 & 0.070 & 0.969 & 0.095 \\ 
		& & $\mplmc$ & -0.001 & -0.001 & 0.023 & 0.023 & 0.016 & 1.001 & 0.951 \\  
		\hline
		500 & 4 & $l_P(\psi)$ & -0.295 & -0.295 & 0.026 & 0.297 & 0.295 & 0.861 & 0.000 \\ 
		& & $\mplmc$& 0.020 & 0.019 & 0.044 & 0.048 & 0.032 & 0.956 & 0.925 \\ 
		
		& 8 & $l_P(\psi)$ & -0.144 & -0.144 & 0.020 & 0.145 & 0.144 & 0.943 & 0.000 \\ 
		& & $\mplmc$& 0.001 & 0.001 & 0.024 & 0.025 & 0.017 & 1.006 & 0.954 \\ 
		
		& 16 & $l_P(\psi)$ & -0.069 & -0.069 & 0.015 & 0.071 & 0.069 & 0.962 & 0.006 \\ 
		& & $\mplmc$ & 0.000 & 0.000 & 0.017 & 0.017 & 0.011 & 0.993 & 0.945 \\ 
		\hline
		1000 & 4 & $l_P(\psi)$ & -0.296 & -0.296 & 0.018 & 0.297 & 0.296 & 0.875 & 0.000 \\ 
		& & $\mplmc$& 0.021 & 0.021 & 0.031 & 0.037 & 0.026 & 0.972 & 0.903 \\ 
		
		& 8 & $l_P(\psi)$ & -0.144 & -0.144 & 0.015 & 0.144 & 0.144 & 0.928 & 0.000 \\ 
		& & $\mplmc$& 0.003 & 0.002 & 0.018 & 0.018 & 0.012 & 0.988 & 0.942 \\ 
		
		& 16 & $l_P(\psi)$ & -0.070 & -0.070 & 0.011 & 0.070 & 0.070 & 0.981 & 0.000 \\ 
		& & $\mplmc$ & 0.000 & 0.000 & 0.012 & 0.012 & 0.008 & 1.012 & 0.952 \\ 
		\hline
	\end{tabular}
	
\end{table}

\section{Probit regression model with missing response}
\label{sec:probit}
Suppose now that specifications (8)--(10) in Section~4.2 of the paper hold with $F(\cdot)=\Phi(\cdot)$ and $G(\cdot)=\mathrm{logit}^{-1}(\cdot)$, where $\Phi(\cdot)$ is the CDF of the standard normal random variable. Even in probit regressions for complete clustered binary data $y_{it}$ $(i=1,\dots,N,\,t=1,\dots,T)$ an explicit formulation for Severini's adjustment exists, and can be computed on the available units under the MCAR assumption. Denoting by $\phi(\cdot)$ the probability density function of the $N(0,1)$, expectation (16) in Section~4.2 becomes
\begin{equation}\label{i-MCARprob}
I_{\lambda_i\lambda_i}(\hat{\theta};\hat{\theta}_{\beta})= \sum_{t :\, y_{it}\in y^{obs}}
\dfrac{\phi\big(\hat{\lambda}_{i\beta}+\beta^\T x_{it}\big)\phi\big(\hat{\lambda}_{i}+\hat{\beta}^\T x_{it}\big)}{\big\{1-\Phi\big(\hat{\lambda}_{i\beta}+\beta^\T x_{it}\big)\big\}\Phi\big(\hat{\lambda}_{i\beta}+\beta^\T x_{it}\big)}\,,\qquad i=1,\dots,N.
\end{equation}
Under these hypotheses, the $i$th score function may be expressed as
\begin{equation}\label{score-MCARprob}
l_{\lambda_i}(\theta)= \sum_{t :\, y_{it}\in y^{obs}}
\dfrac{\big\{y_{it}-\Phi\big(\lambda_{i}+\beta^\T x_{it}\big)\big\}\phi\big(\lambda_{i}+\beta^\T x_{it}\big)}{\Phi\big(\lambda_{i}+\beta^\T x_{it}\big)\big\{1-\Phi\big(\lambda_{i}+\beta^\T x_{it}\big)\big\}}\,,\qquad i=1,\dots,N,
\end{equation}
and $j_{\lambda_i\lambda_i}(\theta)$ is readily derived by changing sign to its first derivative with respect to $\lambda_i$. Using (\ref{i-MCARprob}) and (\ref{score-MCARprob}), it is possible to obtain both $l_{M}(\beta)$ in closed form and $l_{M^*}(\beta)$ as described in (17). In the present probit framework, the formula of the standard profile log-likelihood $l_P(\beta)$ follows in fact from (15) with $\pi_{it}=\Phi(\lambda_i + \beta^\T x _{it})$.

When we conjecture that incompleteness of the data originates from a nonignorable process, Monte Carlo simulation serves to approximate the unconditional expected value $I_{\lambda_i\lambda_i}(\hat{\varphi};\hat{\varphi}_{\psi})$, whose exact formulation is hardly retrievable. The expression of $\mplmc$ in the probit setting may be obtained by multiple substitution of $\Phi(\lambda_i + \beta^\T x _{it})$ for $\pi_{it}$, $\phi(\lambda_i + \beta^\T x _{it})$ for $f_{it}$ and $-(\lambda_i + \beta^\T x _{it})\phi(\lambda_i + \beta^\T x _{it})$ for $f'_{it}$ in equations (11)--(14) of the paper.

The numerical optimization methods employed for the various functions correspond to those of the logistic case. We also recall that exclusion of the non-informative clusters by the dataset must take place prior to the fitting phase.

The basic structure of the studies performed in Section~4.3 is held unchanged, yet here a probit link between the response and the unique predictor is considered. In the first experiment, missing observations are chosen according to an MCAR mechanism with $\gamma_1=2.5$; in the second, the true missingness generation process is MNAR with $\gamma_1=5$ and $\gamma_2=1$. The covariate is again simulated from the $N(-0.35,1)$ distribution. 
Exploiting the well-known relation between the logistic and normal distributions \citep{ame81}, in order to obtain data and quantity of informative groups comparable to the logistic setting, the complete simulated samples are generated by fixing $\beta=1/1.6=0.625$ and intercepts $\lambda_i$ $(i=1,\dots,N)$ are independently generated as $N(-0.22,0.39)$

The corresponding value of the parameter for $x_{it}$ under a marginal model with generalized estimating equations (GEE) is $\beta_m=\beta/\sqrt{1+\sigma^2_\lambda}$, where $\sigma^2_\lambda$ is the variance of the random effects' distribution \citep[Section 9.4.1]{agr15}. Here $\sigma^2_\lambda=0.39$, and therefore $\beta_m=0.625/\sqrt{1.39}=0.53$.  

Tables~\ref{tab:MCARprob} and \ref{tab:MCARprob2} summarize 
the usual measures of estimation accuracy for $\beta$ based on $S=2000$ simulations of the study regarding MCAR data. Relative behaviours of the three MCAR log-likelihoods illustrated by Table \ref{tab:MCARprob} do not significantly differentiate from those viewed in Table~1 of Section~4.3 for the logit link. The defective performance of $l_P(\beta)$ is greatly corrected by the adjustment proposed by Severini, from any relevant inferential perspective and for all possible couples $(T,N)$ with $T=4,6,10$ and $N=50,100,250$. Results achieved by $l_{M}(\beta)$ and $l_{M^*}(\beta)$ are still very similar, thanks to the validity of the MCAR hypothesis.

\begin{table}[t!]\centering
	\caption{Inference on $\beta=1/1.6$ in the probit regression for MCAR longitudinal data. The compared functions are the MCAR profile log-likelihood $l_P(\beta)$, Severini's exact MCAR MPL $l_{M}(\beta)$, and the MCAR MCMPL $l_{M^*}(\beta)$ computed with $R=500$. Results based on a simulation study with 2000 trials.}
	\label{tab:MCARprob}
	\medskip	
	\begin{tabular}{cccccccccc}
		\hline
		$N$ & $T$ &Method & B & MB & SD & RMSE & MAE & SE/SD & 0.95 CI\\ 
		\hline
		50 & 4 & $l_P(\beta)$ & 0.555 & 0.473 & 0.579 & 0.802 & 0.478 & 0.663 & 0.725 \\ 
		& & $l_{M}(\beta)$ & 0.142 & 0.131 & 0.283 & 0.317 & 0.201 & 1.018 & 0.967 \\ 
		& & $l_{M^*}(\beta)$ & 0.142 & 0.128 & 0.301 & 0.333 & 0.203 & 0.956 & 0.960 \\
		
		& 6 & $l_P(\beta)$  & 0.296 & 0.268 & 0.319 & 0.435 & 0.280 & 0.760 & 0.774 \\ 
		& & $l_{M}(\beta)$ & 0.074 & 0.065 & 0.213 & 0.226 & 0.141 & 0.960 & 0.956 \\ 
		& & $l_{M^*}(\beta)$ & 0.067 & 0.057 & 0.215 & 0.225 & 0.140 & 0.945 & 0.955 \\
		
		& 10 & $l_P(\beta)$ & 0.162 & 0.147 & 0.181 & 0.243 & 0.158 & 0.852 & 0.821 \\ 
		& & $l_{M}(\beta)$ & 0.039 & 0.029 & 0.145 & 0.151 & 0.095 & 0.960 & 0.940 \\ 
		& & $l_{M^*}(\beta)$ & 0.035 & 0.025 & 0.145 & 0.149 & 0.093 & 0.960 & 0.940 \\ 
		\hline
		100 & 4 & $l_P(\beta)$ & 0.460 & 0.421 & 0.358 & 0.582 & 0.421 & 0.707 & 0.585 \\ 
		& & $l_{M}(\beta)$ & 0.108 & 0.102 & 0.197 & 0.225 & 0.148 & 1.005 & 0.947 \\ 
		& & $l_{M^*}(\beta)$ & 0.104 & 0.095 & 0.204 & 0.229 & 0.147 & 0.970 & 0.941 \\
		
		& 6 & $l_P(\beta)$ & 0.284 & 0.270 & 0.211 & 0.354 & 0.270 & 0.805 & 0.636 \\ 
		& & $l_{M}(\beta)$ & 0.072 & 0.066 & 0.147 & 0.164 & 0.102 & 0.985 & 0.938 \\ 
		& & $l_{M^*}(\beta)$ & 0.064 & 0.056 & 0.148 & 0.161 & 0.100 & 0.975 & 0.941 \\
		
		& 10 & $l_P(\beta)$ & 0.155 & 0.148 & 0.127 & 0.200 & 0.149 & 0.867 & 0.703 \\ 
		& & $l_{M}(\beta)$ & 0.034 & 0.029 & 0.103 & 0.109 & 0.073 & 0.971 & 0.942 \\ 
		& & $l_{M^*}(\beta)$ & 0.030 & 0.025 & 0.102 & 0.107 & 0.072 & 0.972 & 0.944 \\ 
		\hline
		250 & 4 & $l_P(\beta)$ & 0.413 & 0.401 & 0.208 & 0.462 & 0.401 & 0.754 & 0.283 \\ 
		& & $l_{M}(\beta)$ & 0.095 & 0.091 & 0.124 & 0.156 & 0.105 & 1.017 & 0.907 \\ 
		& & $l_{M^*}(\beta)$ & 0.087 & 0.084 & 0.127 & 0.154 & 0.101 & 0.990 & 0.909 \\ 
		
		& 6 & $l_P(\beta)$ & 0.260 & 0.253 & 0.130 & 0.291 & 0.253 & 0.818 & 0.340 \\ 
		& & $l_{M}(\beta)$ & 0.058 & 0.055 & 0.092 & 0.109 & 0.074 & 0.989 & 0.915 \\ 
		& & $l_{M^*}(\beta)$ & 0.049 & 0.046 & 0.092 & 0.105 & 0.071 & 0.982 & 0.921 \\
		
		& 10 & $l_P(\beta)$ & 0.145 & 0.143 & 0.077 & 0.164 & 0.143 & 0.904 & 0.462 \\ 
		& & $l_{M}(\beta)$ & 0.027 & 0.025 & 0.062 & 0.068 & 0.045 & 1.012 & 0.941 \\ 
		& & $l_{M^*}(\beta)$ & 0.022 & 0.020 & 0.062 & 0.066 & 0.043 & 1.014 & 0.947 \\ 
		\hline
	\end{tabular}
\end{table}

\begin{table}[t!]\centering
	\caption{Inference on $\beta=1/1.6$ in the probit regression for MCAR longitudinal data. The compared methods are the MNAR profile log-likelihood $l_P(\psi)$, the MNAR MCMPL $l_{M^*}(\psi)$ computed with $R=500$ and GEE. Results based on a simulation study with 2000 trials.}
	\label{tab:MCARprob2}
	\medskip	
	\begin{tabular}{cccccccccc}
		\hline
		$N$ & $T$ &Method & B & MB & SD & RMSE & MAE & SE/SD & 0.95 CI\\ 
		\hline
		50 & 4 & $l_P(\psi)$ & 0.419 & 0.340 & 0.643 & 0.767 & 0.424 & 0.630 & 0.784 \\ 
		&  & $l_{M^*}(\psi)$ & 0.121 & 0.099 & 0.296 & 0.320 & 0.183 & 0.996 & 0.981 \\ 
		&  & GEE & 0.026 & 0.020 & 0.182 & 0.184 & 0.120 & 0.987 & 0.955 \\ 
		
		& 6 & $l_P(\psi)$ & 0.213 & 0.186 & 0.350 & 0.409 & 0.250 & 0.720 & 0.809 \\ 
		&  & $l_{M^*}(\psi)$ & 0.044 & 0.031 & 0.213 & 0.218 & 0.136 & 0.978 & 0.968 \\ 
		&  & GEE & 0.014 & 0.010 & 0.152 & 0.153 & 0.099 & 0.971 & 0.949 \\ 
		
		& 10 & $l_P(\psi)$ & 0.111 & 0.109 & 0.217 & 0.244 & 0.156 & 0.728 & 0.814 \\ 
		&  & $l_{M^*}(\psi)$ & 0.026 & 0.016 & 0.144 & 0.146 & 0.093 & 1.001 & 0.955 \\ 
		&  & GEE & 0.012 & 0.010 & 0.112 & 0.113 & 0.074 & 1.000 & 0.974 \\
		\hline
		
		100 & 4 & $l_P(\psi)$ & 0.335 & 0.302 & 0.400 & 0.522 & 0.332 & 0.691 & 0.731 \\ 
		&  & $l_{M^*}(\psi)$ & 0.080 & 0.067 & 0.196 & 0.212 & 0.130 & 1.029 & 0.964 \\ 
		&  & GEE & 0.011 & 0.009 & 0.127 & 0.127 & 0.084 & 0.993 & 0.957 \\ 
		
		& 6 & $l_P(\psi)$ & 0.178 & 0.183 & 0.266 & 0.320 & 0.222 & 0.649 & 0.715 \\ 
		&  & $l_{M^*}(\psi)$ & 0.035 & 0.028 & 0.147 & 0.151 & 0.093 & 1.007 & 0.955 \\ 
		&  & GEE & 0.009 & 0.009 & 0.103 & 0.103 & 0.070 & 1.021 & 0.954 \\ 
		
		& 10 & $l_P(\psi)$ & 0.101 & 0.114 & 0.175 & 0.202 & 0.142 & 0.637 & 0.737 \\ 
		&  & $l_{M^*}(\psi)$ & 0.021 & 0.016 & 0.101 & 0.103 & 0.067 & 1.025 & 0.957 \\ 
		&  & GEE & 0.007 & 0.003 & 0.081 & 0.080 & 0.053 & 0.992 & 0.957 \\
		\hline
		
		250  & 4 & $l_P(\psi)$ & 0.283 & 0.275 & 0.238 & 0.370 & 0.279 & 0.727 & 0.598 \\ 
		&  & $l_{M^*}(\psi)$ & 0.060 & 0.054 & 0.120 & 0.134 & 0.087 & 1.061 & 0.950 \\ 
		&  & GEE & 0.005 & 0.003 & 0.080 & 0.080 & 0.054 & 1.001 & 0.949 \\
		
		& 6 & $l_P(\psi)$ & 0.143 & 0.167 & 0.205 & 0.250 & 0.197 & 0.518 & 0.548 \\ 
		&  & $l_{M^*}(\psi)$ & 0.022 & 0.018 & 0.091 & 0.093 & 0.061 & 1.020 & 0.955 \\ 
		&  & GEE & 0.003 & 0.001 & 0.065 & 0.065 & 0.043 & 1.004 & 0.951 \\
		
		& 10 & $l_P(\psi)$ & 0.079 & 0.108 & 0.154 & 0.173 & 0.130 & 0.447 & 0.563 \\ 
		&  & $l_{M^*}(\psi)$ & 0.013 & 0.011 & 0.061 & 0.063 & 0.040 & 1.061 & 0.962 \\ 
		&  & GEE & 0.001 & 0.001 & 0.049 & 0.049 & 0.034 & 1.008 & 0.950 \\
		\hline
	\end{tabular}
\end{table}

Inferences on $\beta$ drawn using the MNAR functions and the GEE method for the same MCAR samples are displayed by Table \ref{tab:MCARprob2}. In this probit regression setting the empirical properties of $\mplmc$ are much more favourable than those of the corresponding unmodified function. This is well reflected by all bias and Wald coverage indicators. We observe that, contrary to expectations, for fixed $T$ the bias of the ML estimator decreases with $N$. 
In addition, the MNAR log-likelihoods seem to supply better point estimation but less trustworthy confidence intervals compared to their MCAR counterparts. 
Comments pertaining the performance of the approach based on GEE are along the same lines as those made about Table~2 in Section 4.3 of the paper, dedicated to the logistic regression.
Even with the probit link, $\mplmc$ is more reliable than GEE as the group size grows up and generally succeeds in detecting the underlying ignorable missingness process, which represents a reduced form of the full MNAR model presupposed by that MCMPL.

\begin{table}[t!]\centering
	\caption{Inference on $\beta=1/1.6$ in the probit regression for MNAR longitudinal data. The compared methods are the MNAR profile log-likelihood $l_P(\psi)$, Severini's exact MCAR MPL $l_{M}(\beta)$, the MNAR MCMPL $l_{M^*}(\psi)$ computed with $R=500$ and GEE. Results based on a simulation study with 2000 trials.}
	\label{tab:MNARprob}
	\medskip	
	\begin{tabular}{cccccccccc}
		\hline
		$N$ & $T$ &Method & B & MB & SD & RMSE & MAE & SE/SD & 0.95 CI\\ 
		\hline
		50 & 4 & $l_P(\psi)$ & 0.301 & 0.235 & 0.623 & 0.692 & 0.383 & 0.687 & 0.863 \\ 
		& & $l_{M}(\beta)$ & 0.008 & -0.011 & 0.349 & 0.349 & 0.230 & 0.987 & 0.963 \\ 
		& & $\mplmc$ & 0.012 & -0.022 & 0.405 & 0.405 & 0.233 & 0.857 & 0.964 \\ 
		& & GEE & -0.135 & -0.142 & 0.204 & 0.244 & 0.175 & 1.002 & 0.891 \\
		
		& 6 & $l_P(\psi)$ & 0.163 & 0.131 & 0.351 & 0.387 & 0.230 & 0.775 & 0.884 \\ 
		& & $l_{M}(\beta)$ & -0.072 & -0.091 & 0.245 & 0.255 & 0.175 & 0.973 & 0.928 \\ 
		& & $\mplmc$ & -0.041 & -0.063 & 0.253 & 0.256 & 0.167 & 0.956 & 0.941 \\ 
		& & GEE & -0.140 & -0.146 & 0.164 & 0.215 & 0.162 & 1.015 & 0.854 \\
		
		& 10 & $l_P(\psi)$ & 0.079 & 0.087 & 0.244 & 0.256 & 0.155 & 0.714 & 0.865 \\ 
		& & $l_{M}(\beta)$ & -0.113 & -0.116 & 0.163 & 0.199 & 0.146 & 0.986 & 0.870 \\ 
		& & $\mplmc$ & -0.035 & -0.042 & 0.166 & 0.170 & 0.116 & 0.990 & 0.937 \\ 
		& & GEE & -0.143 & -0.144 & 0.124 & 0.190 & 0.148 & 1.025 & 0.784 \\
		
		\hline
		100 & 4 & $l_P(\psi)$ & 0.254 & 0.240 & 0.390 & 0.466 & 0.287 & 0.729 & 0.834 \\ 
		& & $l_{M}(\beta)$ & -0.018 & -0.024 & 0.237 & 0.238 & 0.157 & 0.976 & 0.939 \\ 
		& & $\mplmc$ & -0.028 & -0.033 & 0.240 & 0.242 & 0.161 & 0.974 & 0.940 \\ 
		& & GEE & -0.142 & -0.149 & 0.134 & 0.195 & 0.154 & 1.062 & 0.829 \\
		
		& 6 & $l_P(\psi)$ & 0.140 & 0.134 & 0.257 & 0.293 & 0.187 & 0.742 & 0.827 \\ 
		& & $l_{M}(\beta)$ & -0.074 & -0.085 & 0.169 & 0.185 & 0.131 & 0.997 & 0.920 \\ 
		& & $\mplmc$ & -0.050 & -0.063 & 0.171 & 0.178 & 0.124 & 1.002 & 0.941 \\ 
		& & GEE & -0.145 & -0.151 & 0.114 & 0.184 & 0.153 & 1.046 & 0.749 \\
		
		& 10 & $l_P(\psi)$ & 0.053 & 0.089 & 0.236 & 0.242 & 0.132 & 0.519 & 0.791 \\ 
		& & $l_{M}(\beta)$ & -0.119 & -0.119 & 0.116 & 0.166 & 0.127 & 0.994 & 0.788 \\ 
		& & $\mplmc$ & -0.037 & -0.042 & 0.119 & 0.124 & 0.085 & 0.990 & 0.928 \\ 
		& & GEE & -0.150 & -0.151 & 0.088 & 0.174 & 0.151 & 1.018 & 0.598 \\
		
		\hline
		250 & 4 & $l_P(\psi)$ & 0.197 & 0.190 & 0.229 & 0.302 & 0.208 & 0.781 & 0.769 \\ 
		& & $l_{M}(\beta)$ & -0.046 & -0.048 & 0.147 & 0.154 & 0.107 & 1.008 & 0.933 \\ 
		& & $\mplmc$ & -0.063 & -0.066 & 0.146 & 0.159 & 0.113 & 1.022 & 0.924 \\ 
		& & GEE & -0.150 & -0.150 & 0.085 & 1.173 & 0.150 & 1.056 & 0.607 \\
		
		& 6 & $l_P(\psi)$ & 0.035 & 0.099 & 0.262 & 0.264 & 0.163 & 0.427 & 0.727 \\ 
		& & $l_{M}(\beta)$ & -0.094 & -0.094 & 0.103 & 0.139 & 0.102 & 1.027 & 0.853 \\ 
		& & $\mplmc$ & -0.070 & -0.071 & 0.105 & 0.127 & 0.088 & 1.017 & 0.890 \\ 
		& & GEE & -0.156 & -0.158 & 0.070 & 0.171 & 0.158 & 1.044 & 0.412 \\
		
		& 10 & $l_P(\psi)$ & -0.022 & 0.081 & 0.282 & 0.283 & 0.130 & 0.260 & 0.665 \\ 
		& & $l_{M}(\beta)$ & -0.132 & -0.132 & 0.069 & 0.149 & 0.132 & 1.051 & 0.545 \\ 
		& & $\mplmc$ & -0.045 & -0.045 & 0.070 & 0.083 & 0.057 & 1.060 & 0.912 \\ 
		& & GEE & -0.161& -0.163 & 0.053 & 0.170 & 0.163 & 1.053 & 0.179 \\ 
		\hline
	\end{tabular}
	
\end{table}

An account of the last simulation experiment is given in Table~\ref{tab:MNARprob}, which is referred to incomplete datasets with MNAR units.
Similarly to what emerged by Table~3 in the paper under the MNAR logistic framework, $\mplmc$ appears to retain higher inferential precision than the analytical MPL which ignores the missingness model, the only exception being the case with $T=4$. 
Thus, when groups are not extremely small, taking the true nonignorable missing-data mechanism into consideration via Monte Carlo simulation is practically translated into improved bias and coverage properties of the estimator based on the MNAR MCMPL. Note that the performance of the MCAR MCMPL is not displayed, like for the logit link function, since it does not significantly differ from that of Severini's $l_{M}(\beta)$.
Furthermore, as expected, the GEE approach is found inconsistent when fitting a binary regression with MNAR response. 
In closing, the use of Monte Carlo approximation to compute the MNAR MPL proves to be generally recommendable in the presence of binary missing observations even under the probit link assumption, thanks to its robustness to the true mechanism of missingness.

Finally, we consider a slightly different simulation setting, in which the incidental parameters are generated in such a way that they are intrinsically correlated with the covariate in the model. 
Specifically, we assume $\lambda_i=\sum_{t=1}^T x_{it}/T+u_i$, where $u_i\sim N(0,1)$ $(i=1,\dots,N)$. In this case, a random effects model would not be correctly specified and thus neither a corresponding marginal model would be available for comparing inference based on GEE.  
Tables \ref{tab:MCARlog.2}--\ref{tab:MNARlog.2} show  for the logistic link function that results based on the fixed effects approach and achieved via the MCMPL are, as expected, as accurate as those obtained under the alternative simulations in Section 4.3 of the paper. The same pattern of performance can be found in Tables \ref{tab:MCARprob.2}--\ref{tab:MNARprob.2} concerning the probit specification.

\section{Weibull regression model for right-censored data}
\label{sec:surv}

Tables~\ref{tab:surv02b1}--\ref{tab:surv04b2} summarize inference on $\beta = (\beta_1, \beta_2)$ obtained through the simulation studies described in Section~5.4 of the paper. In particular, Tables~\ref{tab:surv02b1} and \ref{tab:surv02b2} refer to the first experiment considering an average censoring probability $P_c=0.2$, while Tables~\ref{tab:surv04b1} and \ref{tab:surv04b2} refer to the case $P_c=0.4$.
The results attest the sufficient adequacy of $l_P(\psi)$ in drawing conclusions about $\beta$. However, due to better estimation of the standard errors of $\hat{\beta}_{M^*}=(\hat{\beta}_{1M^*}, \hat{\beta}_{2M^*})$, the MCMPL is still superior in terms of appropriateness of confidence intervals. Hence, 
one can conclude that 
the Monte Carlo adjustment is valuable to further improve the quality of standard ML procedures even when making inference on the regression coefficients under the Weibull modelling framework of Section~5.3.

\bibliographystyle{chicago}
\bibliography{MPLref}

\begin{table}[t]\centering
	\caption{Inference on $\beta=1$ in the logistic regression for MCAR longitudinal data. The compared functions are the MCAR profile log-likelihood $l_P(\beta)$, Severini's exact MCAR MPL $l_{M}(\beta)$, and the MCAR MCMPL $l_{M^*}(\beta)$ computed with $R=500$. Results based on a simulation study with 2000 trials, where $\lambda_i=\sum_{t=1}^T x_{it}/T+u_i$, $u_i\sim N(0,1)$.}
	\label{tab:MCARlog.2}
	\medskip	
	\begin{tabular}{cccccccccc}
		\hline
		$N$ & $T$ &Method & B & MB & SD & RMSE & MAE & SE/SD & 0.95 CI\\ 
		\hline
		50 & 4 & $l_P(\beta)$ & 0.872 & 0.679 & 1.232 & 1.509 & 0.716 & 0.575 & 0.806 \\
		& & $l_{M}(\beta)$ & 0.209 & 0.160 & 0.841 & 0.867 & 0.344 & 0.601 & 0.960 \\
		& & $l_{M^*}(\beta)$ & 0.211 & 0.163 & 0.828 & 0.854 & 0.347 & 0.611 & 0.960 \\ 
		
		& 6 & $l_P(\beta)$ & 0.473 & 0.426 & 0.571 & 0.741 & 0.460 & 0.765 & 0.829 \\ 
		& & $l_{M}(\beta)$ & 0.108 & 0.097 & 0.383 & 0.398 & 0.255 & 0.944 & 0.949 \\
		& & $l_{M^*}(\beta)$ & 0.110 & 0.099 & 0.383 & 0.399 & 0.256 & 0.944 & 0.950 \\ 
		
		& 10 & $l_P(\beta)$ & 0.227 & 0.201 & 0.308 & 0.382 & 0.237 & 0.892 & 0.877 \\ 
		& & $l_{M}(\beta)$ & 0.036 & 0.018 & 0.250 & 0.252 & 0.163 & 0.993 & 0.956 \\
		& & $l_{M^*}(\beta)$ & 0.037 & 0.020 & 0.250 & 0.253 & 0.161 & 0.992 & 0.956 \\ 
		\hline
		100 & 4 & $l_P(\beta)$ & 0.715 & 0.639 & 0.631 & 0.954 & 0.641 & 0.723 & 0.674 \\ 
		& & $l_{M}(\beta)$ & 0.146 & 0.129 & 0.345 & 0.375 & 0.239 & 0.995 & 0.955 \\
		& & $l_{M^*}(\beta)$ & 0.148 & 0.131 & 0.346 & 0.376 & 0.240 & 0.993 & 0.955 \\
		
		& 6 & $l_P(\beta)$ & 0.430 & 0.397 & 0.376 & 0.571 & 0.400 & 0.808 & 0.725 \\
		& & $l_{M}(\beta)$ & 0.088 & 0.073 & 0.262 & 0.276 & 0.174 & 0.971 & 0.950 \\
		& & $l_{M^*}(\beta)$ & 0.089 & 0.073 & 0.262 & 0.277 & 0.174 & 0.971 & 0.948 \\
		
		& 10 & $l_P(\beta)$ & 0.213 & 0.198 & 0.216 & 0.304 & 0.206 & 0.902 & 0.812 \\
		& & $l_{M}(\beta)$ & 0.027 & 0.016 & 0.176 & 0.178 & 0.116 & 1.002 & 0.954 \\
		& & $l_{M^*}(\beta)$ & 0.028 & 0.017 & 0.176 & 0.178 & 0.117 & 1.002 & 0.954 \\ 
		\hline
		250 & 4 & $l_P(\beta)$ & 0.625 & 0.595 & 0.370 & 0.727 & 0.595 & 0.764 & 0.435 \\
		& & $l_{M}(\beta)$ & 0.117 & 0.109 & 0.221 & 0.250 & 0.160 & 0.995 & 0.934 \\ 
		& & $l_{M^*}(\beta)$ & 0.118 & 0.110 & 0.221 & 0.250 & 0.160 & 0.994 & 0.932 \\ 
		
		& 6 & $l_P(\beta)$ & 0.395 & 0.388 & 0.228 & 0.456 & 0.388 & 0.817 & 0.454 \\
		& & $l_{M}(\beta)$ & 0.071 & 0.067 & 0.162 & 0.177 & 0.116 & 0.971 & 0.930 \\
		& & $l_{M^*}(\beta)$ & 0.072 & 0.068 & 0.162 & 0.177 & 0.117 & 0.970 & 0.929 \\
		
		& 10 & $l_P(\beta)$ & 0.214 & 0.210 & 0.134 & 0.253 & 0.210 & 0.906 & 0.583 \\
		& & $l_{M}(\beta)$ & 0.027 & 0.024 & 0.109 & 0.113 & 0.075 & 1.003 & 0.950 \\
		& & $l_{M^*}(\beta)$ & 0.027 & 0.024 & 0.109 & 0.113 & 0.075 & 1.003 & 0.948 \\ 
		\hline
	\end{tabular}
	
\end{table}

\begin{table}[t]\centering
	\caption{Inference on $\beta=1$ in the logistic regression for MCAR longitudinal data. The compared methods are the MNAR profile log-likelihood $l_P(\psi)$, the MNAR MCMPL $l_{M^*}(\psi)$ computed with $R=500$. Results based on a simulation study with 2000 trials, where $\lambda_i=\sum_{t=1}^T x_{it}/T+u_i$, $u_i\sim N(0,1)$.}
	\label{tab:MCARlog2.2}
	\medskip	
	\begin{tabular}{cccccccccc}
		\hline
		$N$ & $T$ &Method & B & MB & SD & RMSE & MAE & SE/SD & 0.95 CI\\ 
		\hline
		50 & 4 & $l_P(\psi)$ & 0.567 & 0.408 & 1.235 & 1.359 & 0.713 & 0.599 & 0.822 \\ 
		&  & $l_{M^*}(\psi)$ & -0.021 & -0.030 & 0.720 & 0.720 & 0.454 & 0.816 & 0.890 \\ 
		
		& 6 & $l_P(\psi)$& 0.357 & 0.326 & 0.615 & 0.711 & 0.435 & 0.757 & 0.859 \\ 
		&  & $l_{M^*}(\psi)$ & 0.004 & 0.018 & 0.452 & 0.452 & 0.275 & 0.913 & 0.934 \\ 
		
		& 10 & $l_P(\psi)$& 0.195 & 0.178 & 0.336 & 0.389 & 0.242 & 0.874 & 0.890 \\ 
		&  & $l_{M^*}(\psi)$ & 0.014 & 0.004 & 0.277 & 0.278 & 0.182 & 0.992 & 0.957 \\ 
		
		\hline
		100 & 4 & $l_P(\psi)$ & 0.467 & 0.422 & 0.727 & 0.864 & 0.537 & 0.687 & 0.789 \\ 
		& & $l_{M^*}(\psi)$ & -0.064 & -0.046 & 0.470 & 0.475 & 0.303 & 0.903 & 0.899 \\ 
		
		& 6 & $l_P(\psi)$& 0.323 & 0.293 & 0.408 & 0.520 & 0.331 & 0.800 & 0.815 \\ 
		&  & $l_{M^*}(\psi)$ & -0.004 & -0.004 & 0.313 & 0.313 & 0.194 & 0.951 & 0.952 \\ 
		
		& 10& $l_P(\psi)$& 0.185 & 0.178 & 0.233 & 0.298 & 0.195 & 0.892 & 0.847 \\ 
		&  & $l_{M^*}(\psi)$ & 0.009 & 0.005 & 0.191 & 0.191 & 0.128 & 1.013 & 0.959 \\ 
		
		\hline
		250 & 4 & $l_P(\psi)$& 0.362 & 0.350 & 0.436 & 0.567 & 0.393 & 0.730 & 0.732 \\ 
		& & $l_{M^*}(\psi)$ & -0.140 & -0.097 & 0.357 & 0.384 & 0.226 & 0.857 & 0.875 \\ 
		
		& 6 & $l_P(\psi)$& 0.289 & 0.284 & 0.248 & 0.381 & 0.290 & 0.805 & 0.674 \\ 
		&  & $l_{M^*}(\psi)$ & -0.011 & -0.006 & 0.188 & 0.189 & 0.117 & 0.972 & 0.950 \\ 
		
		& 10 & $l_P(\psi)$& 0.124 & 0.118 & 0.153 & 0.197 & 0.135 & 0.911 & 0.844 \\ 
		& & $l_{M^*}(\psi)$ & -0.037 & -0.042 & 0.127 & 0.132 & 0.091 & 1.010 & 0.940 \\ 
		\hline
	\end{tabular}
	
\end{table}

\begin{table}[t!]\centering
	\caption{Inference on $\beta=1$ in the logistic regression for MNAR longitudinal data. The compared methods are the MNAR profile log-likelihood $l_P(\psi)$, Severini's exact MCAR MPL $l_{M}(\beta)$, the MNAR MCMPL $l_{M^*}(\psi)$ computed with $R=500$. Results based on a simulation study with 2000 trials, where $\lambda_i=\sum_{t=1}^T x_{it}/T+u_i$, $u_i\sim N(0,1)$.}
	\label{tab:MNARlog.2}
	\medskip	
	\begin{tabular}{cccccccccc}
		\hline
		$N$ & $T$ &Method & B & MB & SD & RMSE & MAE & SE/SD & 0.95 CI\\ 
		\hline
		50 & 4 & $l_P(\psi)$ & 0.330 & 0.190 & 1.207 & 1.251 & 0.621 & 0.638 & 0.892 \\
		& & $l_{M}(\beta)$ & -0.062 & -0.121 & 0.758 & 0.761 & 0.431 & 0.793 & 0.944 \\
		& & $\mplmc$ & -0.106 & -0.147 & 0.826 & 0.833 & 0.449 & 0.750 & 0.933 \\
		
		& 6 & $l_P(\psi)$ & 0.217 & 0.163 & 0.592 & 0.631 & 0.366 & 0.819 & 0.926 \\
		& & $l_{M}(\beta)$ & -0.164 & -0.186 & 0.439 & 0.469 & 0.316 & 0.941 & 0.898 \\
		& & $\mplmc$ & -0.061 & -0.082 & 0.432 & 0.436 & 0.279 & 0.979 & 0.945 \\ 
		
		& 10 & $l_P(\psi)$ & 0.126 & 0.106 & 0.346 & 0.368 & 0.233 & 0.916 & 0.923 \\
		& & $l_{M}(\beta)$ & -0.234 & -0.250 & 0.283 & 0.367 & 0.279 & 1.002 & 0.850 \\
		& & $\mplmc$ & -0.036 & -0.052 & 0.287 & 0.290 & 0.193 & 1.014 & 0.958 \\
		
		\hline
		100 & 4 & $l_P(\psi)$ & 0.291 & 0.228 & 0.670 & 0.730 & 0.412 & 0.755 & 0.886 \\ 
		& & $l_{M}(\beta)$ & -0.107 & -0.133 & 0.418 & 0.431 & 0.291 & 0.951 & 0.922 \\
		& & $\mplmc$ & -0.106 & -0.123 & 0.431 & 0.444 & 0.283 & 0.969 & 0.936 \\
		
		& 6 & $l_P(\psi)$ & 0.191 & 0.164 & 0.395 & 0.439 & 0.269 & 0.865 & 0.905 \\
		& & $l_{M}(\beta)$ & -0.193 & -0.211 & 0.293 & 0.350 & 0.260 & 0.996 & 0.879 \\
		& & $\mplmc$ & -0.071 & -0.087 & 0.294 & 0.302 & 0.207 & 1.013 & 0.947 \\
		
		& 10 & $l_P(\psi)$ & 0.125 & 0.116 & 0.244 & 0.274 & 0.181 & 0.925 & 0.919 \\
		& & $l_{M}(\beta)$ & -0.242 & -0.250 & 0.197 & 0.312 & 0.255 & 1.025 & 0.750 \\
		& & $\mplmc$ & -0.035 & -0.040 & 0.202 & 0.205 & 0.139 & 1.025 & 0.955 \\ 
		
		\hline
		250 & 4 & $l_P(\psi)$ & 0.202 & 0.172 & 0.418 & 0.464 & 0.293 & 0.775 & 0.859 \\
		& & $l_{M}(\beta)$ & -0.160 & -0.178 & 0.274 & 0.317 & 0.231 & 0.942 & 0.872 \\ 
		& & $\mplmc$ & -0.142 & -0.153 & 0.281 & 0.315 & 0.223 & 0.972 & 0.918 \\
		
		& 6 & $l_P(\psi)$ & 0.159 & 0.162 & 0.238 & 0.286 & 0.197 & 0.879 & 0.869 \\
		& & $l_{M}(\beta)$ & -0.215 & -0.213 & 0.180 & 0.280 & 0.217 & 0.996 & 0.761 \\
		& & $\mplmc$ & -0.086 & -0.083 & 0.177 & 0.197 & 0.132 & 1.026 & 0.917 \\ 
		
		& 10 & $l_P(\psi)$ & 0.124 & 0.118 & 0.153 & 0.197 & 0.135 & 0.911 & 0.844 \\
		& & $l_{M}(\beta)$ & -0.258 & -0.259 & 0.123 & 0.286 & 0.259 & 1.010 & 0.458 \\
		& & $\mplmc$ & -0.037 & -0.042 & 0.127 & 0.132 & 0.091 & 1.010 & 0.940 \\
		\hline
	\end{tabular}
\end{table}

\begin{table}[t]\centering
	\caption{Inference on $\beta=1/1.6$ in the probit regression for MCAR longitudinal data. The compared functions are the MCAR profile log-likelihood $l_P(\beta)$, Severini's exact MCAR MPL $l_{M}(\beta)$, and the MCAR MCMPL $l_{M^*}(\beta)$ computed with $R=500$. Results based on a simulation study with 2000 trials, where $\lambda_i=\left(\sum_{t=1}^T x_{it}/T+u_i\right)\!/1.6$, $u_i\sim N(0,1)$.}
	\label{tab:MCARprob.2}
	\medskip	
	\begin{tabular}{cccccccccc}
		\hline
		$N$ & $T$ &Method & B & MB & SD & RMSE & MAE & SE/SD & 0.95 CI\\ 
		\hline
		50 & 4 & $l_P(\beta)$ & 0.592 & 0.484 & 0.693 & 0.911 & 0.495 & 0.605 & 0.752 \\
		& & $l_{M}(\beta)$ & 0.148 & 0.138 & 0.304 & 0.338 & 0.222 & 1.026 & 0.967 \\
		& & $l_{M^*}(\beta)$ & 0.152 & 0.132 & 0.336 & 0.369 & 0.225 & 0.931 & 0.955 \\
		
		& 6 & $l_P(\beta)$  & 0.321 & 0.296 & 0.343 & 0.470 & 0.305 & 0.757 & 0.784 \\ 
		& & $l_{M}(\beta)$ & 0.087 & 0.081 & 0.227 & 0.243 & 0.155 & 0.961 & 0.954 \\
		& & $l_{M^*}(\beta)$ & 0.080 & 0.073 & 0.229 & 0.242 & 0.152 & 0.949 & 0.950 \\
		
		& 10 & $l_P(\beta)$ & 0.162 & 0.146 & 0.180 & 0.242 & 0.157 & 0.904 & 0.831 \\
		& & $l_{M}(\beta)$ & 0.037 & 0.027 & 0.145 & 0.150 & 0.097 & 1.015 & 0.959 \\ 
		& & $l_{M^*}(\beta)$ & 0.033 & 0.022 & 0.144 & 0.148 & 0.095 & 1.015 & 0.960 \\ 
		\hline
		100 & 4 & $l_P(\beta)$ & 0.498 & 0.449 & 0.393 & 0.634 & 0.450 & 0.698 & 0.577 \\ 
		& & $l_{M}(\beta)$ & 0.123 & 0.114 & 0.206 & 0.240 & 0.157 & 1.030 & 0.953 \\ 
		& & $l_{M^*}(\beta)$ & 0.121 & 0.108 & 0.219 & 0.250 & 0.157 & 0.966 & 0.946 \\
		
		& 6 & $l_P(\beta)$ & 0.295 & 0.279 & 0.223 & 0.370 & 0.279 & 0.809 & 0.650 \\ 
		& & $l_{M}(\beta)$ & 0.075 & 0.069 & 0.154 & 0.171 & 0.107 & 0.995 & 0.944 \\ 
		& & $l_{M^*}(\beta)$ & 0.066 & 0.060 & 0.154 & 0.168 & 0.106 & 0.986 & 0.944 \\
		
		& 10 & $l_P(\beta)$ & 0.149 & 0.141 & 0.128 & 0.196 & 0.143 & 0.902 & 0.749 \\ 
		& & $l_{M}(\beta)$ & 0.029 & 0.024 & 0.104 & 0.108 & 0.068 & 1.011 & 0.956 \\ 
		& & $l_{M^*}(\beta)$ & 0.025 & 0.020 & 0.103 & 0.106 & 0.068 & 1.012 & 0.957 \\ 
		\hline
		250 & 4 & $l_P(\beta)$ & 0.429 & 0.414 & 0.221 & 0.483 & 0.414 & 0.763 & 0.322 \\ 
		& & $l_{M}(\beta)$ & 0.103 & 0.100 & 0.130 & 0.166 & 0.115 & 1.041 & 0.914 \\ 
		& & $l_{M^*}(\beta)$ & 0.095 & 0.090 & 0.133 & 0.164 & 0.109 & 1.011 & 0.911 \\ 
		
		& 6 & $l_P(\beta)$ & 0.270 & 0.263 & 0.135 & 0.302 & 0.263 & 0.817 & 0.329 \\
		& & $l_{M}(\beta)$ & 0.063 & 0.060 & 0.095 & 0.114 & 0.077 & 0.991 & 0.913 \\ 
		& & $l_{M^*}(\beta)$ & 0.054 & 0.052 & 0.095 & 0.110 & 0.072 & 0.983 & 0.920 \\ 
		
		& 10 & $l_P(\beta)$ & 0.147 & 0.146 & 0.080 & 0.167 & 0.146 & 0.897 & 0.472 \\ 
		& & $l_{M}(\beta)$ & 0.027 & 0.027 & 0.065 & 0.070 & 0.047 & 1.003 & 0.939 \\ 
		& & $l_{M^*}(\beta)$ & 0.022 & 0.022 & 0.064 & 0.068 & 0.046 & 1.005 & 0.943 \\ 
		\hline
	\end{tabular}
\end{table}

\begin{table}[t]\centering
	\caption{Inference on $\beta=1/1.6$ in the probit regression for MCAR longitudinal data. The compared methods are the MNAR profile log-likelihood $l_P(\psi)$, the MNAR MCMPL $l_{M^*}(\psi)$ computed with $R=500$. Results based on a simulation study with 2000 trials, where $\lambda_i=\left(\sum_{t=1}^T x_{it}/T+u_i\right)\!/1.6$, $u_i\sim N(0,1)$.}
	\label{tab:MCARprob2.2}
	\medskip	
	\begin{tabular}{cccccccccc}
		\hline
		$N$ & $T$ &Method & B & MB & SD & RMSE & MAE & SE/SD & 0.95 CI\\ 
		\hline
		50 & 4 & $l_P(\psi)$ & 0.422 & 0.331 & 0.760 & 0.869 & 0.466 & 0.573 & 0.786 \\ 
		&  & $l_{M^*}(\psi)$ & 0.137 & 0.107 & 0.346 & 0.372 & 0.202 & 0.924 & 0.977 \\ 
		
		& 6 & $l_P(\psi)$ & 0.233 & 0.213 & 0.371 & 0.438 & 0.279 & 0.733 & 0.827 \\ 
		&  & $l_{M^*}(\psi)$ & 0.056 & 0.049 & 0.224 & 0.231 & 0.144 & 0.994 & 0.965 \\ 
		
		& 10 & $l_P(\psi)$ & 0.113 & 0.120 & 0.215 & 0.243 & 0.160 & 0.781 & 0.844 \\ 
		&  & $l_{M^*}(\psi)$ & 0.027 & 0.018 & 0.143 & 0.145 & 0.093 & 1.066 & 0.970 \\ 
		\hline
		
		100 & 4 & $l_P(\psi)$ & 0.359 & 0.318 & 0.446 & 0.573 & 0.362 & 0.670 & 0.729 \\ 
		&  & $l_{M^*}(\psi)$ & 0.099 & 0.085 & 0.208 & 0.230 & 0.140 & 1.037 & 0.967 \\ 
		
		& 6 & $l_P(\psi)$ & 0.180 & 0.180 & 0.266 & 0.321 & 0.218 & 0.695 & 0.746 \\ 
		&  & $l_{M^*}(\psi)$ & 0.033 & 0.025 & 0.150 & 0.154 & 0.098 & 1.041 & 0.962 \\ 
		
		& 10 & $l_P(\psi)$ & 0.095 & 0.109 & 0.178 & 0.201 & 0.136 & 0.664 & 0.774 \\ 
		&  & $l_{M^*}(\psi)$ & 0.017 & 0.011 & 0.101 & 0.103 & 0.067 & 1.067 & 0.970 \\ 
		\hline
		
		250  & 4 & $l_P(\psi)$ & 0.282 & 0.274 & 0.257 & 0.381 & 0.281 & 0.740 & 0.646 \\ 
		&  & $l_{M^*}(\psi)$ & 0.069 & 0.065 & 0.126 & 0.143 & 0.090 & 1.092 & 0.949 \\ 
		
		& 6 & $l_P(\psi)$ & 0.145 & 0.173 & 0.216 & 0.260 & 0.205 & 0.509 & 0.542 \\ 
		&  & $l_{M^*}(\psi)$ & 0.025 & 0.021 & 0.094 & 0.097 & 0.062 & 1.021 & 0.944 \\ 
		
		& 10 & $l_P(\psi)$ & 0.098 & 0.116 & 0.136 & 0.168 & 0.130 & 0.538 & 0.581 \\ 
		&  & $l_{M^*}(\psi)$ & 0.013 & 0.012 & 0.063 & 0.064 & 0.044 & 1.062 & 0.963 \\ 
		\hline
	\end{tabular}
\end{table}

\begin{table}[t]\centering
	\caption{Inference on $\beta=1/1.6$ in the probit regression for MNAR longitudinal data. The compared methods are the MNAR profile log-likelihood $l_P(\psi)$, Severini's exact MCAR MPL $l_{M}(\beta)$, the MNAR MCMPL $l_{M^*}(\psi)$ computed with $R=500$. Results based on a simulation study with 2000 trials, where $\lambda_i=\left(\sum_{t=1}^T x_{it}/T+u_i\right)\!/1.6$, $u_i\sim N(0,1)$.}
	\label{tab:MNARprob.2}
	\medskip	
	\begin{tabular}{cccccccccc}
		\hline
		$N$ & $T$ &Method & B & MB & SD & RMSE & MAE & SE/SD & 0.95 CI\\ 
		\hline
		50 & 4 & $l_P(\psi)$ & 0.309 & 0.217 & 0.802 & 0.860 & 0.436 & 0.592 & 0.851 \\ 
		& & $l_{M}(\beta)$ & 0.011 & -0.013 & 0.456 & 0.456 & 0.265 & 0.846 & 0.957 \\ 
		& & $\mplmc$ & 0.012 & -0.028 & 0.517 & 0.517 & 0.265 & 0.749 & 0.957 \\
		
		& 6 & $l_P(\psi)$ & 0.167 & 0.134 & 0.369 & 0.405 & 0.244 & 0.799 & 0.898 \\
		& & $l_{M}(\beta)$ & -0.069 & -0.082 & 0.266 & 0.275 & 0.186 & 0.964 & 0.924 \\
		& & $\mplmc$ & -0.045 & -0.060 & 0.264 & 0.268 & 0.180 & 0.986 & 0.945 \\
		
		& 10 & $l_P(\psi)$ & 0.070 & 0.077 & 0.247 & 0.256 & 0.153 & 0.745 & 0.875 \\
		& & $l_{M}(\beta)$ & -0.118 & -0.123 & 0.167 & 0.204 & 0.148 & 1.027 & 0.889 \\
		& & $\mplmc$ & -0.043 & -0.052 & 0.168 & 0.173 & 0.118 & 1.039 & 0.948 \\
		
		\hline
		100 & 4 & $l_P(\psi)$ & 0.261 & 0.220 & 0.429 & 0.503 & 0.299 & 0.728 & 0.829 \\
		& & $l_{M}(\beta)$ & -0.014 & -0.023 & 0.257 & 0.258 & 0.171 & 0.986 & 0.950 \\
		& & $\mplmc$ & -0.025 & -0.037 & 0.258 & 0.259 & 0.169 & 0.993 & 0.952 \\ 
		
		& 6 & $l_P(\psi)$ & 0.130 & 0.132 & 0.270 & 0.300 & 0.192 & 0.752 & 0.846 \\
		& & $l_{M}(\beta)$ & -0.085 & -0.092 & 0.177 & 0.197 & 0.140 & 1.019 & 0.918 \\
		& & $\mplmc$ & -0.061 & -0.070 & 0.179 & 0.189 & 0.131 & 1.026 & 0.937 \\
		
		& 10 & $l_P(\psi)$ & 0.034 & 0.078 & 0.247 & 0.249 & 0.133 & 0.519 & 0.837 \\
		& & $l_{M}(\beta)$ & -0.126 & -0.128 & 0.118 & 0.173 & 0.135 & 1.038 & 0.808 \\
		& & $\mplmc$ & -0.046 & -0.051 & 0.121 & 0.129 & 0.089 & 1.031 & 0.930 \\ 
		
		\hline
		250 & 4 & $l_P(\psi)$ & 0.192 & 0.172 & 0.262 & 0.325 & 0.207 & 0.761 & 0.792 \\ 
		& & $l_{M}(\beta)$ & -0.051 & -0.059 & 0.170 & 0.177 & 0.121 & 0.974 & 0.930 \\
		& & $\mplmc$ & -0.071 & -0.081 & 0.170 & 0.184 & 0.130 & 0.984 & 0.921 \\ 
		
		& 6 & $l_P(\psi)$ & 0.081 & 0.115 & 0.221 & 0.235 & 0.151 & 0.551 & 0.768 \\
		& & $l_{M}(\beta)$ & -0.100 & -0.100 & 0.108 & 0.147 & 0.106 & 1.024 & 0.844 \\
		& & $\mplmc$ & -0.078 & -0.079 & 0.110 & 0.135 & 0.093 & 1.025 & 0.885 \\
		
		& 10 & $l_P(\psi)$ & 0.019 & 0.087 & 0.251 & 0.252 & 0.119 & 0.314 & 0.708 \\
		& & $l_{M}(\beta)$ & -0.138 & -0.141 & 0.074 & 0.157 & 0.141 & 1.013 & 0.531 \\
		& & $\mplmc$ & -0.049 & -0.050 & 0.075 & 0.089 & 0.064 & 1.034 & 0.901 \\
		\hline
	\end{tabular}
	
\end{table}

\begin{table}[t]\centering
	\caption{Inference on $\beta_1=-1$ in the stratified Weibull regression model for right-censored survival data and probability of censoring $P_c=0.2$. The compared functions are the profile log-likelihood $l_P(\psi)$ and the MCMPL $l_{M^*}(\psi)$ computed with $R=500$. Results based on a simulation study with 2000 trials.}
	\label{tab:surv02b1}
	\medskip	
	\begin{tabular}{cccccccccc}
		\hline
		$N$ & $T$ &Method & B & MB & SD & RMSE & MAE & SE/SD & 0.95 CI\\ 
		\hline
		50 & 4 & $l_P(\psi)$ & -0.005 & -0.002 & 0.122 & 0.123 & 0.081 & 0.823 & 0.894 \\ 
		&  & $\mplmc$ & 0.004 & 0.005 & 0.121 & 0.121 & 0.080 & 0.970 & 0.937 \\ 
		
		& 6 & $l_P(\psi)$ & -0.003 & -0.003 & 0.096 & 0.096 & 0.066 & 0.880 & 0.906 \\ 
		& & $\mplmc$ & 0.002 & 0.001 & 0.095 & 0.095 & 0.066 & 0.983 & 0.941 \\ 
		
		& 10 & $l_P(\psi)$ & 0.001 & 0.000 & 0.072 & 0.072 & 0.051 & 0.921 & 0.935 \\ 
		&  & $\mplmc$ & 0.003 & 0.004 & 0.072 & 0.072 & 0.052 & 0.982 & 0.951 \\ 
		\hline
		100 & 4& $l_P(\psi)$ & -0.010 & -0.008 & 0.087 & 0.087 & 0.059 & 0.822 & 0.894 \\ 
		&  & $\mplmc$ & -0.002 & -0.001 & 0.086 & 0.086 & 0.058 & 0.971 & 0.943 \\ 
		
		& 6 & $l_P(\psi)$ & -0.004 & -0.004 & 0.068 & 0.068 & 0.046 & 0.885 & 0.915 \\ 
		& & $\mplmc$ & 0.001 & 0.001 & 0.067 & 0.067 & 0.045 & 0.988 & 0.948 \\ 
		
		& 10 & $l_P(\psi)$ & -0.004 & -0.003 & 0.051 & 0.052 & 0.035 & 0.912 & 0.925 \\ 
		&  & $\mplmc$ & -0.002 & -0.001 & 0.051 & 0.051 & 0.035 & 0.976 & 0.942 \\ 
		\hline
		250 & 4 & $l_P(\psi)$ & -0.006 & -0.007 & 0.054 & 0.054 & 0.036 & 0.845 & 0.893 \\ 
		& & $\mplmc$ & 0.002 & 0.002 & 0.053 & 0.053 & 0.035 & 0.995 & 0.943 \\ 
		
		& 6 & $l_P(\psi)$ & -0.006 & -0.006 & 0.043 & 0.043 & 0.029 & 0.893 & 0.917 \\ 
		& & $\mplmc$ & -0.000 & -0.001 & 0.042 & 0.042 & 0.028 & 1.000 & 0.949 \\ 
		
		& 10& $l_P(\psi)$ & -0.002 & -0.002 & 0.031 & 0.031 & 0.022 & 0.946 & 0.939 \\ 
		&  & $\mplmc$ & 0.001 & 0.000 & 0.031 & 0.031 & 0.021 & 1.012 & 0.959 \\ 
		\hline
	\end{tabular}
\end{table}

\begin{table}[t]\centering
	\caption{Inference on $\beta_2=1$ in the Weibull regression model for in the stratified Weibull regression model for right-censored survival data and probability of censoring $P_c=0.2$. The compared functions are the profile log-likelihood $l_P(\psi)$ and the MCMPL $l_{M^*}(\psi)$ computed with $R=500$. Results based on a simulation study with 2000 trials.}
	\label{tab:surv02b2}
	\medskip	
	\begin{tabular}{cccccccccc}
		\hline
		$N$ & $T$ &Method & B & MB & SD & RMSE & MAE & SE/SD & 0.95 CI\\ 
		\hline
		50 & 4 & $l_P(\psi)$& 0.010 & 0.009 & 0.078 & 0.079 & 0.053 & 0.833 & 0.901 \\ 
		& & $\mplmc$ & 0.003 & 0.002 & 0.078 & 0.078 & 0.053 & 0.976 & 0.948 \\
		
		& 6 & $l_P(\psi)$ & 0.006 & 0.004 & 0.054 & 0.055 & 0.038 & 0.887 & 0.920 \\ 
		&  & $\mplmc$ & 0.002 & 0.000 & 0.054 & 0.054 & 0.037 & 0.989 & 0.951 \\ 
		
		& 10 & $l_P(\psi)$ & 0.002 & 0.003 & 0.039 & 0.039 & 0.026 & 0.934 & 0.932 \\ 
		&  & $\mplmc$ & -0.000 & -0.000 & 0.039 & 0.039 & 0.026 & 0.999 & 0.946 \\ 
		\hline
		100 & 4 & $l_P(\psi)$ & 0.007 & 0.007 & 0.052 & 0.053 & 0.035 & 0.820 & 0.891 \\ 
		&  & $\mplmc$ & -0.001 & -0.001 & 0.051 & 0.051 & 0.036 & 0.971 & 0.943 \\ 
		
		& 6& $l_P(\psi)$ & 0.005 & 0.004 & 0.040 & 0.040 & 0.027 & 0.876 & 0.911 \\ 
		&  & $\mplmc$ & -0.000 & -0.000 & 0.039 & 0.039 & 0.027 & 0.981 & 0.945 \\ 
		
		& 10 & $l_P(\psi)$ & 0.003 & 0.003 & 0.028 & 0.028 & 0.018 & 0.936 & 0.926 \\ 
		& & $\mplmc$ & 0.001 & 0.000 & 0.028 & 0.028 & 0.018 & 1.002 & 0.944 \\ 
		\hline
		250 & 4 & $l_P(\psi)$ & 0.009 & 0.009 & 0.034 & 0.035 & 0.024 & 0.824 & 0.886 \\ 
		&  & $\mplmc$ & -0.000 & -0.001 & 0.033 & 0.033 & 0.023 & 0.979 & 0.944 \\ 
		
		& 6 & $l_P(\psi)$ & 0.004 & 0.004 & 0.025 & 0.025 & 0.017 & 0.873 & 0.900 \\ 
		&  & $\mplmc$ & -0.001 & -0.002 & 0.024 & 0.024 & 0.016 & 0.981 & 0.941 \\ 
		
		& 10 & $l_P(\psi)$ & 0.002 & 0.002 & 0.018 & 0.018 & 0.013 & 0.910 & 0.926 \\ 
		&  & $\mplmc$ & -0.000 & -0.000 & 0.018 & 0.018 & 0.013 & 0.972 & 0.948 \\
		\hline
	\end{tabular}
\end{table}

\begin{table}[t]\centering
	\caption{Inference on $\beta_1=-1$ in the stratified Weibull regression model for right-censored survival data and probability of censoring $P_c=0.4$. The compared functions are the profile log-likelihood $l_P(\psi)$ and the MCMPL $l_{M^*}(\psi)$ computed with $R=500$. Results based on a simulation study with 2000 trials.}
	\label{tab:surv04b1}
	\medskip	
	\begin{tabular}{cccccccccc}
		\hline
		$N$ & $T$ &Method & B & MB & SD & RMSE & MAE & SE/SD & 0.95 CI\\ 
		\hline
		50 & 4 & $l_P(\psi)$ & -0.014 & -0.014 & 0.149 & 0.150 & 0.100 & 0.793 & 0.876 \\ 
		&  & $\mplmc$ & 0.001 & 0.003 & 0.146 & 0.146 & 0.096 & 0.974 & 0.940 \\ 
		
		& 6 & $l_P(\psi)$ & -0.008 & -0.008 & 0.116 & 0.116 & 0.078 & 0.860 & 0.908 \\ 
		& & $\mplmc$1 & 0.002 & 0.004 & 0.115 & 0.115 & 0.078 & 0.984 & 0.947 \\ 
		
		& 10 & $l_P(\psi)$ & -0.001 & 0.000 & 0.085 & 0.085 & 0.059 & 0.917 & 0.930 \\ 
		&  & $\mplmc$ & 0.004 & 0.006 & 0.084 & 0.084 & 0.059 & 0.995 & 0.951 \\ 
		\hline
		100 & 4 & $l_P(\psi)$ & -0.016 & -0.016 & 0.107 & 0.108 & 0.073 & 0.786 & 0.878 \\ 
		&  & $\mplmc$ & -0.001 & 0.000 & 0.104 & 0.104 & 0.072 & 0.971 & 0.949 \\ 
		
		& 6 & $l_P(\psi)$ & -0.009 & -0.007 & 0.082 & 0.082 & 0.056 & 0.867 & 0.909 \\ 
		&  & $\mplmc$ & 0.002 & 0.003 & 0.081 & 0.081 & 0.055 & 0.996 & 0.949 \\ 
		
		& 10 & $l_P(\psi)$ & -0.006 & -0.006 & 0.061 & 0.062 & 0.040 & 0.902 & 0.915 \\ 
		&  & $\mplmc$ & -0.002 & -0.000 & 0.061 & 0.061 & 0.040 & 0.981 & 0.943 \\ 
		\hline
		250 & 4 & $l_P(\psi)$ & -0.014 & -0.013 & 0.066 & 0.067 & 0.045 & 0.815 & 0.881 \\ 
		&  & $\mplmc$ & 0.002 & 0.002 & 0.064 & 0.064 & 0.042 & 1.004 & 0.948 \\ 
		
		& 6 & $l_P(\psi)$ & -0.010 & -0.009 & 0.051 & 0.052 & 0.035 & 0.873 & 0.897 \\ 
		&  & $\mplmc$ & 0.001 & 0.002 & 0.051 & 0.051 & 0.033 & 1.007 & 0.945 \\ 
		
		& 10 & $l_P(\psi)$ & -0.005 & -0.004 & 0.038 & 0.038 & 0.025 & 0.925 & 0.922 \\ 
		&  & $\mplmc$ & 0.000 & 0.001 & 0.038 & 0.038 & 0.025 & 1.005 & 0.951 \\
		\hline
	\end{tabular}
\end{table}

\begin{table}[t]\centering
	\caption{Inference on $\beta_2=1$ in the stratified Weibull regression model for right-censored survival data and probability of censoring $P_c=0.4$. The compared functions are the profile log-likelihood $l_P(\psi)$ and the MCMPL $l_{M^*}(\psi)$ computed with $R=500$. Results based on a simulation study with 2000 trials.}
	\label{tab:surv04b2}
	\medskip	
	\begin{tabular}{cccccccccc}
		\hline
		$N$ & $T$ &Method & B & MB & SD & RMSE & MAE & SE/SD & 0.95 CI\\ 
		\hline
		50 & 4 & $l_P(\psi)$ & 0.018 & 0.014 & 0.099 & 0.100 & 0.065 & 0.801 & 0.877 \\ 
		&  & $\mplmc$ & 0.003 & -0.000 & 0.097 & 0.097 & 0.064 & 0.982 & 0.946 \\ 
		
		& 6 & $l_P(\psi)$ & 0.011 & 0.010 & 0.070 & 0.071 & 0.048 & 0.855 & 0.907 \\ 
		& & $\mplmc$ & 0.002 & 0.001 & 0.069 & 0.069 & 0.047 & 0.983 & 0.946 \\ 
		
		& 10 & $l_P(\psi)$ & 0.004 & 0.002 & 0.048 & 0.048 & 0.031 & 0.932 & 0.928 \\ 
		&  & $\mplmc$ & -0.001 & -0.002 & 0.048 & 0.048 & 0.030 & 1.010 & 0.948 \\
		\hline
		100 & 4 & $l_P(\psi)$ & 0.015 & 0.014 & 0.066 & 0.067 & 0.045 & 0.799 & 0.880 \\ 
		&  & $\mplmc$ & 0.000 & -0.002 & 0.064 & 0.064 & 0.045 & 0.985 & 0.946 \\ 
		
		& 6 & $l_P(\psi)$ & 0.010 & 0.011 & 0.050 & 0.051 & 0.035 & 0.851 & 0.899 \\ 
		&  & $\mplmc$ & 0.000 & 0.000 & 0.049 & 0.049 & 0.034 & 0.977 & 0.944 \\ 
		
		& 10 & $l_P(\psi)$ & 0.006 & 0.006 & 0.034 & 0.035 & 0.024 & 0.935 & 0.929 \\ 
		&  & $\mplmc$ & 0.001 & 0.001 & 0.034 & 0.034 & 0.023 & 1.015 & 0.953 \\ 
		\hline
		250 & 4 & $l_P(\psi)$ & 0.017 & 0.017 & 0.042 & 0.045 & 0.030 & 0.810 & 0.864 \\ 
		&  & $\mplmc$ & 0.000 & 0.000 & 0.041 & 0.041 & 0.027 & 1.011 & 0.950 \\ 
		
		& 6 & $l_P(\psi)$ & 0.009 & 0.008 & 0.031 & 0.032 & 0.021 & 0.851 & 0.896 \\ 
		&  & $\mplmc$ & -0.002 & -0.003 & 0.030 & 0.030 & 0.020 & 0.982 & 0.943 \\ 
		
		& 10& $l_P(\psi)$ & 0.005 & 0.005 & 0.022 & 0.023 & 0.016 & 0.911 & 0.919 \\ 
		& & $\mplmc$ & -0.000 & -0.001 & 0.022 & 0.022 & 0.015 & 0.986 & 0.942 \\  
		\hline
	\end{tabular}
\end{table}